\documentclass[8.5pt,twoside,twocolumn]{article}
\usepackage[super,sort&compress,comma]{natbib} 
\usepackage{mhchem}
\usepackage{times}
\numberwithin{equation}{section} 
\usepackage{textcomp} 
\usepackage{amssymb} 
\usepackage{gensymb} 
\usepackage{sectsty}
\usepackage{balance}
\usepackage{graphicx} 
\usepackage{lastpage}
\usepackage[format=plain,justification=raggedright,singlelinecheck=false,font=small,labelfont=bf,labelsep=space]{caption} 
\usepackage{fancyhdr}
\begin{document}

\thispagestyle{plain}
\fancypagestyle{plain}{}
\renewcommand{\thefootnote}{\fnsymbol{footnote}}
\renewcommand\footnoterule{\vspace*{1pt}%
\hrule width 3.4in height 0.4pt \vspace*{5pt}} 
\setcounter{secnumdepth}{5}

\makeatletter 
\def\subsubsection{\@startsection{subsubsection}{3}{10pt}{-1.25ex plus -1ex minus -.1ex}{0ex plus 0ex}{\normalsize\bf}} 
\def\paragraph{\@startsection{paragraph}{4}{10pt}{-1.25ex plus -1ex minus -.1ex}{0ex plus 0ex}{\normalsize\textit}} 
\renewcommand\@biblabel[1]{#1}            
\renewcommand\@makefntext[1]%
{\noindent\makebox[0pt][r]{\@thefnmark\,}#1}
\makeatother 
\renewcommand{\figurename}{\small{Fig.}~}
\sectionfont{\large}
\subsectionfont{\normalsize} 

\fancyfoot{}
\fancyhead{}
\renewcommand{\headrulewidth}{1pt} 
\renewcommand{\footrulewidth}{1pt}
\setlength{\arrayrulewidth}{1pt}
\setlength{\columnsep}{6.5mm}
\setlength\bibsep{1pt}

\twocolumn[
  \begin{@twocolumnfalse}
\noindent\LARGE{\textbf{Gap and channelled plasmons in tapered grooves: a review}}
\vspace{0.6cm}

\noindent\large{\textbf{C. L. C. Smith,\textit{$^{a}$} N. Stenger,\textit{$^{b,c}$} A. Kristensen,\textit{$^{a}$} N. A. Mortensen,\textit{$^{b,c}$} and
S. I. Bozhevolnyi,\textit{$^{d}$}}}\vspace{0.5cm}

\vspace{0.6cm}

\noindent \normalsize{Tapered metallic grooves have been shown to support plasmons -- electromagnetically coupled oscillations of free electrons at metal-dielectric interfaces -- across a variety of configurations and V-like profiles. Such plasmons may be divided into two categories: gap-surface plasmons (GSPs) that are confined laterally between the tapered groove sidewalls and propagate either along the groove axis or normal to the planar surface, and channelled plasmon polaritons (CPPs) that occupy the tapered groove profile and propagate exclusively along the groove axis. Both GSPs and CPPs exhibit an assortment of unique properties that are highly suited to a broad range of cutting-edge nanoplasmonic technologies, including ultracompact photonic circuits, quantum-optics components, enhanced lab-on-a-chip devices, efficient light-absorbing surfaces and advanced optical filters, while additionally affording a niche platform to explore the fundamental science of plasmon excitations and their interactions. In this Review, we provide a research status update of plasmons in tapered grooves, starting with a presentation of the theory and important features of GSPs and CPPs, and follow with an overview of the broad range of applications they enable or improve. We cover the techniques that can fabricate tapered groove structures, in particular highlighting wafer-scale production methods, and outline the various photon- and electron-based approaches that can be used to launch and study GSPs and CPPs. We conclude with a discussion of the challenges that remain for further developing plasmonic tapered-groove devices, and consider the future directions offered by this select yet potentially far-reaching topic area.}
\vspace{0.5cm}
 \end{@twocolumnfalse}
  ]

\section{Introduction}
\label{sec:Introduction}
\footnotetext{\textit{$^{a}$~Department of Micro- and Nanotechnology, Technical University of Denmark, DK-2800, Kgs. Lyngby, Denmark. E-mail: cameron.smith@nanotech.dtu.dk; anders.kristensen@nanotech.dtu.dk }}
\footnotetext{\textit{$^{b}$~Department of Photonics Engineering, Technical University of Denmark, DK-2800, Kgs. Lyngby, Denmark. E-mail: niste@fotonik.dtu.dk; asger@mailaps.org }}
\footnotetext{\textit{$^{c}$~Center for Nanostructured Graphene (CNG), Technical University of Denmark, DK-2800, Kgs. Lyngby, Denmark. }}
\footnotetext{\textit{$^{d}$~Institute of Technology and Innovation (ITI), University of Southern Denmark, DK-5230, Odense M, Denmark. E-mail: seib@iti.sdu.dk }}

Plasmonics represents a burgeoning subdiscipline of nanophotonics that considers the control of light at the nanoscale based on the nature of plasmons.\cite{Maier2001,Zia2006,Lal2007,Schuller2010,Han2013} Plasmons, considered here in the form of propagating surface-plasmon polaritons (SPPs), are oscillations of the free electron gas at metal-dielectric interfaces coupled to electromagnetic (EM) fields.\cite{Raether1988} These EM fields decay exponentially into the neighbouring media, and through various techniques their distribution may be controlled to facilitate the concentration of light beyond its diffraction limit.\cite{Takahara1997,Nerkararyan,Barnes2003,Gramotnev2010,Gramotnev2013} Realisations of sub-diffraction-limited light have given rise to a host of scientifically and technologically significant optical components, including integrable ultra-compact photonic circuits,\cite{Bozhevolnyi2006c,Chen2006,Pyayt2008,Papaioannou2011,Kauranen2012} subwavelength nanolasers,\cite{Bergman2003,Hill2007,Oulton2009,Noginov2009,Khajavikhan2012} quantum photonics devices,\cite{Altewischer2002a,Akimov2007a,Kumar2013,Tame2013} enhanced filters and structurally coloured surfaces,\cite{Martin-Moreno2001,Genet2007,Laux2008,Roberts2014,Clausen2014} nanoscale volume and single-molecule sensors,\cite{Kneipp2002,Anker2008,Mayer2010,Chung2011a,Feng2012,Zijlstra2012} super-resolution nanoscopes and nanoprobes,\cite{Fang2005,Hell2007,Xiong2007,Liu2011,Schuck2013} and near-field traps for the manipulation of single molecules, atoms, and nano-objects.\cite{Grigorenko2008,Juan2011,Erickson2011,Pang2012a,Marago2013}

The achievements rendered by plasmonics have been made possible by advances in fabrication techniques that can now routinely synthesise and pattern metals on the nanoscale.\cite{Maier2005,Barnes2006,Ebbesen2008,Boltasseva2009} This is due to the fact that the topology and configuration of plasmonic nanostructures strongly determine the degree of EM field enhancement enabled by plasmons.\cite{Moreno2006a,Gramotnev2010,Gramotnev2013} In particular, the proximity between metal surfaces,\cite{Economou1969,Wang2004,Veronis2005,Tanaka2005,Dionne2006,Nerkararyan2011} the sharpness at the ends of tapered points,\cite{Nerkararyan,Babadjanyan2000,Volkov2009,Gramotnev2011,Desiatov2011} adiabatic scale transitions\cite{Stockman2004,Pile2006a,Issa2006,Verhagen2009,Bozhevolnyi2010} and the constituent materials involved\cite{Oulton2008,West2010,Goykhman2010,Yang2011,Naik2013} all govern the degree of light confinement in space. Accordingly, a diverse set of structure types are pursued in order to nanofocus light, including nanoparticles,\cite{Quinten1998,Maier2003,Anker2008,Chung2011a} nanoholes,\cite{Genet2007,Laux2008,Xu2010,Pang2012a} nanopatches,\cite{Henzie2007,Puscasu2008,Qu2013} nanoantennas,\cite{Bryant2008,Kawata2009,Schnell2009,Novotny2011} nanowires,\cite{Akimov2007a,Pyayt2008,Verhagen2009,Huck2011} and an extensive list of planar waveguiding geometries.\cite{Bozhevolnyi2006c,Lal2007,Han2013,Gramotnev2013} 

\begin{figure*}[t!] 
   \centering
    \includegraphics[width=17.4cm,height=7.83cm]{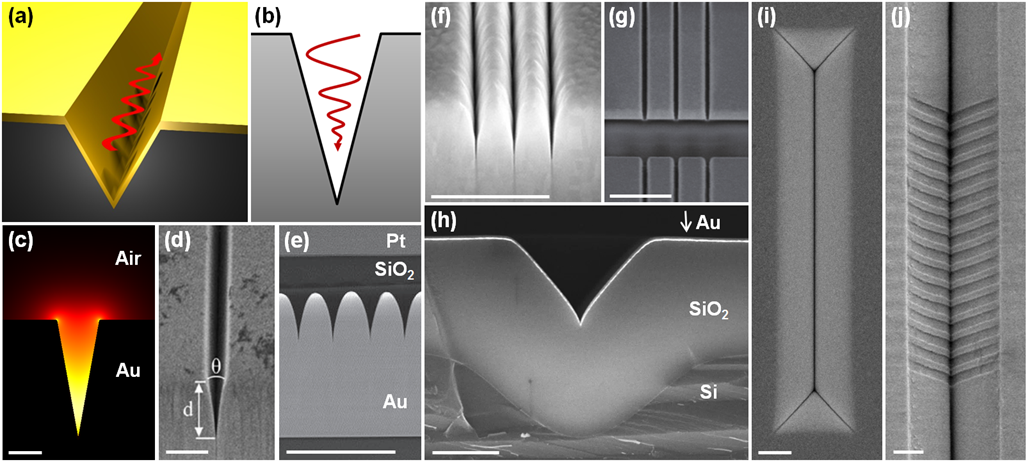}
    \captionsetup{justification=justified}
    \caption{A variety of plasmonic tapered-groove geometries. (a) Illustration of a plasmonic tapered-groove, where a V-like channel is introduced into an otherwise planar metal surface. The red oscillating arrow depicts the propagation of a plasmon along the groove axis. (b) Schematic representation of a GSP propagating normal to the planar surface. (c) The normalized electric field distribution of a CPP mode solved by the finite element method at the free space telecommunications wavelength of {1.55~$\mu$m}. (d)-(j) Scanning electron micrographs of assorted gold tapered-groove structures. (d) A FIB-milled tapered-groove with its end-facet at the edge of a gold slab.\cite{Zenin2011a} (e) A lamellar consisting of a set of tapered grooves used for studying extreme confinement behaviour under electron energy-loss spectroscopy.\cite{Raza2014a} (f) Efficient broadband light-absorbers given by a series of convex-shaped grooves.\cite{Skovsen2013a} (g) Tapered periodic slits in a gold film for enhancing extraordinary optical transmission.\cite{Beermann2011} (h) A tailored tapered-groove profile realised by oxidation of a anisotropically-defined silicon V-groove.\cite{Smith2014} (i) A similar device as shown in (h); waveguide termination mirrors form innate to the anisotropic process. (j) A Bragg grating filter in a plasmonic tapered-groove produced by nanoimprint lithography.\cite{Smith2012} All scale bars are $1~\mu\text{m}$. (Adapted with permission from refs.\cite{Zenin2011a,Raza2014a,Skovsen2013a,Beermann2011,Smith2014,Smith2012} Copyright \textcopyright{} 2011 Optical Society of America, \textcopyright{} 2014 Macmillan Publishers Ltd, \textcopyright{} 2013 American Institute of Physics, \textcopyright{} 2011 IOP Publishing \& Deutsche Physikalische Gesellschaft, \textcopyright{} 2014 American Chemical Society, \textcopyright{} 2012 Optical Society of America).}
    \label{fgr:Intro}
\end{figure*}

In this Review, we selectively consider the assortment of geometries and configurations of metallic tapered grooves that have been shown to support SPP-type modes;\cite{Volkov2007a,Nielsen2008a,Volkov2009,Radko2011a,Søndergaard2012} a sample of structures is displayed in Fig.~\ref{fgr:Intro}.\cite{Zenin2011a,Raza2014a,Skovsen2013a,Beermann2011,Smith2014,Smith2012} Figure~\ref{fgr:Intro}(a) illustrates a representative tapered-groove -- a V-shaped profile introduced into an otherwise planar metallic surface -- with a plasmon propagating along the groove axis. The ensemble of plasmons supported by tapered grooves may be divided into two principal categories. First, gap-surface plasmons (GSPs),\cite{Søndergaard2010b,Søndergaard2010c,Søndergaard2012,Raza2014a} which are laterally confined modes between the tapered-groove sidewalls and may propagate either along the groove axis [Fig.~\ref{fgr:Intro}(a)] or normal to the planar surface [Fig.~\ref{fgr:Intro}(b)]. Second, channelled plasmon polaritons (CPPs), whose EM fields occupy the tapered profile and propagate along the groove axis;\cite{Pile2004a,Bozhevolnyi2005a,Bozhevolnyi2006c,Bozhevolnyi2007a} Fig.~\ref{fgr:Intro}(c) shows an example of a CPP mode obtained by a finite-element simulation (COMSOL Multiphysics) at telecommunications wavelengths. Figures~\ref{fgr:Intro}(d)-(j) highlight the variety of configurations of metallic tapered grooves so-far available, with the selection shown here offering new opportunies for strongly confined photonic circuitry,\cite{Zenin2011a} scientific studies of extreme EM field confinement,\cite{Raza2014a} broadband light absorbers and polarizers,\cite{Skovsen2013a} enhanced extraordinary optical transmission,\cite{Beermann2011} UV-defined nanoplasmonic components with efficient plasmon excitation,\cite{Smith2014} and sophisticated nanoimprinted plasmonic waveguides.\cite{Smith2012} 

The aim of this Review is to introduce readers to the collection of plasmonic tapered-groove structure types so-far established and provide a research status update of their fundamentals, applications, synthesis and operation methods. We start in section~\ref{sec:Plasmons_in_V-shaped_grooves} with a brief presentation of the theory and important features of GSPs and CPPs, highlighting the latest developments\cite{Bozhevolnyi2010,Søndergaard2012} in the context of initial results.\cite{Lu1990,Novikov2002a,Moreno2006a,Bozhevolnyi2006b,Bozhevolnyi2007a,Vernon2008a,Bozhevolnyi2008a} We outline the potential applications afforded by GSP and CPP characteristics in section~\ref{sec:Applications}, emphasising recent reports on effective nanophotonic circuitry,\cite{Dintinger2009a,Zenin2011a,Lee2011a,Smith2012,Zenin2012a,Bian2013a,Burgos2014} efficient quantum emitter to plasmonic mode coupling,\cite{Vesseur2010,Martin-Cano2010,Martin-Cano2011,Gonzalez-Tudela2011,Vernon2012a} adiabatic nanofocusing,\cite{Søndergaard2012,Skovsen2013a,Beermann2013a,Søndergaard2013a,Beermann2014} enhanced extraordinary optical transmission,\cite{Søndergaard2010c,Beermann2011,Søndergaard2011} resonant absorption\cite{Zhang2011,Zhang2012b} and field enhancement,\cite{Søndergaard2010b,Søndergaard2010c,Beermann2011,Søndergaard2013a,Rose2014,Odgaard2014} and nano-opto-mechanics.\cite{Shalin2014a} We follow in section~\ref{sec:Design_&_Fabrication} with an overview of the methods to fabricate plasmonic tapered grooves, concentrating on key results that now enable low cost, wafer-scale production of exceptional quality devices.\cite{Fernandez-Cuesta2007a,Nielsen2008a,Smith2012,Smith2014} We review in section~\ref{sec:Excite_&_Characterise} the existing techniques used to excite and study GSPs and CPPs, making a particular mention of direct, normal illumination arrangements that yield high plasmon launch efficiencies using waveguide termination mirrors,\cite{Radko2011a,Smith2014} and electron-based approaches that have yielded new insights into optically dark modes of nano-scale structures via electron energy-loss spectroscopy (EELS).\cite{Raza2014a} We discuss in section~\ref{sec:Challenges} the challenges that remain for further developing tapered-groove-based devices, noting that recent work elucidating non-local effects\cite{Toscano2012,Toscano2013} and the presence of antisymmetric modes\cite{Raza2014a} represent potentially crucial factors that should be accounted for in device design. The Review concludes in section~\ref{sec:Conclusion_&_Outlook} with an outlook of the future directions offered by this rapidly advancing topic area.

\section{Plasmons in tapered grooves: fundamentals}
\label{sec:Plasmons_in_V-shaped_grooves}
In this section we present the fundamental physics and important features of SPP-type modes in tapered grooves. We consider two principal categories: GSPs and CPPs, for which both may be described as EM modes that exist within the interior of the metallic V-like profiles. Of these, we first cover GSPs -- plasmons in metal-insulator-metal (MIM) type configurations -- which may either propagate along the tapered-groove axis [Fig.~\ref{fgr:Intro}(a)], or normal to the planar surface [Fig.~\ref{fgr:Intro}(b)]. We then follow with CPPs, a subset of GSP eigenmodes propagating along the tapered-groove axis, which consist of EM fields occupying the V-like profile [Fig.~\ref{fgr:Intro}(c)]. We distinguish between GSPs and CPPs propagating along the groove by their EM distributions: GSPs may be \textit{locally} excited such that the difference in spacing of the gap in the vertical direction is negligible, whereas CPP modes necessarily experience the gap-width variation of the tapered-groove shape. Nevertheless, GSPs and CPPs are both SPP-type modes, so before proceeding we summarise the relevant elements of SPPs. 

\subsection{Surface-plasmon polaritons}

\begin{figure}[t!] 
   \centering
    \includegraphics[width=8.6cm,height=3.58cm]{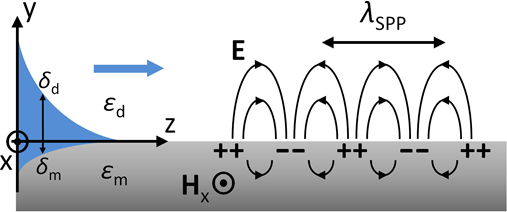}
    \captionsetup{justification=justified}
    \caption{Illustration of the field components of a SPP supported by a metal-dielectric interface. The mode is transverse magnetic, exhibiting electric field components normal to the interface ($y$-axis) and along the propagation direction ($z$-axis). The mode profile (blue) represents the magnitude of the corresponding magnetic field, itself oriented along the $x$-axis, as a function of $y$.}
    \label{fgr:SPP}
\end{figure}

SPPs are modes bound to and propagating along a metal-dielectric interface consisting of transverse magnetic EM waves (electric field oriented along the $y$-axis) that decay exponentially into both the metal and dielectric media\cite{Barnes2003,Maier2007} (Fig.~\ref{fgr:SPP}). The tangential field components (along the $z$-axis) are continuous across the interface and correspond to the wavevector $k_z$ such that the SPP electric fields can be written as follows:
\begin{subequations} 
\begin{align}
\label{eq:EzSPPd}
\textit{\textbf{E}}(y > 0) &= 
\begin{pmatrix}
0 \\ 
E_y^\text{d} \\
E_z^0
\end{pmatrix}
{e^{{i(k_{\text{SPP}}z-{\omega}t)}}}{e^{-y\sqrt{k_{\text{SPP}}^2-{\varepsilon}_{\text{d}}k_0^2}}}, \\ 
\label{eq:EzSPPm}
\textit{\textbf{E}}(y < 0) &= 
\begin{pmatrix}
0 \\ 
E_y^\text{m} \\
E_z^0
\end{pmatrix}
{e^{i(k_{\text{SPP}}z-{\omega}t)}}{e^{y\sqrt{k_{\text{SPP}}^2-{\varepsilon}_{\text{m}}k_0^2}}}, 
\end{align} 
\end{subequations}
where $E_z^0$ and $E_y^\text{d(m)}$ are the amplitudes of the electric field components in the dielectric (metal) medium, $\varepsilon_{\text{m}}$ and $\varepsilon_{\text{d}}$ are the dielectric constants of the metal and dielectric, respectively, $k_0$ is the free-space light wavenumber ($k_0$ = $\omega$/$c$ = 2$\pi$/$\lambda$, where $\omega$ is the angular frequency and $\lambda$ is the light wavelength in free space) and the SPP propagation constant, $k_{\text{SPP}}$ = $k_z$, is governed by the SPP dispersion relation, $k_{\text{SPP}}(\omega)$, yet to be determined. For reference, throughout this review we define the $z$-axis to be either the direction of SPP propagation for flat-surface structures or the groove axis for V-like structures, and the $y$-axis to be the direction normal to the planar surface. 

The SPP EM waves are coupled to the free electrons of the metal surface and give rise to collectively oscillating charge separations that correspond to the EM fields. Since the electric field components should satisfy Coulomb's law $\nabla{\,}{\cdot}{\,}\textit{\textbf{E}} = 0$, transverse electric EM waves (electric field oriented along the $x$-axis) cannot couple to the tangential electron oscillations along the $z$-axis at the metal surface and form SPPs. The normal field components (along the $y$-axis) can then be written in terms of the mutual tangential component:
\begin{subequations} 
\begin{align}
\label{eq:EzExd}
E_y^\text{d} &= \frac{ik_{\text{SPP}}}{\sqrt{k_{\text{SPP}}^2-\varepsilon_{\text{d}}k_0^2}}E_z^0, \\
\label{eq:EzExm}
E_y^\text{m} &= -\frac{ik_{\text{SPP}}}{\sqrt{k_{\text{SPP}}^2-\varepsilon_{\text{m}}k_0^2}}E_z^0.
\end{align} 
\end{subequations}
By using the boundary condition $\varepsilon_{\text{d}}E_x^\text{d} =  \varepsilon_{\text{m}}E_x^\text{m}$, the SPP dispersion relation can be found: 
\begin{equation}
\label{eq:kSPP}
k_{\text{SPP}}(\omega) = k_0(\omega)\sqrt{\frac{\varepsilon_{\text{m}}(\omega)\varepsilon_{\text{d}}(\omega)}{\varepsilon_{\text{m}}(\omega)+\varepsilon_{\text{d}}(\omega)}},
\end{equation}
where $\varepsilon_{\text{m}}(\omega)$ and $\varepsilon_{\text{d}}(\omega)$ denote the individual metal and dielectric material dispersions respectively, and it is evident that $k_{\text{SPP}}(\omega) \gg k_0(\omega)$ for EM waves of the same frequency propagating in the dielectric. 

The SPP propagation constant relates to the effective refractive index of the SPP mode by $N_{\text{eff}} = k_{\text{SPP}}\lambda/(2\pi)$, with the real and imaginary parts determining the SPP wavelength, $\lambda_{\text{SPP}} = 2\pi/\text{Re}\{\textit{k}_{\text{SPP}}\}$, and the intensity-based propagation length, $L_{\text{SPP}}$ = 1/\text{Im}\{2\textit{k}$_{\text{SPP}}\}$. We can see that the SPP wavelength is shorter than the light wavelength in the dielectric, and using the free-electron (Drude) approximation for the metal permittivity, $\varepsilon_{\text{m}}(\omega) = 1-(\omega_{\text{p}}/\omega(\omega+i\gamma))^2$ where $\omega_{\text{p}}$ is the plasma frequency and $\gamma$ is the damping coefficient of the electron oscillations, the SPP wavelength approaches zero at the high frequency (short wavelength) limit of the dispersion relation, i.e. when $\omega \rightarrow \omega_{\text{p}}/\sqrt{1+\varepsilon_{\text{d}}}$.\cite{Economou1969,Raether1988,Maier2007} The propagation length is also an important factor for SPP-based devices since it is significantly limited by $\gamma$ as a result of Ohmic dissipation. This leads to short propagation lengths that restrict the available length-scales for realising SPP-based devices, and may otherwise generate unwanted heat. Nevertheless, there are major efforts to address plasmonic losses,\cite{Noginov2008,Leon2010} (together with critical discussions on the topic)\cite{Khurgin2011,Khurgin2014} and the efficient conversion of EM fields to heat via plasmons has been harnessed to great effect in various configurations.\cite{Hirsch2003,Ndukaife2014}

The SPP amplitudes of the normal and tangential electric field components in the dielectric and metal can be explicitly related as follows [equations (\ref{eq:EzExd}) and (\ref{eq:EzExm})]: 
\begin{subequations} 
\begin{align}
\label{eq:EzExdAmp}
E_y^\text{d} &= i\sqrt{-\varepsilon_{\text{m}}/\varepsilon_{\text{d}}}E_z^0, \\
\label{eq:EzExmAmp}
E_y^\text{m} &= -i\sqrt{-\varepsilon_{\text{d}}/\varepsilon_{\text{m}}}E_z^0.
\end{align} 
\end{subequations}
Generally, metals have large magnitudes of the dielectric permittivity such that $| \varepsilon_{\text{m}} | \gg \varepsilon_{\text{d}}$, resulting in the normal component being greater in the dielectric and the tangential component being greater in the metal. This is representative of the hybrid nature of SPPs, which consist of EM waves propagating through the dielectric and free-electron oscillations in the metals.\cite{Bozhevolnyi2009} We remark here that since the SPP damping is caused by ohmic dissipation, it is therefore the tangential electric field component that generally determines the SPP losses.\cite{Han2013}

The penetration depth of the SPP electric fields (distance from the interface where the field amplitude decreases to 1/$e$ of its value at $y = 0$) into both the dielectric $\delta_{\text{d}}$ and metal $\delta_{\text{m}}$ are different, as given by equations (\ref{eq:EzSPPd}) and (\ref{eq:EzSPPm}). Together with the dispersion relation [equation (\ref{eq:kSPP})], $\delta_{\text{d(m)}}$ is found to be:
\begin{equation}
\label{eq:deltadm}
\delta_{\text{d(m)}} = | k_y^{\text{d(m)}} |^{-1} = \frac{\lambda}{2\pi}{\,}{\cdot}{\,}\sqrt{\left| \frac{\varepsilon_{\text{m}}+\varepsilon_{\text{d}}}{\varepsilon_{\text{d(m)}}^2} \right|},
\end{equation}
with $\varepsilon_{\text{d}}\delta_{\text{d}} = |\varepsilon_{\text{m}}|\delta_{\text{m}}$. Accordingly, the penetration depth is limited to tens of nanometres into the metal while for the dielectric it is larger and highly dependent on the wavelength. Substituting the Drude approximation into equation (\ref{eq:deltadm}) results in the following expressions for the long-wavelength limit:\cite{Bozhevolnyi2009}
\begin{subequations} 
\begin{align}
\label{eq:deltadLW}
\delta_{\text{d}} &\approx (\lambda/2\pi)^2{\,}{\cdot}{\,}(\omega_{\text{p}}/c{\,}\varepsilon_{\text{d}}), \\
\label{eq:deltamLW}
\delta_{\text{m}} &\approx c/\omega_{\text{p}}.
\end{align} 
\end{subequations}
The above relations [equations (\ref{eq:deltadm}) and (\ref{eq:deltadLW}, \ref{eq:deltamLW})] can be used to estimate the fraction of the SPP electric field in the dielectric as being proportional to $(| \varepsilon_{\text{m}} |/\varepsilon_{\text{d}})^2$,\cite{Han2013} which at a first glance may seem surprising from a photonics perspective, since typically light experiences greater confinement within dielectric regions of higher refractive index. This trait of SPPs indicates that the electric field increases its overlap with the metal for shorter wavelengths, and therefore dielectrics of higher refractive index result in a shorter SPP propagation length and a reduced penetration depth, $\varepsilon_{\text{d}}$.

We note here that this brief subsection on SPPs has only summarised the details relevant to the topics of the Review. More detailed information can be found throughout the comprehensive set of literature available.\cite{Barnes2003,Lal2007,Maier2007,Pitarke2007,Oulton2008a,Gramotnev2010,Stockman2011,Han2013,Gramotnev2013}

\subsection{Gap-surface plasmons}

\begin{figure}[t!] 
   \centering
    \includegraphics[width=8.6cm,height=4.22cm]{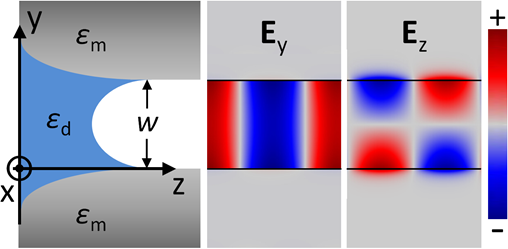}
    \captionsetup{justification=justified}
    \caption{Illustration and numerically-solved field components of a symmetric GSP mode propagating along the $z$-axis supported by two metal surfaces with a dielectric gap of width $w$ between. The corresponding electric field has its normal component, $E_y$, maintaining the same sign across the gap with an even-symmetric profile, whereas the tangential component, $E_z$, consequently exhibits an odd-symmetric profile. The calculated electric fields were obtained with the aid of finite-element method simulations (scale normalized for clarity) using gold surfaces with an air gap of width $w = 0.5~{\mu}\text{m}$ at $\lambda = 775~\text{nm}$ $(n_\text{m} = 0.174+4.86i)$.\cite{Palik}}
    \label{fgr:GSPdiag}
\end{figure}

We now consider SPP modes in the MIM configuration consisting of two close metal surfaces of the same material separated by a dielectric gap of width $w$ (Fig.~\ref{fgr:GSPdiag}). The SPPs associated with the individual metal surfaces interact with each other, and the SPP-type mode surviving all values of the gap width, $w$, is the symmetric GSP, which exhibits odd symmetry of the tangential electric field component, $E_z$, and an even symmetry of the normal field component, $E_y$. This configuration, often called the plasmonic slot waveguide, has sparked an extensive research interest owing to the possibility it offers for strong mode confinement while achieving relatively low Ohmic dissipation\cite{Dionne2006,Dionne2006a} and bending losses,\cite{Pile2005} making them well-suited to planar photonic integration.\cite{Bozhevolnyi2005a,Bozhevolnyi2006c,Khurgin2012}

\begin{figure}[t!] 
   \centering
    \includegraphics[width=8.6cm,height=5.8cm]{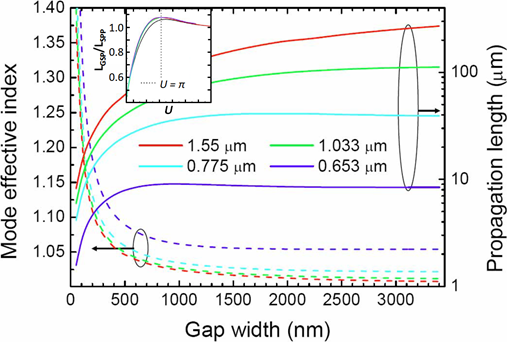}
    \captionsetup{justification=justified}
    \caption{The GSP mode effective refractive index and its propagation length as a function of the width, $w$, of the gap region (air) between the metal (gold) surfaces for several light wavelengths.\cite{Bozhevolnyi2006b} Under certain conditions, the propagation length may \textit{increase} for \textit{decreasing} gap width. Inset: The ratio of the GSP and SPP propagation lengths, $L_{\text{GSP}}/L_{\text{SPP}}$, as a function of the normalised gap width, $u = k_y^dw$.\cite{Bozhevolnyi2007a} (Adapted with permission from refs.\cite{Bozhevolnyi2006b,Bozhevolnyi2007a} Copyright \copyright{} 2006 Optical Society of America, \copyright{} 2007 Springer).}
    \label{fgr:GSPeim}
\end{figure}

The GSP mode effective refractive indices and their corresponding propagation lengths were numerically solved (Fig.~\ref{fgr:GSPeim}) as a function of gap width for light wavelengths selected between the visible and telecommunications.\cite{Bozhevolnyi2006b} A major feature of the symmetric GSP is that its propagation constant, $k_{\text{GSP}}$, increases indefinitely as the gap width approaches zero, indicating that the mode can be indefinitely squeezed in its lateral cross-section, albeit at the cost of generally increasing losses. It is interesting to note that the GSP propagation length first increases when the gap width decreases from large values that correspond to uncoupled separate SPPs. This may seem counterintuitive at first, since it implies that \textit{GSP modes with greater confinement may experience longer propagation lengths}.\cite{Bozhevolnyi2006b} The inset of Fig.~\ref{fgr:GSPeim} shows that the longest propagation length, $L_{\text{GSP}}$, is ${\sim}8\%$ larger than the individual SPP propagation length, $L_{\text{SPP}}$, $(w \rightarrow \infty)$. The optimal gap width in this respect can be explained as the two electric field components of the GSP, $E_y$ and $E_z$, approaching the electrostatic (capacitor) mode, where the larger component, $E_y$, is constant across the gap. Here, the fraction of the electric field energy located in the gap region first increases (lossy component in the metal decreases) as $w$ decreases, before reaching its maximum and then decreasing as the field is squeezed from the gap into the metal. As such, the GSP configuration optimally exploits the available dielectric space of the gap region to minimise dissipative losses in the metal.

\begin{table*}[t!]
\small
  \caption{\ Summary of analytical approximations for the GSP propagation constant, $k_{\text{GSP}}$, under different conditions of gap width, $w$\cite{Bozhevolnyi2007a,Bozhevolnyi2008a} } 
  \label{tbl:kGSPapproximations}
  \begin{tabular*}{1\textwidth}{@{\extracolsep{\fill}} l p{2.7cm} l l}  
    \hline\noalign{\smallskip}
    Gap width & Condition            & Approximation                 & Analytical Expression \\
    \noalign{\smallskip}\hline\noalign{\smallskip}
    Small     & $w \rightarrow 0$    & $\text{tanh}x \approx x$      & 
$\begin{aligned}
k_{\text{GSP}} \approx k_0\sqrt{
\begin{aligned}
 & \varepsilon_{\text{d}} + 0.5(k_{\text{GSP}}^0/k_0)^2  + \text{...} \\
 & \sqrt{(k_{\text{GSP}}^0/k_0)^2[\varepsilon_{\text{d}} - \varepsilon_{\text{m}} + 0.25(k_{\text{GSP}}^0/k_0)^2]}
\end{aligned}} \\
\end{aligned}$ \\[0.7cm]  
    Large     & Individual SPPs start to interact 
                                     & $\tanh x \approx 1-2\exp(2x)$ & 
$\begin{aligned}
k_{\text{GSP}} \approx k_\text{SPP}\sqrt{1-\frac{4\varepsilon_{\text{m}}\varepsilon_{\text{m}}}{\varepsilon_{\text{m}}^2-\varepsilon_{\text{d}}^2}\exp(-k_y^dw)} \\
\end{aligned}$ \\[0.6cm] 
    Moderate  & $w > (\lambda\varepsilon_\text{d})/(\pi|\varepsilon_\text{m}|)$  
                                     & $\text{tanh}x \approx x$      & 
$\begin{aligned}
k_{\text{GSP}} \approx k_0\sqrt{\varepsilon_\text{d}+\frac{2\varepsilon_\text{d}\sqrt{\varepsilon_\text{d}-\varepsilon_\text{m}}}{(-\varepsilon_\text{m})k_0w}}
\end{aligned}$ \\[0.5cm] 
    \hline
  \end{tabular*}
\end{table*}

Applying the appropriate boundary conditions of the electric field components for SPPs and the configurations of symmetry described above, the dispersion relation for $k_{\text{GSP}}$ may be derived\cite{Economou1969,Zia2004,Raza2013}
\begin{equation} 
\begin{split}
\label{eq:dispRelGSP}
\tanh \left(\frac{k_y^{\text{d}}w}{2}\right) &= -\frac{\varepsilon_{\text{d}}k_y^{\text{m}}}{\varepsilon_{\text{m}}k_y^{\text{d}}}(1 + \delta_{\text{nl}}),{\;}{\;}{\;}{\;}{\;}\text{with} \\ 
k_y^{\text{d(m)}} &= \sqrt{k_\text{GSP}^2 - \varepsilon_{\text{d(m)}}k_0^2},
\end{split} 
\end{equation}
where $\delta_{\text{nl}}$ is a corrective term accounting for non-local effects which will be discussed in more detail in section~\ref{sec:Challenges}.\cite{Raza2013} For now, the local response approximation is considered $(\delta_{\text{nl}} \rightarrow 0)$. Equation~(\ref{eq:dispRelGSP}) implies that, unlike the case for SPPs, $k_\text{GSP}$ cannot be expressed in an explicit form. However, it is highly desirable to find an explicit expression for the GSP propagation constant in order to design and assess a variety of GSP configurations. 

\begin{figure}[t!] 
   \centering
    \includegraphics[width=8.6cm,height=5.82cm]{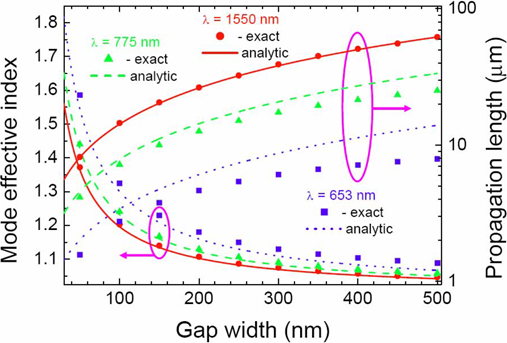}
    \captionsetup{justification=justified}
    \caption{The GSP mode effective refractive index and its propagation length as a function of the width of the gap region (air) between the metal (gold) surfaces for several light wavelengths calculated exactly [equation~(\ref{eq:dispRelGSP})] and using the analytical moderate-gap approximation (Table~\ref{tbl:kGSPapproximations}). (Reprinted with permission from ref.\cite{Bozhevolnyi2008a} Copyright \copyright{} 2008 Optical Society of America).}
    \label{fgr:GSPexVan}
\end{figure}

Table~\ref{tbl:kGSPapproximations} summarises several analytical approximations for $k_{\text{GSP}}$ under different conditions of the gap width, $w$.\cite{Bozhevolnyi2007a,Bozhevolnyi2008a} For the small-gap approximation, $k_{\text{GSP}}^0 = -(2\varepsilon_{\text{d}})/(w\varepsilon_{\text{m}})$, which represents the GSP propagation constant for vanishing gaps $(w \rightarrow 0)$. Here, the real part of the corresponding effective refractive index, $\text{Re}\{k_{\text{GSP}}\lambda/(2\pi)\}$, becomes much larger than the dielectric refractive index, while the imaginary part also increases. For larger gaps where the two SPP modes barely interact, the first-order corrected expression is given, where higher-order corrections may be carried out by iterating the values for $k_{\text{GSP}}$ into the normal wavevector component $k_y^{\text{d}} = \sqrt{k_\text{SPP}^2-\varepsilon_\text{d}k_0^2}$ and recalculating $k_{\text{GSP}}$.\cite{Bozhevolnyi2007a}

Using the large-gap approximation and evaluating $k_{\text{GSP}}$ with only the major electric field components, the normalised fraction of the electric field energy inside the gap can be expressed as a function of the normalised gap width, $u = k_y^{\text{d}}w$:
\begin{equation} 
\begin{split}
\label{eq:normGapWidth}
G(u) &\propto \frac{\exp{(-k_yw)}[2k_yw + \exp{(k_yw)}-\exp{(-k_yw)}]}{(k_y/k_y^{\text{d}})(1 + \exp{(-k_yw)})^2} \\
\text{with}{\:}{\:}{\:} k_y &= k_y^{\text{d}}\sqrt{1 + 4\exp{(-k_y^{\text{d}}w)}}.
\end{split} 
\end{equation}
This function reaches its maximum value, $6\%$ larger than $w \rightarrow \infty$, at $u \approx 3.48$, resulting in the following expression for the optimum gap width: $w_\text{opt} \approx 0.5\lambda|\text{Re}\{\varepsilon_\text{m}\}+1|^{0.5}$. It is interesting to note that the condition ensuring longest GSP propagation length is close to $u = \pi$ (inset of Fig.~\ref{fgr:GSPeim}), which resembles the resonator condition where $k_y^{\text{d}}$ is the wave-vector component in the dielectric pointing across the gap.\cite{Bozhevolnyi2007a}

If the different gap-dependent terms in the small-gap approximation are considered, then the GSP propagation constant may be approximated for moderate gap widths, defined as $|k_\text{GSP}^0|<k_0 \Leftrightarrow w>(\lambda\varepsilon_\text{d})/(\pi|\varepsilon_\text{m}|)$. The GSP mode effective refractive indices and propagation lengths are plotted as a function of the gap width in Fig.~\ref{fgr:GSPexVan}. As before, several wavelengths spanning from visible to telecommunications regimes are considered, using both the exact dispersion relation from equation (\ref{eq:dispRelGSP}) and the moderate-gap approximation.\cite{Bozhevolnyi2008a} The validity of the moderate-gap approximation and its applicability for analysing various GSP configurations has been demonstrated in work exploring the plasmonic analogs of black holes\cite{Nerkararyan2011} and, as will be discussed below, the adiabatic nanofocusing of radiation in narrowing gaps.\cite{Bozhevolnyi2009b,Bozhevolnyi2010,Bozhevolnyi2010a}

\subsubsection{Adiabatic nanofocusing of GSPs. }
\begin{figure}[b!] 
   \centering
    \includegraphics[width=8.6cm,height=4cm]{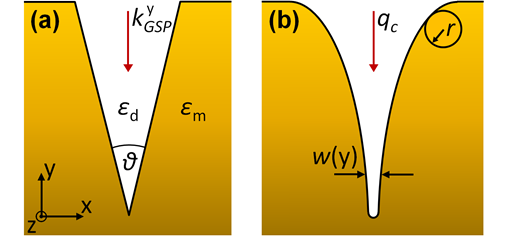}
    \captionsetup{justification=justified}
    \caption{Parameters defining plasmonic V-grooves consisting of (a) linear (b) convex-shaped tapers for GSPs propagating normal to the planar surface (along the $y$-axis). ((b) is adapted from ref.\cite{Søndergaard2012} Copyright \copyright{} 2012 Macmillan Publishers Ltd.)}
    \label{fgr:VandConvex}
\end{figure}

The physical processes of concentrating GSPs are strongly influenced by how rapidly the structure's gap width, $w$, varies in space, given by its taper angle, $\theta$ (Fig.~\ref{fgr:VandConvex}). For sufficiently small $\theta$, the adiabatic approximation becomes valid,\cite{Stockman2004,Gramotnev2005a} which represents the condition where the mode does not ``feel'' the structural taper and simply adjusts its properties according to the local $w$. Here, reflection and scattering become negligible due to the taper and consequently the losses are primarily Ohmic.

The adiabatic parameter may be described mathematically as\cite{Stockman2004,Gramotnev2005a} 
\begin{equation}\label{eq:adiabCond}
\delta(y) = \left| \frac{\text{d}(\text{Re}\{k_{\text{GSP}}^y\})^{-1}}{\text{d}y} \right| \ll 1
\end{equation}
where the $y$-axis is parallel to the taper direction [Fig.~\ref{fgr:VandConvex}(a)], $\text{Re}\{k_{\text{GSP}}^y\}$ is the $y$-component of the real part of the GSP propagation constant, and weak GSP dissipation is assumed: $\text{Re}\{k_{\text{GSP}}\} \gg \text{Im}\{k_{\text{GSP}}\}$. The adiabatic condition signifies that a small change of $k_{\text{GSP}}$ occurs within the length of a GSP period. It has been further shown that the condition of (\ref{eq:adiabCond}) for tapered waveguides can usually be relaxed to\cite{Pile2006a}
\begin{equation}\label{eq:adiabCondRelax}
\delta(y) \lesssim 1.
\end{equation}
In the continuous-matter approximation, the adiabatic parameter $\delta(y)$ reaches its maximum (finite) at the (infinitely) sharp tip of a linear taper.\cite{Gramotnev2005a,Gramotnev2007,Vernon2007a} Accordingly, if the adiabatic condition is held at the tip, it will be satisfied everywhere.\cite{Gramotnev2005a,Gramotnev2007} Assuming the above relation (\ref{eq:adiabCondRelax}) is valid, and using asymptotic analysis for the GSP propagation near the structure's tapered end (tip), the following analytical expression can be found for the critical taper angle:\cite{Gramotnev2005a,Gramotnev2007}
\begin{equation}\label{eq:adiabCritAngle}
\theta_c \approx -2\varepsilon_{\text{d}}/\text{Re}\{\varepsilon_{\text{m}}\}.
\end{equation}
For tapered geometries consisting of a taper angle below the critical angle, $\theta<\theta_c$, the adiabatic condition holds.

As an example, let us consider a GSP propagating vertically towards the bottom of a gold V-groove in air with a taper angle $\theta$. Using equation (\ref{eq:adiabCritAngle}), a light wavelength of $\lambda = 775~\text{nm}$ results in a critical taper angle of $\theta_c \approx 5\degree$, and for $\lambda = 1550~\text{nm}$ it is $\theta_c \approx 1\degree$. Such narrow V-grooves with straight sidewalls can be difficult to produce in practice, however, and results on V-grooves of larger taper angles $(\theta>10\degree)$ report \textit{resonant} GSP nanofocusing, where reflection from the groove sidewalls is significant (i.e. not adiabatic) and becomes an important aspect of the local field enhancement.\cite{Søndergaard2009a,Søndergaard2010b,Beermann2011} 

In order to explore the adiabatic condition further, the moderate-gap approximation can be rewritten in the following form:\cite{Bozhevolnyi2009a}
\begin{equation}\label{eq:dispGSPmodQc}
\begin{split}
q_{\text{c}}(w) &\approx \sqrt{k^2 + \frac{2k\chi}{w}};{\:}{\:}{\:}\text{with} \\
k &= k_0\sqrt{\varepsilon_{\text{d}}},{\:}{\:}{\:}\chi = \frac{\sqrt{\varepsilon_{\text{d}}(\varepsilon_{\text{d}}-\varepsilon_{\text{m}})}}{-\varepsilon_{\text{m}}},
\end{split}
\end{equation}
where $q_{\text{c}}$ is the complex GSP propagation constant. As $w(y)$ decreases towards zero at the bottom of the groove (taper), both the real and imaginary parts of $q_{\text{c}}$ increase indefinitely, representing the slow down of the GSP mode and its subsequent escalation of dissipation by Ohmic losses. We will discuss this further in section~\ref{sec:Applications} regarding the design of efficient light-absorbing materials. Following the form of equation~\ref{eq:dispGSPmodQc}, the adiabatic condition can similarly be rewritten as
\begin{equation}\label{eq:adiabCondq}
\delta(y) = \left| \frac{\text{d}(q^{-1})}{\text{d}y} \right| = \frac{1}{q^2}\left|\frac{\text{d}q}{\text{d}w}\right|{\cdot}\left|\frac{\text{d}w}{\text{d}y}\right| \ll 1,
\end{equation}
where $q$ is the real part of the GSP propagation constant that can be used for metals with $\varepsilon' \ll \varepsilon''$ and $\varepsilon' \ll \varepsilon_{\text{d}}$ according to
\begin{equation}\label{eq:dispGSPmodq}
q(w) \approx \sqrt{k^2 + \frac{2k\chi'}{w}};{\:}{\:}{\:}\text{with}{\:}{\:}{\:}\chi' = \sqrt{\frac{\varepsilon_{\text{d}}}{\varepsilon'}}.
\end{equation}
By using equation (\ref{eq:dispGSPmodq}) for the adiabatic condition expressed in equation (\ref{eq:adiabCondq}) for the limit of $w(y) \rightarrow 0$ (where variations of $q$ become strong), the final representation of the adiabatic condition can be obtained, given as\cite{Søndergaard2012}
\begin{equation}\label{eq:adiabCondqFinal}
\delta(y) = \left| \frac{\text{d}(q^{-1})}{\text{d}y} \right| \approx \frac{1}{4}\sqrt{\frac{\lambda\sqrt{\varepsilon'}}{\pi\varepsilon_{\text{d}}w(y)}}\left|\frac{\text{d}w}{\text{d}y}\right| \ll 1.
\end{equation}
This relation implies that, in the case of linear V-grooves [Fig.~\ref{fgr:VandConvex}(a)] where $\text{d}w/\text{d}y = \text{constant}$, the adiabatic condition must inevitably break down near the groove bottom as the groove width becomes sufficiently small, causing resonant behaviour and GSP reflection.\cite{Søndergaard2009a,Søndergaard2010b,Beermann2011} However, the condition may be satisfied for the entire groove depth under the configuration of convex-shaped grooves [Fig.~\ref{fgr:VandConvex}(b)], where the groove walls become parallel at the contact point much like the case of touching cylinders.\cite{Nerkararyan2011} In the case of cylinders of equal radii $R$, the adiabatic condition reduces to\cite{Søndergaard2012} 
\begin{equation}\label{eq:adiabCondqCylinder}
\delta(y) \approx \sqrt{\frac{\lambda\sqrt{\varepsilon'}}{4\pi{}R\varepsilon_{\text{d}}}} \ll 1,
\end{equation}
which is met in practice for radii much larger than the light wavelength. The importance of this result harkens to the possibility of achieving \textit{non}resonant light absorption, where the sensitivity to the light's wavelength and angle of incidence are greatly reduced. Furthermore, the focusing of light via cross-sections that are large relative to the geometry significantly improves the absorption efficiency\cite{Søndergaard2012,Beermann2013a,Søndergaard2013a,Skovsen2013a,Odgaard2014} and is a notable advantage with respect to other nonresonant geometries such as sub-wavelength cylinders\cite{Aubry2010} or spheres.\cite{Fernandez-Dominguez2010}

\subsection{Channel-plasmon polaritons}

\begin{figure}[t!] 
   \centering
    \includegraphics[width=8.6cm,height=5.74cm]{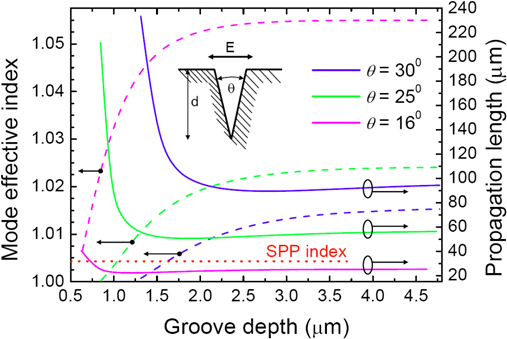}
    \captionsetup{justification=justified}
    \caption{The effective refractive indices of fundamental CPP modes at $\lambda = 1.55~{\mu}\text{m}$ and their propagation lengths as a function of the groove depth for different groove angles, $\theta$. The insert shows the groove configuration and dominant orientation of the CPP electric field. The planar SPP effective refractive index is also included. (Reprinted with permission from ref.\cite{Bozhevolnyi2006b} Copyright \copyright{} 2006 Optical Society of America.)}
    \label{fgr:VgrEIM}
\end{figure}

We now consider the class of GSP eigenmodes that propagate along the tapered-groove axis: CPPs. From an Effective Index Method (EIM) perspective,\cite{Bozhevolnyi2006b} the tapered groove can be represented as a stack of infinitesimally thick MIM waveguides, each with a positional-dependent insulator width, $w(y)$, between a top dielectric layer and bottom metal layer. The local GSP effective refractive index of each stack is determined by its $w(y)$, increasing towards the bottom of the groove and resulting in confinement of the CPP mode to within the structure. 

The CPP effective refractive index increases and the propagation length correspondingly decreases as the groove angle, $\theta$, is reduced (Fig.~\ref{fgr:VgrEIM}).\cite{Hocker1977,Bozhevolnyi2005a,Bozhevolnyi2006b} The CPP mode effective refractive index, $N_{\text{eff}}$, determines the mode confinement since the mode penetration depth in the dielectric is given by $\delta_{\text{d}} = (\lambda/2\pi)(N_{\text{eff}}^2-1)^{-0.5}$. Similar to the SPP, the CPP mode is more tightly confined to the metal (bottom of the groove) when $N_{\text{eff}}$ is larger, and also when the groove angle is narrower. The result of this means \textit{yet further mode confinement for progressively smaller tapered-groove cross-sections}\cite{Bozhevolnyi2007a} -- a feature made possible by the nature of plasmons -- although at the cost of larger ohmic losses. One can also see that the CPP propagation length starts to rapidly increase when the decreasing groove depth reaches a certain (cut-off) value, below which no CPP modes exist. Here, the CPP mode effective refractive index approaches that of the dielectric and corresponds to the extension of the CPP mode profile progressively out of the groove.

\begin{figure}[t!] 
   \centering
    \includegraphics[width=8.6cm,height=3.99cm]{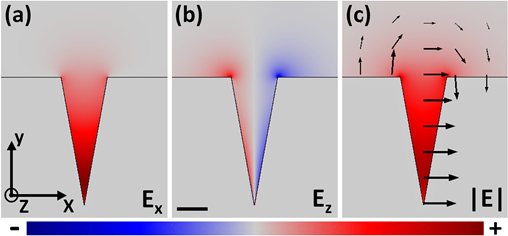}
    \captionsetup{justification=justified}
    \caption{A CPP mode calculated via finite-element method for a gold V-groove in air at $\lambda = 1.55~{\mu}\text{m}$. (a) The dominant electric field component, $E_x$, points across the groove. The scale bar is $1~{\mu}\text{m}$. (b) The electric field component oriented along the direction of propagation, $E_z$, highlights the antisymmetric (i.e. capacitor-like) charge distribution of the CPP mode. The field amplitude scales for $E_x$ and $E_z$ in (a) and (b) are normalised for clarity; the true maximum of $E_x$ in (a) is significantly greater. (c) The magnitude of the normalised electric field component, $|E|$, represents the resulting CPP. Here, the mode closely resembles the distribution of $E_x$, although regions located near the opening corners (wedges) are comprised of significant $E_z$ (and $E_y$) components. The arrows represent the electric field lines in the $xy$ plane.} 
    \label{fgr:VgrCPP}
\end{figure}

Using the finite-element method (COMSOL Multiphysics), we numerically solve the CPP mode for a gold V-groove in air (tip rounding curvature radius: $5~\text{nm}$) at a light wavelength of $\lambda = 1.55~{\mu}\text{m}$ $(n_{\text{m}} = 0.55+11.5i)$ and plot various electric field components in Fig.~\ref{fgr:VgrCPP}. The dominant component of the electric field, $E_x$, is essentially perpendicular to the metal sidewalls which corresponds to the expected behaviour of individual SPPs on planar surfaces [Fig.~\ref{fgr:VgrCPP}(a)]. Like SPPs and GSPs, CPPs also possess a component in the direction of propagation, $E_z$, where Fig.~\ref{fgr:VgrCPP}(b) shows an antisymmetric profile about the V-groove geometry and represents the capacitor-like charge distribution of the mode. The overall normalised magnitude of the CPP electric field, $|E|$, is given in Fig.~\ref{fgr:VgrCPP}(c) together with the logarithmically-scaled electric field strengths (black arrows) in the $xy$-plane. One can readily see that the whole CPP mode closely resembles the distribution of $E_x$, although locations near the opening corners (rounding curvature radii: $50~\text{nm}$) are comprised of significant $E_z$ (and $E_y$) components since $E_x$ is not supported by the parallel surface plane.\cite{Han2013}

The EIM approach has been shown to provide sound qualitative and quantitative insights into the behaviour of CPPs,\cite{Polemi2011} offering the advantage of simplicity over preceding work (i.e. Green's theorem first in the electrostatic limit\cite{Lu1990} and then later for propagating modes).\cite{Novikov2002a} Taking this further, and in conjunction with the explicit expression for the propagation constant of the moderate-gap approximation (Table \ref{tbl:kGSPapproximations}), the EIM framework was used to derive useful \textit{analytic} formulas for CPP mode distributions\cite{Bozhevolnyi2009a,Bozhevolnyi2009b} in the form of
\begin{equation}\label{eq:analytCPPform}
E_x(x,y,z) = X(x,y)Y(y)\exp{(ik_{\text{CPP}}z)},
\end{equation}
where $X(x,y)$ and $Y(y)$ represent factors of the CPP mode distributions, $k_{\text{CPP}}$ is the CPP propagation constant and $E_x$ is the dominant electric field orientation. Since equation (\ref{eq:dispGSPmodQc}) constitutes the well-known GSP field distribution, $X[x,w(y)]$, it then becomes a case of using several approximations\cite{Bozhevolnyi2009a} to arrive at the remaining components of the analytic solution:
\begin{equation} \label{eq:analytCPP-k_n}
k_{\text{CPP},n} = k\sqrt{1 + \left[\frac{\chi}{\theta{}(n + 1)}\right]^2},{\:}{\:}{\:}n = 0,1,2,{\:}...{\:};
\end{equation}
for which the mode field distributions may then be found recursively:

\begin{figure}[b!] 
   \centering
    \includegraphics[width=8.6cm,height=2.17cm]{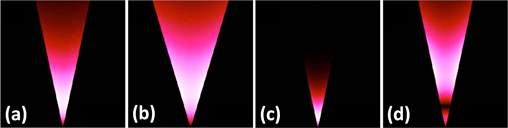}
    \captionsetup{justification=justified}
    \caption{(a)-(c) Fundamental and (d) second CPP mode field magnitude distributions for gold V-grooves in air obtained using equations (\ref{eq:analytCPP-k_n}) and (\ref{eq:analytCPP-recurs}) for different groove angles: (a),(c),(d) $\theta = 25 \degree$ and (b) $\theta = 35 \degree$, and wavelengths: (a),(b) $\lambda = 1.55~{\mu}\text{m}$ and (c),(d) $\lambda = 1.033~{\mu}\text{m}$. All panels have a lateral size of $6~{\mu}\text{m}$. (Adapted with permission from ref.\cite{Bozhevolnyi2009a} Copyright \copyright{} 2009 Optical Society of America.)} 
    \label{fgr:CPPanalyt}
\end{figure}

\begin{equation} \label{eq:analytCPP-recurs}
\begin{split}
Y_n(y) &= \exp{\left[-\frac{k|\chi|y}{\theta{}(n + 1)}\right]}\sum_{\upsilon=0}^{n}{\alpha_{\upsilon}y^{\upsilon+1}},\\
\text{with}{\:}{\:}{\:}\alpha_{\upsilon+1} &= 2\alpha_{\upsilon}\frac{k|\chi|}{\theta}\frac{(\upsilon-n)}{(\upsilon+1)(\upsilon+2)(n+1)},
\end{split}
\end{equation}
where $\alpha_{\upsilon}$ are normalisation constants. 

\begin{figure}[t!] 
   \centering
    \includegraphics[width=8.6cm,height=3.6cm]{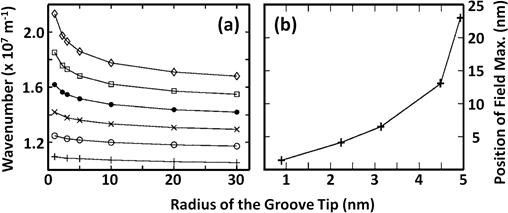}
    \captionsetup{justification=justified}
    \caption{CPP mode dependencies for a silver V-groove in air at $\lambda = 632.8~\text{nm}$, $\varepsilon_{\text{m}} = -16.22+0.52i$, as a function of tip radius. (a) The CPP wavenumber for a groove angle of $\theta = 30 \degree$ for different values of the dielectric permittivity in the groove: $\varepsilon_{\text{d}} = 1$ $(+)$, $\varepsilon_{\text{d}} = 1.21$ $(\circ)$, $\varepsilon_{\text{d}} = 1.44$ $(\times)$, $\varepsilon_{\text{d}} = 1.69$ $(\bullet)$, $\varepsilon_{\text{d}} = 1.96$ $(\Box)$, and $\varepsilon_{\text{d}} = 2.25$ $(\diamond)$. (b) The location at which the maximum of the CPP field occurs away from the tip into the dielectric (along the $z$ coordinate) for a groove angle of $\theta = 45 \degree$. (Adapted with permission from ref.\cite{Vernon2008a} Copyright \copyright{} 2008 American Institute of Physics.)} 
    \label{fgr:TipRad}
\end{figure}

Equations (\ref{eq:analytCPP-k_n}) and (\ref{eq:analytCPP-recurs}) have been used to calculate the CPP mode distributions for gold V-grooves in air (Fig.~\ref{fgr:CPPanalyt}),\cite{Bozhevolnyi2009a} reflecting the influence of groove angle and light wavelength $(n_{\text{m}} = 0.272+7.07i{\;}\text{at}{\;}1.033~\mu{}\text{m})$, and bearing a close resemblence to those calculated numerically.\cite{Gramotnev2004a,Moreno2006a,Yan2007a} The CPP mode profiles and propagation constants can, on one hand, expect to be similar for finite-depth grooves provided that the planar surface occurs sufficiently far away from the mode maximum. On the other hand, a cut-off depth where CPP modes are no longer supported can be evaluated as a function of the position of the mode maximum for the fundamental CPP mode, using the following condition as a normalised depth coordinate $\xi_{\text{co}} = k|\chi|d/~\theta>1$, where $d$ is the depth of the groove. Alternatively, this condition may be conveniently expressed by making use of the normalised waveguide parameter, $V_{\text{CPP}}$:\cite{Bozhevolnyi2008a,Bozhevolnyi2009a,Bozhevolnyi2009b}
\begin{equation}\label{eq:Vparam}
V_{\text{CPP}} = 4\sqrt{\frac{\pi{}d\varepsilon_{\text{d}}\sqrt{|\varepsilon_{\text{d}}-\varepsilon_{\text{m}}|}}{\lambda\theta|\varepsilon_{\text{m}}|}} > 2\sqrt{2},
\end{equation}
where single-moded operation occurs within the criteria: $0.5\pi<V_{\text{CPP}}(w,d)<1.5\pi$.\cite{Bozhevolnyi2008a} 

The dependence of wavelength on the fundamental CPP mode distribution is represented in Figs.~\ref{fgr:CPPanalyt}(a) and \ref{fgr:CPPanalyt}(c),\cite{Bozhevolnyi2009a} and also extensively covered in ref.\cite{Moreno2006a} Generally, shorter light wavelengths excite modes with stronger mode confinement to the groove bottom, and longer wavelengths lead to shifting of the CPP mode up towards the groove opening. This process leads to hybridisation with the wedge plasmon polariton (WPP) pair -- modes supported at the opening corners of the groove\cite{Pile2005a,Moreno2008,Oulton2008a,Boltasseva2008a,Dintinger2009a,Bian2011a} -- with progressively longer wavelengths resulting in increased portions of the electric field in the WPPs. 

The role of the cladding material filling the V-groove, for which an increase of its dielectric permittivity, $\varepsilon_{\text{d}}$, was shown to result in a rapid increase of the mode localisation to the bottom tip of the groove [Fig.~\ref{fgr:TipRad}(a)] while also leading to larger critical angles where CPP modes can exist,\cite{Vernon2008a} was reported in refs.\cite{Vernon2008a,Srivastava2009a} This behaviour corresponds to the similar physics governing the penetration depths of SPPs, $\delta_{\text{d(m)}}$ in equation (\ref{eq:deltadm}), where a larger $\varepsilon_{\text{d}}$ results in a larger electric field overlap with the metal, $\delta_{\text{m}}$. Although the CPP ohmic losses are greater for larger $\varepsilon_{\text{d}}$, designing tapered grooves with cladding materials other than air represents potentially important findings with regards to e.g. relaxed tapered-groove fabrication criteria, applications related to sensing or other lab-on-a-chip (LOC) devices, and efforts regarding gain-assisted loss compensation.\cite{Leon2010}

A significant parameter to account for is the finite corner rounding at the bottom tip of the groove, which profoundly affects the CPP mode profile and necessarily occurs in all practically fabricated devices. It was shown numerically that less-sharp corners result in less tightly confined modes (Fig.~\ref{fgr:TipRad}),\cite{Vernon2008a} indicating that corner rounding increases the penetration depth into the dielectric for progressively larger radii of curvature. The phenomenon can be related to locally reducing gap widths towards the groove bottom that, as with the case of GSPs, result in larger values for $N_{\text{eff}}$ and therefore tighter mode confinement. The results of ref.\cite{Vernon2008a} stress the importance of having a close estimate of the corner rounding (e.g. ${\sim}5~\text{nm}$)\cite{Smith2014} and also highlight the need for including finite rounding values in numerical models since infinitely sharp points lead to non-physical singularities (within the local-response approximation).\cite{Toscano2013} 

It is worth mentioning here that the position of the electric field maximum for the CPP is not located at the very bottom of the groove [Fig.~\ref{fgr:TipRad}(b)], but is instead located slightly above. This spatial shift is due to the finite corner rounding, which may seem unexpected since the gap still reduces towards the tip. The shift can be explained by successive reflections from the tip of the groove and the turning point (simple caustic),\cite{Gramotnev2005a} which occurs when the plasmon experiences non-adiabatic focusing: i.e. when the groove angle is larger than the critical taper angle, $\theta>\theta_c$,\cite{Gramotnev2005a,Vernon2008a} as discussed earlier.

\subsubsection{Adiabatic nanofocusing of CPPs. }

\begin{figure}[t!] 
   \centering
    \includegraphics[width=6.6cm,height=6.13cm]{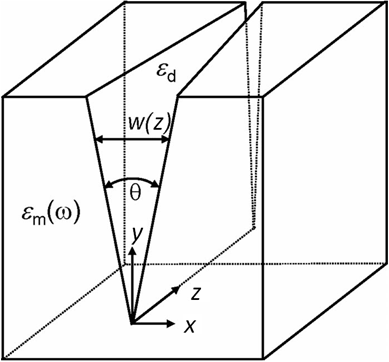}
    \captionsetup{justification=justified}
    \caption{Geometry of a tapered groove whose angle, $\theta$, reduces along the groove axis. (Adapted with permission from ref.\cite{Bozhevolnyi2010} Copyright \copyright{} 2010 Optical Society of America.)} 
    \label{fgr:CPPtaper}
\end{figure}

We now discuss the adiabatic nanofocusing of a CPP propagating along a tapered-groove axis with linearly decreasing groove angle, for which large intensity enhancements have been predicted $({\sim}1200)$\cite{Volkov2009,Volkov2009a,Bozhevolnyi2010} and measured $({\sim}130)$.\cite{Volkov2009,Volkov2009a} The geometry illustrated in Fig.~\ref{fgr:CPPtaper} is considered, with the groove angle being described by $\theta(z) = \theta_0(1-z/L)$, where $\theta_0$ is the groove angle at the start of the taper whose length is $L$. Considering the adiabatic approximation,\cite{Stockman2004,Gramotnev2005a} the evolution of the CPP mode for the dominant electric field component, $E_x$, is given by the following:\cite{Bozhevolnyi2010}
\begin{equation} \label{eq:focusCPPevol}
\begin{split}
&E_x\left(|x|<\frac{w(y)}{2},y,z\right) \\
&\!\begin{aligned}
&\approx A(z)\left(\frac{k|\chi|y}{\theta(z)}\right)\exp{\left(-\frac{k|\chi|y}{\theta(z)}\right)}\exp{\left(i\int_0^zk_{\text{CPP}}(u)\text{d}u\right)}, \\
&\text{with}{\:}{\:}{\:}k_{\text{CPP}}(z) = k\sqrt{1+(\chi/\theta(z))^2},
\end{aligned}
\end{split}
\end{equation}
where the CPP field dependence on width (i.e. the $x$-axis) is neglected for subwavelength $(w<\lambda)$ V-grooves.\cite{Bozhevolnyi2006b} Additionally, the explicit relation for $k_{\text{GSP}}$ used to derive the analytical formulas for CPPs,\cite{Bozhevolnyi2008a} $w>(\lambda\varepsilon_{\text{d}})/(\pi|\varepsilon_{\text{m}}|)$, must be valid at least at the position of the CPP maximum $[y_{\text{max}} = \theta/(k|\chi|): \theta>\theta_{\text{min}} = \sqrt{2}(\varepsilon_{\text{d}}|\varepsilon_{\text{m}}|)(|\varepsilon_{\text{d}}-\varepsilon_{\text{m}}|)^{0.25}]$. For gold V-grooves in air at telecommunications wavelengths, this condition results in $\theta>\theta_{\text{min}} \approx 2\degree$.

\begin{figure}[t!] 
   \centering
    \includegraphics[width=6.6cm,height=6.45cm]{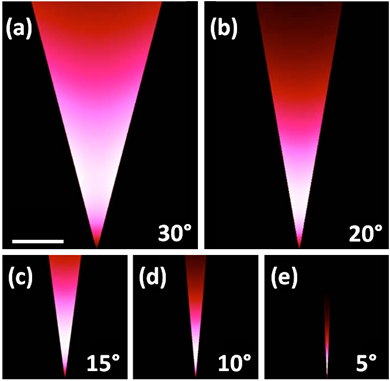}
    \captionsetup{justification=justified}
    \caption{Fundamental CPP mode field magnitude distributions calcaulated for gold V-grooves in air at $\lambda = 1.55~\mu{}\text{m}$ with varying groove angles: $\theta = \text{(a)}{\;}30\degree,{\;}\text{(b)}{\;}20\degree,{\;}\text{(c)}{\;}15\degree,{\;}\text{(d)}{\;}10\degree,{\;}\text{(e)}{\;}5\degree$. The scale bar in (a) represents $1~\mu{}\text{m}$. (Adapted with permission from ref.\cite{Bozhevolnyi2010} Copyright \copyright{} 2010 Optical Society of America.)} 
    \label{fgr:CPPfocus}
\end{figure}

Using the approximations given in ref.\cite{Bozhevolnyi2010} and the CPP power dissipating according to $\gamma(z)$, the CPP intensity, normalised to the value at the beginning of the taper, can be expressed as follows: 
\begin{equation} \label{eq:taperCPPenhance}
\begin{split}
\Gamma(z) &= \left|\frac{A(z)}{A(0)}\right|^2 \\
&= \sqrt{\frac{\theta(z)^2+|\chi|^2}{\theta_0^2+|\chi|^2}}\frac{\theta_0^2(\theta_0^2-\theta_{\text{cr}}^2)}{\theta(z)^2(\theta(z)^2-\theta_{\text{cr}}^2)}\exp{(-\gamma(z))}, \\
&\approx \frac{\theta_0^3}{16|\chi|^3}\left[\frac{2|\chi|}{\theta(z)}\right]^{4+k(\text{Im}\{\varepsilon_{\text{m}}\}/\text{Re}\{\varepsilon_{\text{m}}\})(L|\chi|/\theta_0)}, \\
\text{with}{\;}\theta_{\text{cr}} &= \sqrt{\frac{2}{3}}\frac{\varepsilon_{\text{d}}}{-\text{Re}\{\varepsilon_{\text{m}}\}},
\end{split}
\end{equation}
which describes the local electric field intensity enhancement considered at the same normalised depth coordinate, $\xi = k|\chi|y/\theta$. For the adiabatic approximation to be justified, $\delta = |\text{d}k_{\text{CPP}}^{-1}/\text{d}z|)$ should be sufficiently small: i.e. $\delta \ll 1$. Considering $k_{\text{CPP}}$ at the taper end $(|\chi| \gg \theta)$, the taper length requirement for the adiabatic approximation can be derived: $L \gg L_{\text{min}} = \theta_0/(k|\chi|)$,\cite{Bozhevolnyi2010} which is not a strict condition regarding propagation loss: gold grooves in air at telecommunications wavelengths and $\theta_0 = 30\degree$ require $L \gg L_{\text{min}} \approx 1.5~\mu{}\text{m}$, which is significantly smaller than the corresponding CPP propagation length.\cite{Volkov2007a,Volkov2009} The field enhancements at the taper end here are also consistent with the values calculated by finite element method calculations invoking the adiabatic approximation,\cite{Volkov2009} which predict the possibility of reaching intensity enhancements of {\texttildelow}1200 for realistic groove tapers.\cite{DeAngelis2010} An example of CPP mode focusing mediated by a progressively reduced groove angle, $\theta$, is shown in Fig.~\ref{fgr:CPPfocus}.

The approximate expression of (\ref{eq:taperCPPenhance}) provides straightforward access to a fundamental characteristic of any nanofocusing configuration, namely the critical taper length, $L_{\text{cr}}$, at which the focusing and dissipation effects are balanced. Considering the case described above, useful enhancement of the CPP mode intensity can be achieved for taper lengths $L<L_{\text{cr}} \approx 4(-\text{Re}\{\varepsilon_{\text{m}}\}/\text{Im}\{\varepsilon_{\text{m}}\})(\theta_0/k|\chi|)$,\cite{Bozhevolnyi2010} i.e. $L_{\text{cr}} \approx 62~\mu{\text{m}}$ at $\lambda = 1.55~\mu{\text{m}}$. It is noteworthy that for $\lambda = 1.033~\mu{\text{m}}$, $L_{\text{cr}} \approx 31~\mu{\text{m}}$, signifying the strong wavelength dependence on the relevant length-scales for achieving adiabatically nanofocused CPPs. 

\section{Plasmons in tapered grooves: applications}
\label{sec:Applications}

In this section we cover the potential applications offered by plasmons in tapered grooves. Their properties as described in section~\ref{sec:Plasmons_in_V-shaped_grooves} are often investigated in the context of information-based technologies, where the subwavelength confinement of optical modes offers the possibility to facilitate extremely compact, high performance photonics components\cite{Dintinger2009a,Zenin2011a,Lee2011a,Smith2012,Zenin2012a,Bian2013a} and effective platforms to realise quantum-plasmonics technologies.\cite{Vesseur2010,Martin-Cano2010,Martin-Cano2011,Gonzalez-Tudela2011,Vernon2012a} However, an increasing list of other types of tapered-groove-based devices are also appearing in the literature, such as efficient broadband light absorbers,\cite{Søndergaard2012,Skovsen2013a,Beermann2013a,Søndergaard2013a,Beermann2014} coloured surfaces,\cite{Zhang2011,Zhang2012b} and enhanced extraordinary optical transmission filters,\cite{Søndergaard2010c,Søndergaard2013a} while new paths towards state-of-the-art LOC systems are additionally being reported.\cite{Søndergaard2010b,Søndergaard2010c,Beermann2011,Søndergaard2013a,Rose2014} We therefore aim to provide a bird's-eye overview of the applications offered by tapered plasmonic grooves, focusing particularly on recent developments.

\subsection{Ultra-compact photonic circuits}

As the demand for increased information processing rates continues unabated, photonic circuits based on plasmons represent a promising \textit{enabling technology} to integrate photonic and electronic physical phenomena together on the nanoscale.\cite{Ozbay2006,Atwater2007} Importantly, plasmonics-based components have been shown to carry information with unprecedented bandwidth while simultaneously offering subwavelength confinement of the optical energy suitable for sufficient device miniaturisation.\cite{Dicken2008,Cai2009a} The opportunity encompassed here is difficult to overstate and has served as a major impetus behind the rapid expansion of the field of plasmonics.\cite{Brongersma2010a} In this light, CPPs propagating in tapered-groove waveguides exhibit useful subwavelength confinement,\cite{Pile2004a} relatively low propagation loss,\cite{Bozhevolnyi2005a,Nielsen2008a} single-moded operation,\cite{Gramotnev2004a} efficient transmission around sharp and gradual bends,\cite{Pile2005,Volkov2006,Volkov2006a} and have since been regarded as one of the most well-suited waveguiding geometries to optimise the trade-off between propagation loss and confinement. Subsequently, a series of CPP-based subwavelength components have been demonstrated, including Mach-Zender interferometers, ring resonators, Y-splitters,\cite{Bozhevolnyi2006c} add-drop multiplexers and Bragg grating filters.\cite{Volkov2007a} 

\begin{figure}[t!] 
   \centering
    \includegraphics[width=8.6cm,height=9.2cm]{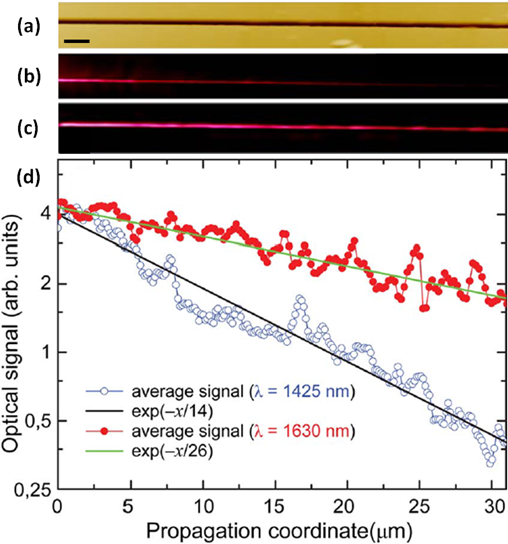}
    \captionsetup{justification=justified}
    \caption{CPP propagation length estimation using NSOM measurements. (a) Topographical optical image of the tapered-groove under test. Scale bar is $2~\mu\text{m}$. (b),(c) NSOM images of the tapered-groove supporting CPP modes for excitation wavelengths (b) $\lambda = 1425~\text{nm}$ and (c) $\lambda = 1630~\text{nm}$. (d) Measured optical signal of the CPP modes in (b),(c) averaged transverse to the propagation direction; the fitted exponential dependences of the signal are included. (Adapted with permission from ref.\cite{Zenin2011a} Copyright \copyright{} 2011 Optical Society of America.)}
    \label{fgr:CPPnsom}
\end{figure}

\begin{figure*}[t!] 
   \centering
    \includegraphics[width=17cm,height=6.97cm]{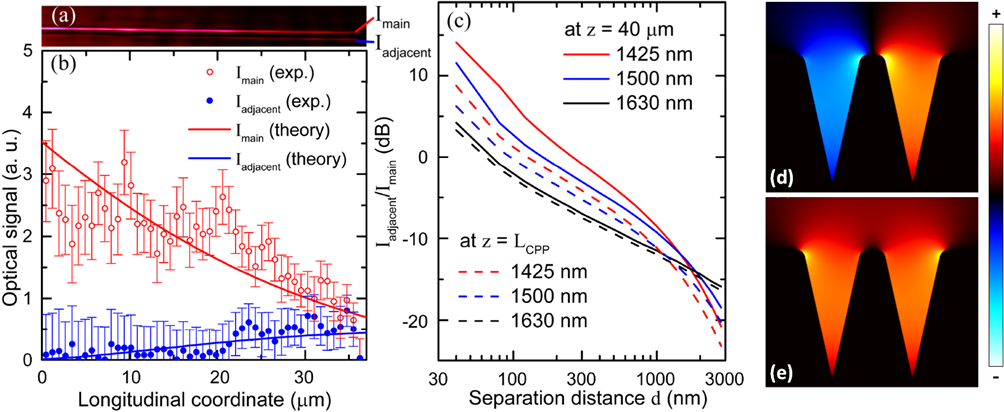}
    \captionsetup{justification=justified}
    \caption{(a) NSOM optical image of the coupling region for a directional coupler with a separation distance of $d = 0.25~\mu\text{m}$ at an excitation wavelength of $\lambda = 1500~\text{nm}$. Experimental values of the optical signal corresponding to the image of (a) for the main (red hollow circles) and adjacent (blue filled circles) waveguides and compared with the results of numerical simulations. (c) Crosstalk $I_{\text{adjacent}}/I_{\text{main}}$ calculated numerically as a function of separation $d$ varied from $40~\text{nm}$ to $3000~\text{nm}$ at different wavelengths: $1425~\text{nm}$ (red lines), $1500~\text{nm}$ (blue lines), and $1635~\text{nm}$ (black lines) and at different longitudinal coordinates along the coupler region: $z = 40~\mu\text{m}$ (solid lines) and $z = L_{\text{CPP}}$ (dashed lines). (d),(e) Distribution of the dominant electric field component, $E_x$, of (d) even and (e) odd supermodes using a light wavelength of $\lambda = 1500~\text{nm}$ calculated with COMSOL. Panel sizes in (d) and (e) are $2.5~\mu\text{m}$. (Adapted with permission from ref.\cite{Zenin2012a} Copyright \copyright{} 2012 Optical Society of America.)}
    \label{fgr:DirectCoupl}
\end{figure*}

Owing to the advantages outlined above, CPP-based circuit components remain an active topic of research, with new devices and paths for their implementation continuing to be explored. V-groove waveguides supporting CPP modes laterally confined to {\texttildelow}$(\lambda/5)$ at telecommunications wavelengths were reported in terms of their dispersion, propagation lengths and confinement.\cite{Zenin2011a} Here the confinement was improved by a factor of more than 3 over previous work,\cite{Bozhevolnyi2005a,Volkov2006,Bozhevolnyi2006c,Volkov2007a} and it was further shown these above-mentioned CPP characteristics may be determined \textit{directly} from near-field scanning optical microscopy (NSOM) measurements (Fig.~\ref{fgr:CPPnsom}). As will be discussed in section \ref{sec:Excite_&_Characterise}, issues inherent to the spatial correlation between the NSOM fiber tip and the tapered-groove geometry were identified that limit the achievable accuracies, but nonetheless these results denote an important step towards the realisation of CPP-based circuitry.

Following from the above results, a directional coupler consisting of two adjacent, parallel V-groove waveguides (Fig.~\ref{fgr:DirectCoupl}) was demonstrated.\cite{Zenin2012a} Here, the proof-of-principle was shown using a separation distance between the waveguides of $d = 0.25~\mu\text{m}$ [Fig.~\ref{fgr:DirectCoupl}(a)-(c)] across telecommunications wavelengths. The error bars in Fig.~\ref{fgr:DirectCoupl}(b) denote the uncertainty related to the CPP mode intensity dependence on the NSOM probe penetration depth inside the groove. The output cross-talk was measured to be close to $0{\,}\text{dB}$ after $40~\mu{}\text{m}$ of coupling length, with minimal dependence on wavelength. Numerical calculations performed by finite element method (COMSOL Multiphysics) showed the existence of even [Fig.~\ref{fgr:DirectCoupl}(d)] and odd [Fig.~\ref{fgr:DirectCoupl}(e)] supermodes for the vector field of $E_x$. Based on the approximately equal power splitting in the measurements, it was understood that the odd mode contributes most to the energy exchange of the coupler, since the even mode with less confinement experiences greater scattering loss from surface roughness. The coupling length, given as $L_{\text{coupling}} = \lambda/[2(N_{\text{odd}}-N_{\text{even}})]$, was found to be around $90~\mu\text{m}$, satisfying the basic requirement of $L_{\text{coupling}} \leq L_{\text{CPP}}$ in this case. The crosstalk was also characterised in this work as a function of light wavelength and $d$ in order to determine the maximum allowed density of non-interacting waveguides, where it was found that no coupling between V-grooves occurred for $d = 2~\mu\text{m}$ and indicating a clear potential for highly miniaturised integration.

\begin{figure}[t!] 
   \centering
    \includegraphics[width=8.6cm,height=6.92cm]{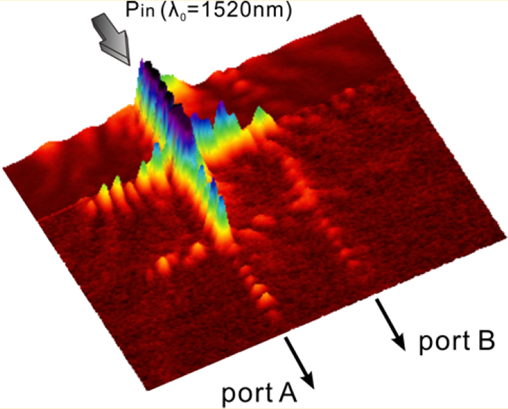}
    \captionsetup{justification=justified}
    \caption{The measured near-field response of a $2\times2$ port RGWN logic device composed of four 15-$\mu\text{m}$-long V-groove waveguides. The plasmonic mode of one of the arms is coupled to from a silicon ridge waveguide with TE-polarised light of $\lambda_0 = 1520~\text{nm}$. The output ports are labelled A and B. (Adapted with permission from ref.\cite{Burgos2014} Copyright \copyright{} 2014 American Chemical Society.)}
    \label{fgr:RGWN}
\end{figure}

As a testament to their integrability, an implementation of plasmonic V-grooves in a $2\times2$ port logic element based on a resonant guided wave network (RGWN) configuration (Fig.~\ref{fgr:RGWN}) was recently reported.\cite{Burgos2014} The operating principle of the device relied on the ability of two intersecting subwavelength plasmonic waveguides to generate four-way power splitting at their junction; an otherwise cumbersome feat to achieve for dielectric photonic structures. Here, highly confined CPPs supported by the subwavelength plasmonic V-grooves could enter the crossing and undergo volumetric mode expansion, resulting in an isotropic in-plane distribution of \textit{k}-vectors and thereby coupling to all four ports. Since the power splitting ratios into the outgoing ports were determined by a resonance condition at the intersection, the RGWN could be tailored to route different wavelengths to separate transmission ports. Moreover, it was suggested that four V-groove waveguides as per the arrangement indicated in Fig.~\ref{fgr:RGWN} potentially enables a facile eight-port device, which cannot be matched by other photonic crystal or plasmonic add/drop filters where only two on/off states are accessible. As such, RGWNs of plasmonic V-grooves represent a manifestly straightforward and effective approach to develop power splitters and logic elements for nanocircuits.

\begin{figure}[t!] 
   \centering
    \includegraphics[width=8.6cm,height=7.81cm]{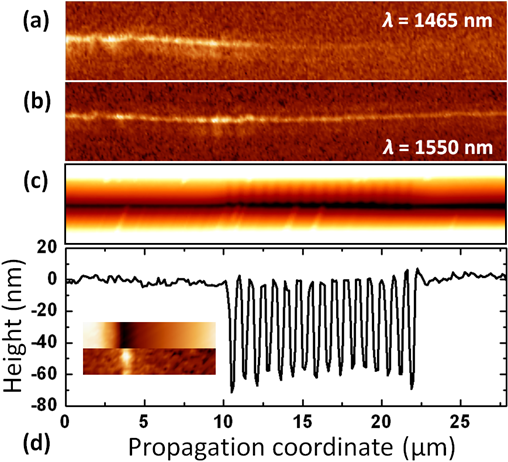}
    \captionsetup{justification=justified}
    \caption{NSOM measurements on a tapered-groove containing a 16-period Bragg grating filter. (a),(b) $5.0 \times 27.9~\mu\text{m}$ panels of NSOM measurements taken along the V-groove for the excitation wavelengths of (a) $\lambda = 1465~\text{nm}$ and (b) $\lambda = 1550~\text{nm}$. (c) Topographical optical image taken by atomic force microscopy for the same region of (a) and (b). (d) Cross-section close to the tapered-groove bottom from (c) with inset: $5 \times 1~\mu\text{m}$ panels showing topography and near-field optical intensity $(\lambda = 1550~\text{nm})$ taken simultaneously by the NSOM probe. (Adapted with permission from ref.\cite{Smith2012} Copyright \copyright{} 2012 Optical Society of America.)}
    \label{fgr:BGFnsom}
\end{figure}

In the pursuit of generally improving plasmonic component quality and production affordability, nanoimprint lithography was adopted to form tapered-groove waveguides consisting of Bragg grating filters.\cite{Smith2012} The wavelength-selective filtering was demonstrated experimentally by NSOM (Fig.~\ref{fgr:BGFnsom}), and supplemented by in-plane optical transmission measurements. The in-plane measurements were made possible by the nature of the samples being \textit{stand-alone}: that is, end-facets at both ends of ${\sim}100~\mu\text{m}$-length waveguides were accessible by lensed optical fibres due to the fabrication method including a UV lithography step that could define narrow yet robust polymer substrates. The results of the two separate experiments on the same device [16 period, $\Lambda = 760~\text{nm}$ ($12.1~\mu\text{m}$-length) grating] compared favourably to previous focused ion beam (FIB)-milled periodic wells,\cite{Volkov2007a} where NSOM scans along the waveguide direction observed an extinction ratio, $I_i/I_o$, of $4.5 \pm 0.9$ [Fig.~\ref{fgr:BGFnsom}(a)] taken before and after the corrugation and within the appropriate wavelength range, while transmission measurements exhibited a maximum extinction ratio of $8.2{\,}\text{dB}$ (centre wavelength, $\lambda_0 = 1454~\text{nm}$) and a $-3{\,}\text{dB}$ bandwidth of $\Delta{}\lambda = 39.9~\text{nm}$. It is intriguing to note that the reported centre wavelengths of the Bragg gratings in both refs.\cite{Volkov2007a,Smith2012} are several percentage points shorter than the expected values given the grating pitches and mode effective refractive indices, and remains a topic of interest that possibly involves more complicated physical phenomena. Nevertheless, the high-throughput nanoimprint lithography technique used to fabricate the devices, to be discussed further in section \ref{sec:Design_&_Fabrication}, consolidates the prospects of achieving high quality and mass-production-compatible nanoplasmonic components.

\begin{figure}[b!] 
   \centering
    \includegraphics[width=8.6cm,height=4.79cm]{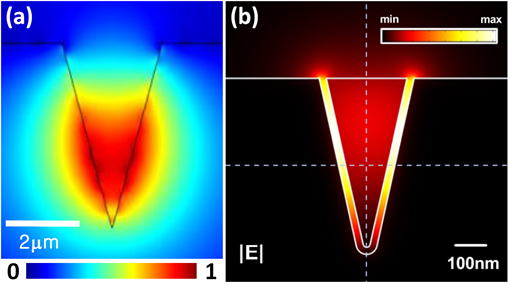}
    \captionsetup{justification=justified}
    \caption{(a) Normalised dominant electric field profile, $E_x$, of a LR-CPP supported by a $10~\text{nm}$ gold film sandwiched between symmetric dielectric media with $\theta = 30\degree$, a groove depth of $5~\mu\text{m}$ and an excitation wavelength of $\lambda = 1550~\text{nm}$.\cite{Lee2011a} (b) Electric field distribution of the fundamental hybrid-plasmonic waveguide mode with a \ce{SiO2} gap thickness of $t = 20~\text{nm}$ between a \ce{Si} wedge and an \ce{Ag} tapered-groove at an excitation wavelength of $\lambda = 1550~\text{nm}$.\cite{Bian2013a} (Adapted with permission from refs.\cite{Lee2011a,Bian2013a} Copyright \copyright{} 2011 Optical Society of America, \copyright{} 2013 IOP Publishing Ltd.)}
    \label{fgr:HybridCPP}
\end{figure}

Alternative tapered-groove geometries have been considered with the aim of combining the advantages of other plasmonic waveguide configurations together with those offered by CPPs. The effects of a finite metal thickness have been numerically investigated,\cite{Dintinger2009a,Lee2011a} where short-\cite{Burke1986} and long-range SPPs\cite{Sarid1981,Dionne2005,Boltasseva2005} (SR-SPPs and LR-SPPs) were considered in the context of V-grooves for achieving short- and long-range CPPs (SR-CPPs and LR-CPPs). SR-SPPs and LR-SPPs, which are supported by thin metal films surrounded by dielectric media, have the tangential electric field profile (main damping component) either efficiently occupying the metal with large confinement and Ohmic dissipation (SR-SPPs), or exhibiting a central null inside the metal with minimal confinement and Ohmic dissipation (LR-SPPs). SR-CPPs were shown to exhibit a strong dependence on the metal film thickness, with a decreasing thickness resulting in progressively increased confinement and losses.\cite{Dintinger2009a} This behaviour, in accordance with expected SR-SPP characteristics, suggests a useful method to tailor the CPP mode properties where the metal thickness is the variable parameter. Along similar lines, LR-CPPs of a gold film sandwiched between symmetric dielectric media $(n_1 = n_2 = 2.38)$ were explored\cite{Lee2011a} where the mode distribution [Fig.~\ref{fgr:HybridCPP}(a)] was found to be notably larger than the CPP counterpart, indicating longer propagation lengths -- order of millimetres for telecommunications wavelengths -- at the cost of reduced integration density. The strict requirement for symmetry of the surrounding dielectric media for LR-CPP modes suggests other possibilities, where the high sensitivity to small refractive index changes may be useful to LOC sensing devices.

\begin{figure*}[t!] 
   \centering
    \includegraphics[width=17.5cm,height=5.63cm]{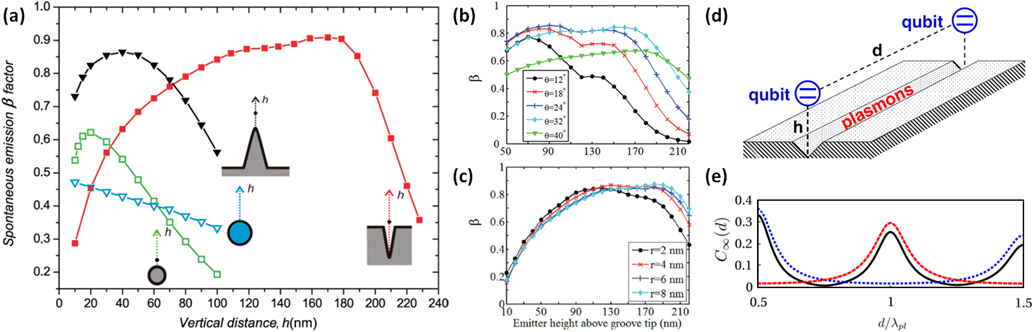}
    \captionsetup{justification=justified}
    \caption{(a) Spontaneous emission $\beta$-factor of a single emitter in the neighbourhood of various silver plasmonic waveguides as a function of the emitter height, $h$, at the excitation wavelength of $\lambda = 600~\text{nm}$.\cite{Martin-Cano2010} (b),(c) Coupling efficiency of an emitter into the plasmonic mode of a gold V-groove at an excitation wavelength of $\lambda = 632.8~\text{nm}$ as a function of vertical emitter height in the groove. Different groove angles, $\theta$, are considered in (b) for a groove depth of $d = 120~\text{nm}$, while in (c) different tip radii are considered for a groove depth of $d = 150~\text{nm}$ at a fixed groove angle, $\theta = 26\degree$.\cite{Vernon2012a} (d) Schematic illustration of two qubits interacting via a plasmonic V-groove waveguide.\cite{Gonzalez-Tudela2011} (e) Steady-state concurrence as a function of separation between two equal qubits with $\beta = 0.94$ and propagation length $L_{\text{CPP}} = 2~\mu\text{m}$ for a silver V-groove ($\theta = 20\degree$, $140~\text{nm}$ groove depth) at an excitation wavelength of around $\lambda = 640~\text{nm}$ under three different laser configurations on the two qubits (see text): asymmetric (solid black curve), symmetric (dotted blue line), and antisymmetric (dashed red line).\cite{Gonzalez-Tudela2011} (Adapted with permission from refs.\cite{Martin-Cano2010,Vernon2012a,Gonzalez-Tudela2011} Copyright \copyright{} 2010 American Chemical Society, \copyright{} 2012 American Institute of Physics, \copyright{} 2011 American Physical Society.)}
    \label{fgr:QPbetahQ}
\end{figure*}

Another alternate tapered-groove geometry somewhat resembling the hybrid-plasmonic configuration\cite{Oulton2008,Oulton2008a} was suggested and numerically investigated.\cite{Bian2013a} A V-groove of silver $(\varepsilon_{\text{m}} = -129 + 3.3i)$, predominantly filled with silicon $(\varepsilon_{\text{d}} = 12.25)$, had a silicon dioxide (\ce{SiO2}) $(\varepsilon_{\text{g,c}} = 2.25)$ gap layer located between the \ce{Si} and \ce{Ag} [$\lambda = 1550~\text{nm}$, Fig.~\ref{fgr:HybridCPP}(b)]. It was shown that the mode energy of the hybrid plasmon-polariton could be strongly confined to the low-index dielectric gap region while maintaining reasonable propagation lengths on the order of tens of microns. A performance comparison between hybrid V-grooves and standard V-grooves in air suggested that modes supported by hybrid geometries can exhibit superior properties in terms of a combined effective mode area reduction and propagation length. Practical aspects of the scheme, such as fabrication and characterisation, warrant further investigation, but make for clear paths to future work since the scheme potentially offers the advantage of seamlessly integrating tapered plasmonic grooves with \ce{Si}-based photonics and electronics.

\subsection{Quantum plasmonics }

Arguably the most rapidly emerging field of research in nanophotonics is the topic of quantum plasmonics, which studies the quantum nature of light and its interaction with matter on the nanoscale. The ability of plasmonic systems to concentrate EM fields beyond the diffraction limit make for new possibilities to achieve quantum-controlled devices, such as single-photon sources, qubit operators and transistors.\cite{Altewischer2002a,Akimov2007a,Kumar2013,Tame2013} As we will discuss in the following, recent work has shown numerically that tapered plasmonic grooves can efficiently collect and enhance the emission from a local quantum emitter\cite{Vesseur2010,Martin-Cano2010,Martin-Cano2011,Gonzalez-Tudela2011,Vernon2012a} as well as facilitate the entanglement between separated qubits,\cite{Martin-Cano2010,Martin-Cano2011,Gonzalez-Tudela2011} making the tapered-groove a well-suited structure type to incorporate into future quantum plasmonic systems.

A particular advantage of tapered plasmonic grooves with respect to emitter-waveguide systems is their combination of subwavelength confinement and \textit{one-dimensional character}, which was proposed as a means to mediate the resonant energy transfer between quantum emitters.\cite{Martin-Cano2010} For such systems, a major parameter is the coupling efficiency, $\beta$, defined as the ratio of spontaneous emission coupled into the plasmonic waveguide mode, $\hat{\gamma}_{\text{plas}}$, to the sum of decay rates into all available paths, $\hat{\gamma}_{\text{tot}}$; i.e. $\beta = \hat{\gamma}_{\text{plas}}/\hat{\gamma}_{\text{tot}}$. The plasmonic V-groove stands out as a competitive structure in terms of promoting a large maximum $\beta$ due to the strong EM confinement it offers. In addition, the CPP mode profile supports a broad range of emitter heights within the structure, $h$, where $\beta$ remains high [Fig.~\ref{fgr:QPbetahQ}(a)]. The influence of groove angle and tip radius on an emitter-groove system was quantified,\cite{Vernon2012a} where $\beta$ was found to be large [Fig.~\ref{fgr:QPbetahQ}(b),(c)] within a small window of parameters. Owing to existing limits on fabrication precision at the nanoscale and the importance of maximising $\beta$ in quantum-optics systems, the sensitivity of emitter-groove coupling on geometrical parameters therefore demands a detailed knowledge of these settings in order to predict the possible range of $\beta$s available. Nevertheless, the results above\cite{Martin-Cano2010,Vernon2012a} highlight the fact that tapered grooves offer a robust scheme to realise strong emitter-waveguide coupling across a broad range of readily-achievable specifications.

As a next step towards the realisation of quantum-operating devices, a numerical study of the entanglement between two emitters mediated by a V-groove waveguide was reported [Fig.~\ref{fgr:QPbetahQ}(d)].\cite{Martin-Cano2011,Gonzalez-Tudela2011} The steady-state entanglement driven by continuous laser pumping was described,\cite{Gonzalez-Tudela2011} where large values of the concurrence (measure of entanglement)\cite{Wootters1998} were shown to be achievable for qubit-qubit distances larger than the operating wavelength. In particular, the steady-state concurrence, $C_\infty$, was investigated as a function of qubit separation, $d/\lambda_{\text{pl}}$ ($\lambda_{\text{pl}}$ the wavelength of the plasmon), under continuous laser pumping on the two qubits [Fig.~\ref{fgr:QPbetahQ}(e)]. Three configurations of the laser-driven Rabi frequencies on the two qubits, $\Omega_1$ and $\Omega_2$, were studied: (1) symmetric with $\Omega_1 = \Omega_2 = 0.1\gamma$ for preparing the $|+\rangle$ state ($\gamma$ denotes the decay rate); (2) antisymmetric with $\Omega_1 = -1\Omega_2 = 0.1\gamma$ for preparing the $|-\rangle$ state; and (3) asymmetric with $\Omega_1 = 0.15\gamma$ and $\Omega_2 = 0$ for preparing a mixture of $|+\rangle$ and $|-\rangle$ states resulting in a changed separation dependence on $C_\infty$. It was further elucidated that the two-qubit entanglement generation is essentially due to the Ohmic dissipation part of the effective qubit-qubit coupling caused by the mediating plasmons.\cite{Martin-Cano2011} The significance of this points to the concept of implementing dissipative engineering of states to achieve quantum-operating devices\cite{Martin-Cano2011,Gonzalez-Tudela2011} where the plasmonic V-groove represents a realistic structure to facilitate sufficient entanglement between the separated qubits. 

\begin{figure}[t!] 
   \centering
    \includegraphics[width=8.6cm,height=8.37cm]{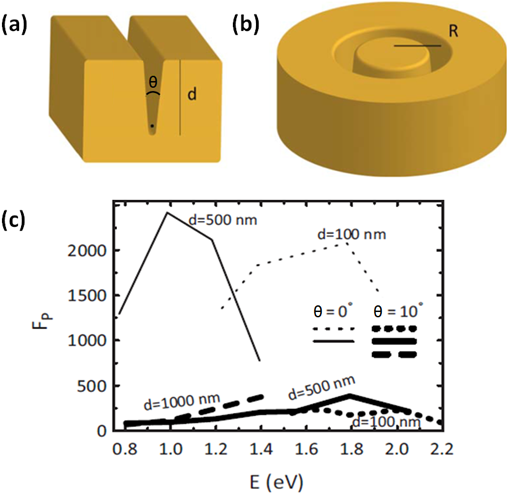}
    \captionsetup{justification=justified}
    \caption{(a),(b) V-groove-based ring resonator geometry. The solid dot in (a) indicates the position at which the LDOS was calculated. (c) Purcell factors of fundamental CPP modes with $m = 1$ azimuthal cavity resonance for different V-groove parameters as a function of energy. The lower three curves are for resonators with $\theta = 10\degree$, where $F_P$ can be more than 350. For the resonators based on thin grooves $(\theta = 0\degree)$, $F_P$ reaches over 2000. (Adapted with permission from ref.\cite{Vesseur2010} Copyright \copyright{} 2010 American Institute of Physics.)}
    \label{fgr:QPring}
\end{figure} 

The V-groove waveguide in a ring resonator configuration [Fig.~\ref{fgr:QPring}(a),(b)] was reported with the aim to achieve strong control over the spontaneous emission of a local emitter.\cite{Vesseur2010} According to Fermi's golden rule, the rate of spontaneous emission of an optical emitter is proportional to the local density of states (LDOS), which may be increased in an optical micro-cavity by the Purcell factor, \cite{Purcell1946}
\begin{equation}
\label{eq:PurcellFactor}
F_P = \frac{3}{4\pi{}^2}\frac{Q}{V}\left(\frac{\lambda}{n_{\text{eff}}}\right)^3,
\end{equation}
where $Q$ is the cavity quality factor (measure of time light is stored in the cavity), $V$ is the mode volume, and $n_{\text{eff}}$ is the effective refractive index of the cavity mode. The $Q$-factors were not found to be especially high in the plasmonic resonators considered here due to Ohmic losses $(Q = 10-50)$, although the extremely confined mode volumes down to $V \approx 0.005\lambda^3$ (i.e. much less than the diffraction limit: $V = 0.125\lambda^3$) give rise to Purcell factors above 350 for feasible device specifications ($\theta = 10\degree$) and reaching over 2000 for very thin grooves ($\theta = 0; 10~\text{nm}$ groove width) [Fig.~\ref{fgr:QPring}(c)]; for comparison, a recent prominent demonstration exploited a Purcell factor of 2 using GSPs in slot waveguides.\cite{Huang2014} The high Purcell factors here were shown to be possible over a large range of energies $(1.0 - 1.8{\,}\text{eV})$ and, due to the moderate $Q$, a broad spectral range $(50 - 100{\,}\text{meV})$, suggesting V-groove ring resonators offer a valuable capability to the tapered-groove-quantum-plasmonics toolbox in addition to the straight tapered-groove waveguide configuration described just above.  

We note here that tapered grooves are especially interesting structures in the context of quantum plasmonics, since their demonstrations as effective nanophotonic circuit elements make them ideal to serve as connectors to and from miniaturised quantum systems. Furthermore, tapered grooves represent a straightforward geometry to fill with and align the relevant quantum emitter materials, pointing to a relative ease of device formation. However, experimental demonstrations of quantum devices consisting of tapered plasmonic grooves are yet to be reported in the literature, thus setting the stage for exciting developments to come in the pursuit of general and groove-based quantum plasmonics.

\subsection{Advanced light absorbers and filters} 

\begin{figure}[t!] 
   \centering
    \includegraphics[width=8.6cm,height=4.97cm]{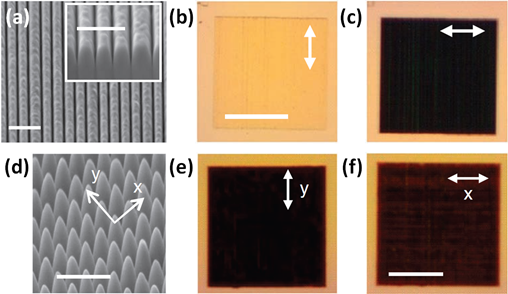}
    \captionsetup{justification=justified}
    \caption{(a) Scanning electron micrograph of a 1D array of ultra-sharp convex grooves in gold (scale bar is $1~\mu\text{m}$). (b),(c) Optical micrographs of the device from (a) illuminated with light polarised along the grooves (b), and across the grooves (c), demonstrating the polarisation dependence on light absorption here (scale bar is $1~\mu\text{m}$). (d) Scanning electron micrograph of a 2D array of ultra-sharp convex grooves in gold (scale bar is $700~\text{nm}$). (e),(f) Optical micrographs of the device from (d) illuminated with orthogonal polarisations, demonstrating the reduced polarisation dependence on the light absorption for the 2D case (scale bar is $7~\mu\text{m}$). (Adapted with permission from ref.\cite{Søndergaard2012} Copyright \copyright{} 2012 Macmillan Publishers Ltd.)}
    \label{fgr:BlackGold1D2D}
\end{figure}

Periodic arrays of grooves, implemented to make practical use of plasmonic nanofocusing in tapered structures, were recently shown to be able to modify metal surfaces from having a highly reflective character (shiny) into exhibiting efficient, broadband absorption (dark).\cite{Søndergaard2012,Beermann2013a,Søndergaard2013a,Skovsen2013a} 
Early blackened metal surfaces, produced by distillation at high pressures\cite{Pfund1933} or intense laser irradiation,\cite{Vorobyev2005,Vorobyev2008} were able to suppress reflection through a combined effect of nano-, micro- and macro-structured surface morphology, however, such surfaces are essentially impossible to design, characterise and reproduce down to all fine details due to the high degree of randomness. Alternatively, approaches utilising resonant interference have achieved efficient absorption properties for metal surfaces via structures such as gratings and micro-cavities, but inevitably come with wavelength, angular or polarisation dependences, thereby reducing their effectiveness e.g. for photovoltaic and thermophotovoltaic applications. In order to overcome the limitations inherent to random- and resonant-based metal light absorbers, so-called ``black gold'' surfaces were realised by broadband \textit{non}resonant adiabatic nanofocusing of GSPs excited by scattering off from the subwavelength-sized wedges of ultra-sharp convex groove arrays\cite{Søndergaard2012,Beermann2013a,Søndergaard2013a,Skovsen2013a,Beermann2014} (phenomena described in section~\ref{sec:Plasmons_in_V-shaped_grooves}). While other determinable and nonresonant light absorption configurations had been demonstrated at the time, such as subwavelength cylinders\cite{Aubry2010} and spheres,\cite{Fernandez-Dominguez2010} these configurations only supported limited absorption efficiencies due to their small absorption cross-sections relative to the geometry. By contrast, convex grooves may be fabricated to occupy large fractions of the surface area [Figs.~\ref{fgr:BlackGold1D2D}(a),(d)], thereby resulting in an efficient scheme to achieve broadband light absorption.

The first proof-of-concept of plasmonic black gold by adiabatic nanofocusing reported an average reflectivity of 4.0\% was reported across the wavelength range $450-850~\text{nm}$.\cite{Søndergaard2012} The influence of period dimension (1D or 2D arrays of grooves) was reported, where the 1D case exhibited a polarisation dependence based on whether the incident electric field corresponded to the dominant component of the GSPs (perpendicular to the groove axis) [Figs.~\ref{fgr:BlackGold1D2D}(b),(c)]. It was shown that the polarisation dependence could be otherwise mitigated by implementing 2D arrays of grooves, which facilitated the efficient excitation of GSPs for polarisations oriented along both groove axes [Figs.~\ref{fgr:BlackGold1D2D}(e),(f)].\cite{Søndergaard2012,Beermann2014} Conversely, the polarisation dependence of the 1D case was exploited for the design of broadband linear polarisers operating in reflection, where a factor of 3.9 suppression for p-polarised light relative to s-polarised light was experimentally achieved.\cite{Skovsen2013a} These devices were additionally shown to induce negligible spectral broadening of ultra-short $(5-10{\,}\text{fs})$ laser pulses, leading to a new class of high-damage-threshold, low-dispersion polarising filters. These demonstrations highlight a key practical advantage of plasmonic black gold: its tailorable reflective properties, which can be selected to suit particular applications.

\begin{figure}[b!] 
   \centering
    \includegraphics[width=8.6cm,height=8.89cm]{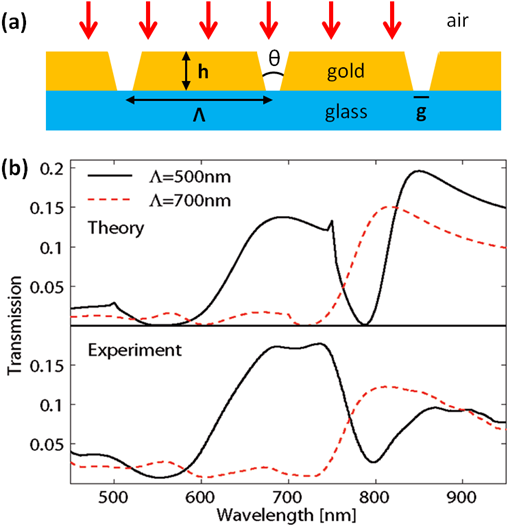}
    \captionsetup{justification=justified}
    \caption{(a) Illustration of a periodic array of tapered slits in a gold film of thickness $h$, period $\Lambda$, taper angle $\theta$ and gap at the bottom, $g$. The gold film resides on a glass substrate. (b) Theoretically calculated and experimentally measured zero-order transmission spectra from normally incident light for a structure with $g = 65~\text{nm}$, $\theta = 20.5\degree$ and periods $\Lambda = 500 \text{ and }700~\text{nm}$. The resonant transmission is as high as ${\sim}0.18$ for a filling ratio of ${\sim}g/\Lambda = 0.13$. (Adapted with permission from ref.\cite{Søndergaard2010c} Copyright \copyright{} 2010 American Chemical Society.)}
    \label{fgr:EnhancedEOT}
\end{figure} 

Methods to further enhance the light absorption in plasmonic groove arrays have been investigated.\cite{Beermann2013a,Søndergaard2013a,Beermann2014,Odgaard2014} The use of other metals with larger plasmonic losses was initially proposed\cite{Søndergaard2012} and later validated using nickel\cite{Beermann2013a} and palladium.\cite{Beermann2014} The potential advantages of using other metals includes lower production costs, higher melting temperatures (${\sim}1550\degree$C in the case of palladium),\cite{Beermann2014} and the possibility to allow for more practical fabrication techniques such as nanoimprint lithography or reactive-ion etching. A report on the spectral response showed a generally increasing reflectivity to occur for longer wavelengths, but with a remarkable dip that could be tuned by precise groove geometries -- a feature potentially useful for selective thermal emitters in thermophotovoltaics.\cite{Beermann2013a} The influence on reflectivity for a wide range of parameters, including groove profile, groove depth, bottom groove width, period, width of flat plateau, choice of metal and general direction of light incidence, have also been quantified.\cite{Søndergaard2013a,Odgaard2014} In particular, the importance of realising the adiabatic condition was highlighted, with nearly parallel groove walls at the groove bottom being required in all cases to sufficiently suppress reflection. However, away from the groove bottom, broader taper angles were found to be otherwise permissible since the adiabatic condition could be met more easily, thereby enabling large fractions of the surface to collect incident light for high absorption efficiencies.\cite{Søndergaard2013a} Calculations for angles of light incidence along the plane containing the groove axis generally showed lower reflectance across the wavelength spectrum than for angles along the plane normal to the groove axis,\cite{Odgaard2014} suggesting that this direction would be preferable for applications pertaining to thermophotovoltaics, concentrated solar power or broadband polarizers where low reflectivity is sought and different angles of incidence might occur. Moreover, we note here that the presence of antisymmetric GSPs caused by slight asymmetries in the groove profiles may also play an important role in the design of further-improved light absorbing metals, since these modes are optically dark (propagation lengths below $10~\text{nm}$) and could therefore provide an extremely efficient channel to dissipate the incident light.\cite{Raza2014a}

\begin{figure}[t!] 
   \centering
    \includegraphics[width=8.6cm,height=5.76cm]{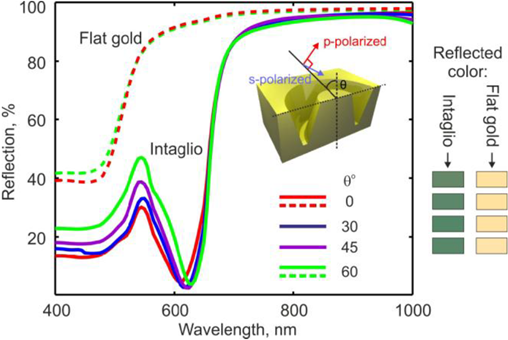}
    \captionsetup{justification=justified}
    \caption{Tapered-groove-based photonic metamaterial for engineering the spectral response of metals in the visible region. The graph depicts the colour invariance with viewing angle: Numerically simulated spectra and associated perceived colours for an array of tapered-groove-rings in gold with $170~\text{nm}$ diameters and a depth of $180~\text{nm}$ viewed at normal incidence and a selection of oblique angles (as labelled, averaged over incident s- and p-polarisations). For comparison, the corresponding perceived colour of an unstructured gold surface is shown for each viewing angle, and reflection spectra for the unstructured metal at $0\degree$ and $60\degree$ angles are plotted. (Reprinted with permission from ref.\cite{Zhang2011} Copyright \copyright{} 2011 Optical Society of America.)}
    \label{fgr:VgrResColor}
\end{figure} 

Plasmonic relief meta-surfaces were experimentally shown to control the perceived colour of light reflected from metals.\cite{Zhang2011,Zhang2012b} The configuration consisted of periodic, tapered-groove-based ring resonators that were engineered to absorb select wavelength components of the visible spectrum and thereby change the spectra reflecting back from the surface (Fig.~\ref{fgr:VgrResColor}).\cite{Zhang2011} The functionality of the design relied on a combination of the resonant properties of the plasmonic rings together with a sub-wavelength periodicity of the array which excluded diffraction effects and rendered the scheme viewing-angle-invariant. The platform, suitable to nanoimprint-based fabrication techniques since the structures are of continuous metal (no perforations), points towards an affordable approach to realise structurally-based, high-resolution coloured surfaces that does not require pigments or other chemical modification.

In contrast with the above configurations that consider reflection, the case of transmission through gold films perforated with holes of tapered gap widths [Fig.~\ref{fgr:EnhancedEOT}(a)] was shown to further boost extraordinary optical transmission (EOT).\cite{Søndergaard2010c,Beermann2011,Søndergaard2011} EOT describes the resonantly-enhanced transmission of light through periodically spaced, subwavelength apertures in otherwise opaque metallic films,\cite{Martin-Moreno2001,Genet2007,Laux2008} which was enhanced here by the use of large entrance apertures that collected the incident light and adiabatically funnelled it as GSPs through the taper to exit smaller apertures. The maximum transmission at resonance was found to occur for taper angles in the range of $6\degree - 10\degree$,\cite{Søndergaard2011} supporting a significant increase in transmission compared to straight slits. For a $180~\text{nm}$ gold film on glass, an enhanced EOT with resonance transmission as high as ${\sim}0.18$ for a filling ratio of ${\sim}g/\Lambda = 0.13$ was experimentally achieved with clear theoretical agreement [Fig.~\ref{fgr:EnhancedEOT}(b)].\cite{Søndergaard2010c} Following this, a numerical study into the effect of higher diffraction and slit resonance orders caused by an increased metal film thickness showed that the maximum transmission occurred when only the fundamental reflection order and at the same time higher transmission orders were available, implying that an optimum metal film thickness exists even though thicker metals would allow for larger collection apertures;\cite{Søndergaard2011} a $250~\text{nm}$ gold film in air was shown to potentially yield a factor of more than 20 times higher transmission relative to the gap-to-period ratio. Accordingly, the use of tapered slits points to numerous approaches for augmenting the range of applications that EOT phenomena are suited for, including subwavelength optics, optoelectronics, chemical sensing and biophysics.\cite{Genet2007}

\subsection{Novel lab-on-a-chip systems} 

Plasmonics is playing an increasing role in LOC devices since the EM field enhancement it offers has led to significant improvements in detection capabilities,\cite{Kneipp2002,Anker2008,Mayer2010,Chung2011a,Feng2012,Zijlstra2012} imaging resolutions\cite{Fang2005,Xiong2007,Verma2009,Liu2011,Schuck2013} and nano-manipulation control.\cite{Grigorenko2008,Juan2011,Erickson2011,Pang2012a,Marago2013} The over-arching advantage of plasmonic components in this respect lies with their ability to achieve highly concentrated EM fields while remaining within the technologically mature and non-obtrusive nature of the optical domain. The prospect of utilising tapered plasmonic grooves in particular for LOC devices in the pursuit of point-like, on-chip light sources and biosensors was initially demonstrated with the report that fluorescent beads within the groove contour could be individually excited by CPPs.\cite{Fernandez-Cuesta2009} Importantly, the tapered-groove profile was shown in this work to provide topography-assisted self-alignment of particles of interest, a unique and highly appropriate feature for LOC devices that acts in addition to the benefit offered by supporting SPPs.

\begin{figure}[t!] 
   \centering
    \includegraphics[width=8.6cm,height=8.47cm]{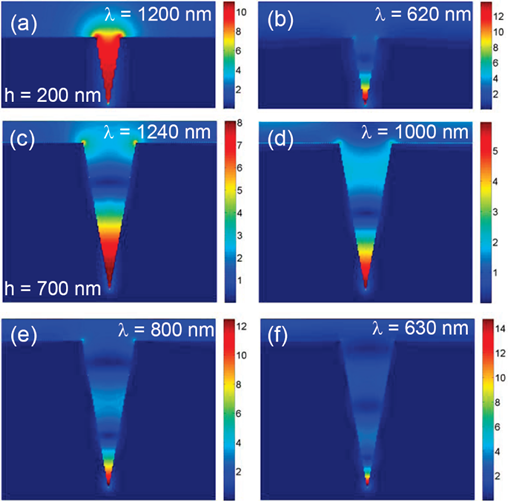}
    \captionsetup{justification=justified}
    \caption{Calculated normalized field magnitude distributions at resonant wavelengths as indicated on images for two $20\degree$ angle grooves with depths $h$ = (a),(b) $200~\text{nm}$ and (c),(e),(f) $700~\text{nm}$. The distributions show the resonant GSP standing-wave patterns of different orders, $m$ = (a) 1, (b),(c) 2, (e) 3 and (f) 4. For comparison, a nonresonant field distribution is shown in (d) for $h = 700~\text{nm}$. (Reprinted with permission from ref.\cite{Søndergaard2010b} Copyright \copyright{} 2010 American Chemical Society.)}
    \label{fgr:ResFieldEnhance}
\end{figure} 

Several different LOC-suitable approaches for achieving radiation nanofocusing in tapered plasmonic grooves have been reported in the literature. Appreciable field concentration $({\sim}\lambda/40)$ in V-shaped grooves was demonstrated by collecting normally incident radiation and compressing it as GSPs into nanoscale gaps located at the groove tips.\cite{Choi2009} Also under normal illumination, \textit{resonant} field enhancement of reflected GSPs in periodic arrays of tapered grooves, caused by a deliberate break down of the adiabatic condition, was shown to exhibit enhancement factors up to $550$\cite{Søndergaard2009a} theoretically and ${\sim}110$ experimentally (see Fig.~\ref{fgr:ResFieldEnhance}).\cite{Søndergaard2010b,Beermann2011} Alternatively, V-grooves tapered along the waveguiding axis as discussed in section~\ref{sec:Plasmons_in_V-shaped_grooves} have demonstrated adiabatic nanofocusing of channelled plasmons,\cite{Volkov2009,Volkov2009a,Bozhevolnyi2010} where the field enhancement factors reached up to ${\sim}1200$ theoretically\cite{Volkov2009,Volkov2009a,Bozhevolnyi2010} and ${\sim}130$ experimentally.\cite{Volkov2009,Volkov2009a} In light of these reports, it is evident that tapered plasmonic grooves not only offer strong field enhancement suitable for LOC devices, but also a broad choice of design options due to the variety of configurations that can achieve it.

\begin{figure}[t!] 
   \centering
    \includegraphics[width=8.6cm,height=6.47cm]{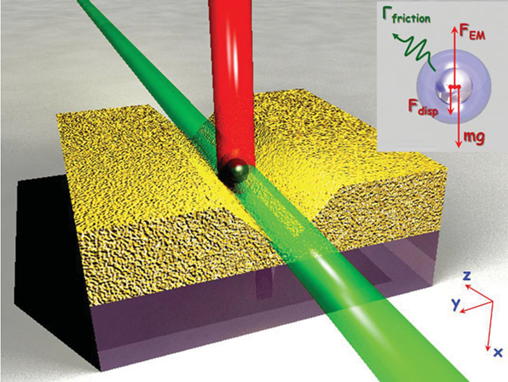}
    \captionsetup{justification=justified}
    \caption{Schematic of the nano-opto-mechanical modulator: a nanoparticle in the V-groove waveguide is driven into auto-oscillatory motion by a control beam (red). The particle moves in and out of the fundamental mode of the waveguide (green) and modulates modal loss. Inset shows the forces acting on a nano-object inside the V-groove. (Reprinted with permission from ref.\cite{Shalin2014a} Copyright \copyright{} 2013 John Wiley and Sons.)}
    \label{fgr:NanoOptoMech}
\end{figure} 

A nano-opto-mechanical device platform consisting of a plasmonic V-groove was theoretically considered as an approach to mediate the auto-oscillatory motion of individual nano-objects via normally incident light (Fig.~\ref{fgr:NanoOptoMech}).\cite{Shalin2014a} The configuration is such that for specific conditions of particle polarisability and V-shaped profile, an electric field distribution forms inside the groove as a result of the control beam $(250~\text{mW}/\mu\text{m}^2)$ and exerts an upwards force on a nano-object at the bottom of the groove -- sufficient to overcome gravity and van der Waals attraction by an order of magnitude. The vertical component of the optical gradient force varies strongly as a function of the particle's position, with its direction reversing beyond a certain equilibrium point in the groove and hence facilitating auto-oscillatory motion. Importantly, the intensity of the control beam governs the magnitude of the up and down accelerations and therefore the particle's round-trip frequency, which, depending on particle and groove specifications, may be varied to achieve a range of $22-40~\text{MHz}$. The configuration could be of interest to the development of low energy all-optical information technologies that do not require nonlinear materials, or LOCs that employ motion-based transduction or operate by means of optical actuation.

\begin{figure}[t!] 
   \centering
    \includegraphics[width=8.6cm,height=7.19cm]{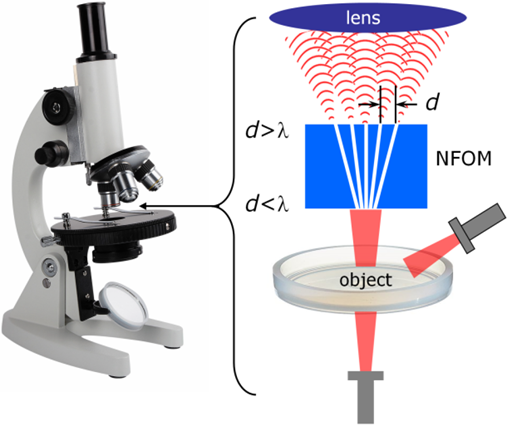}
    \captionsetup{justification=justified}
    \caption{Schematic of imaging with proposed near field optical magnifier comprised of nanowaveguides (white lines) supported by dielectric filler (blue). The subwavelength-size ``object'' is placed at the input ends of the nanowaveguides array (having waveguide separation $d < \lambda$). When illuminated, the image of the object is propagated along the waveguides and reemitted as a magnified image at the output ends of the waveguides (with separation $d > \lambda$). Traditional optics are used to capture the magnified image. (Reprinted with permission from ref.\cite{Rose2014} Copyright \copyright{} 2014 Optical Society of America.)}
    \label{fgr:Nanoscope}
\end{figure} 

The use of divergent tapered nanogrooves was recently suggested as a method to improve the resolution of conventional light microscopes.\cite{Rose2014} Owing to the eminent role of light microscopy in science and its establishment as a routine investigative tool, the possibility to break the optical diffraction barrier without fluorescent labels is tantalising to researchers since a wide variety of fundamental building blocks in the life and physical sciences are of nanoscale dimensions. The scheme (Fig.~\ref{fgr:Nanoscope}) suggests subwavelength features may be collected in real-time with the nanogroove magnifier and remitted as an image to be captured by traditional lenses; calculations indicate a potential resolution of ${\sim}\lambda/32 = 25~\text{nm}$ for $40~\text{nm}$-wide waveguides. In another arrangement that also associates the use of tapered grooves with optical microscopes, termination mirrors at the ends of tapered-groove waveguides have been shown to provide a highly convenient approach to efficiently excite in-plane CPPs via direct illumination from microscope objectives.\cite{Radko2011a,Smith2014} The importance of this result, which we will discuss further in section~\ref{sec:Excite_&_Characterise}, corresponds to enabling new opportunities for sensing and manipulation via LOC devices in microscope settings where the photon-to-plasmon efficiencies would otherwise act as a limiting factor. In sum, we anticipate tapered plasmonic grooves to play an increasing role in optical microscopes and microscope-based LOC devices.

\section{Fabrication}
\label{sec:Design_&_Fabrication}

The increasing prevalence of nanofabrication techniques has led to the rapid growth of the field of plasmonics over the past fifteen years.\cite{Maier2005,Barnes2006,Ebbesen2008,Boltasseva2009} The strong dependence on feature-size for nanofocusing light via plasmons implies that the ability to routinely pattern metals on the nanoscale allows for the advantages of SPPs to be systematically harnessed. Nevertheless, fabrication challenges still remain in terms of achieving high quality and cost-effective nanoplasmonic devices suitable for technological integration (e.g. CMOS compatibility). In this light, tapered grooves represent comparatively convenient planar plasmonic structures to fabricate since they correspond to the innate profiles formed by focused-ion beam (FIB) milling or crystallographic etching of silicon. In the following, we present the variety of lithographic techniques that have been demonstrated to form metallic tapered-groove profiles and discuss their relative merits.

\subsection{Focused-ion beam milling}

Today, the most common procedure for fabricating tapered plasmonic grooves is via the use of FIB milling, where a finely focused beam of (typically gallium) ions sputters material away from a specific location. Owing to the broad flexibility it offers for generating arbitrary metal geometries with feature sizes below $10~\text{nm}$, FIB milling represents the ``go-to'' approach for fast prototyping -- tapered grooves included.\cite{Bozhevolnyi2005a,Bozhevolnyi2006c,Choi2009,Zhang2011,Søndergaard2012,Burgos2014} In addition to the broad selection of structure types that can be realised by FIB milling, modern FIB systems are also commonly available at research facilities and processing centres, making it a relatively accessible technique. 

Despite the diverse set of tapered plasmonic grooves that have been achieved via FIB milling, issues related to either its production rate or quality must be borne in mind. Concerning the production rate, FIB milling is a serial-based technique, which leads to long writing times per device that render it impractical for large-scale fabrication. In terms of production quality, the act of sputtering material away from a region inherently leads to its re-deposition nearby, a problem that results in the deterioration of side-wall smoothness\cite{Bozhevolnyi2005a} and causes damage to already-formed structures in close proximity.\cite{Søndergaard2012} Furthermore, some gallium ions become implanted into the top few nanometres of the metal surface, causing non-negligible effects into the behaviour of the free electrons.\cite{Melngailis1987} Recent demonstrations of helium-based FIB milling make for a viable alternative to gallium sources in terms of producing high quality devices with minimal structural and material degradation.\cite{Melli2013} Helium-based FIB milling therefore represents a compelling avenue to improve upon existing demonstrations and promote new opportunities, although its serial-based aspect still retains the high labour-intensive- and time-related costs inherent to FIB processes.

\subsection{Anisotropic etching}

\begin{figure}[t!] 
   \centering
    \includegraphics[width=8.6cm,height=10.98cm]{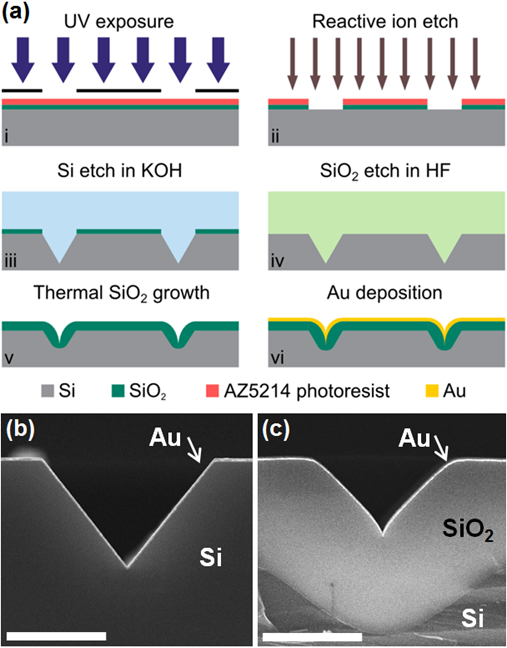}
    \captionsetup{justification=justified}
    \caption{(a)[i-vi] The procedure for fabricating gold V-grooves by photolithography combined with anisotropic etching in KOH. The \ce{SiO2} etch mask is removed by hydrofluoric acid (HF). A thermal oxidation step may be performed to modify the V-groove profile before the gold layer is deposited. (b),(c) SEM images of resulting V-groove cross-sections: (b) without thermal oxidation, and (c) with thermal oxidation. Scale bars are $2~\mu\text{m}$. (Adapted with permission from ref.\cite{Smith2014} Copyright \copyright{} 2014 American Chemical Society.)}
    \label{fgr:VgrvUVlith}
\end{figure} 

V-shaped grooves produced by anisotropic (crystallographic) etching have been shown to support CPPs with propagation lengths exceeding $100~\mu\text{m}$ for excitation wavelengths around $\lambda = 1.5~\mu\text{m}$.\cite{Nielsen2008a} Anisotropic etching involves the use of a wet chemical agent that etches crystalline materials at considerably different rates depending on which crystal face is exposed. In the case of $\langle100\rangle$-oriented silicon with potassium hydroxide (\ce{KOH}) as the etchant, the $\langle111\rangle$ set of silicon crystal planes etch at a much slower rate than all others (e.g. ${\sim}400$ times slower than $\langle100\rangle$). As a consequence, V-shaped grooves with a fixed groove angle of $\theta = 70.6\degree$, corresponding to the crossing of $\langle111\rangle$ silicon crystal planes, can be defined using an etch mask of \ce{SiO2}.\cite{Fernandez-Cuesta2007a,Nielsen2008a,Choi2009} A major advantage of anisotropic etching is its suitability for wafer-scale production, since the etching agent can act over one or more whole wafers simultaneously. Additionally, the groove-formation using this method relies on a chemical process that results in almost atomically-smooth sidewalls and tips with curvature radii sharper than $5~\text{nm}$.\cite{Smith2014}

\begin{figure}[t!] 
   \centering
    \includegraphics[width=8.6cm,height=5.48cm]{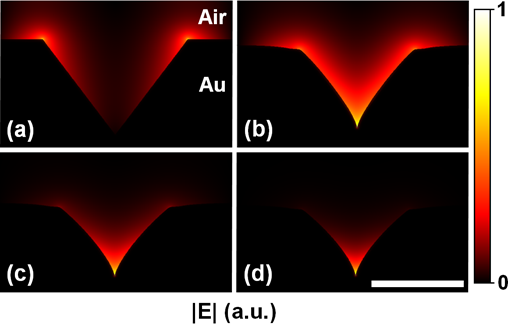}
    \captionsetup{justification=justified}
    \caption{FEM calculations of the magnitude of the electric fields supported by varying V-groove cross-sections in gold for an excitation wavelength of $\lambda = 811~\text{nm}$. (a) Zero oxidation. In this case, only modes located at the wedges exist; (b) $1320~\text{nm}$ oxidation; (c) $1720~\text{nm}$ oxidation; and (d) $2320~\text{nm}$ oxidation. Thicker \ce{SiO2} layers lead to sharper V-shaped profiles near the tip resulting in increased confinement of the electric field distribution towards the bottom of the  waveguide. A clear CPP mode occurs in the case for (d). Scale bar is $2~\mu\text{m}$. (Adapted with permission from ref.\cite{Smith2014} Copyright \copyright{} 2014 American Chemical Society.)}
    \label{fgr:VgrvSiO2modes}
\end{figure} 

\subsubsection{Tailoring tapered-groove profiles. } 

It is important to note that it is highly desirable to modify the otherwise fixed angle of anisotropically-defined grooves in order to render them suitable to their intended application. Such V-grooves in silicon may be tailored by an additional thermal oxidation process in order to modify the geometric profiles (Fig.~\ref{fgr:VgrvUVlith}) and give rise to a striking effect on the resulting plasmonic mode distributions (Fig.~\ref{fgr:VgrvSiO2modes}).\cite{Smith2014} This adjustment of the geometry, here represented by a reduction of $\theta$ at the groove tip from $70.6\degree$ [Fig.~\ref{fgr:VgrvUVlith}(b)] down to ${\sim}50\degree$ [Fig.~\ref{fgr:VgrvUVlith}(c)] for a \ce{SiO2} thickness of $2320~\text{nm}$, is necessary to convert the initially wedge-based modes of unoxidised V-grooves [Fig.~\ref{fgr:VgrvSiO2modes}(a)] into well-confined CPPs [Fig.~\ref{fgr:VgrvSiO2modes}(d)]. Such mode tailoring is consistent with the expected behaviour of CPPs, where the electric field confinement increases for decreasing $\theta$ as described in section~\ref{sec:Plasmons_in_V-shaped_grooves}.\cite{Bozhevolnyi2007a} The reduction of $\theta$ at the groove tip during thermal oxidation is a result of the variation of \ce{SiO2} formation caused by the V-shaped profiles present in the silicon. The thermal oxidation step, in keeping with wafer-scale processes, also affords nano-scale control of the resulting geometric profiles and offers a possible route to ultra-precise plasmonic mode engineering.

Alternatively, we posit that a method to etch higher-index crystal planes in silicon at slower rates than the $\langle111\rangle$ set would potentially allow for the formation of sharper grooves with smaller $\theta$. For example, the crossing of certain planes of $\langle311\rangle$ with respect to $\langle100\rangle$ would result in $\theta = 34.8\degree$. Additives including isopropyl alcohol (IPA) or Triton-x-100 surfactant to etchants such as KOH or tetramethylammonium hydroxide (TMAH) have been shown to change the preferred, slowest-etch-rate crystal plane.\cite{Resnik2005} Standards to select higher index planes are not well-established to the best of our knowledge, but we remark that such an approach could be ideal to make tailored, reproducible V-grooves without additional process steps.

\subsubsection{Photolithography with anisotropic etching. }
 
Photolithography, a high-throughput microfabrication technique that involves the patterning of a resist with UV light through a photomask,\cite{Madou2002} has been used together with anisotropic etching to define nanoplasmonic V-groove devices (Fig.~\ref{fgr:VgrvUVlith}).\cite{Smith2014} The process, hereby summarised [Fig.~\ref{fgr:VgrvUVlith}(a)], starts with photolithography to define openings in a \ce{SiO2} etch mask film that are aligned with the $\langle100\rangle$ planes of the silicon substrate [i-ii]. These openings determine the dimensions of the V-grooves to be formed by subsequent anisotropic etching [iii-iv]. Once the grooves are formed (and possibly tailored by thermal oxidation [v]), the surface is coated with a metal [vi] of a thickness $\geq{}50~\text{nm}$ in order to avoid the coupling of plasmons between the above- and below-interfaces [Fig.~\ref{fgr:VgrvUVlith}(b),(c)]. 

Importantly, the combined procedure here is able to maintain the cost advantages of wafer-scale production while simultaneously yielding high quality plasmonic devices. In addition, propagation-length measurements on such devices have been consistent with curvature radii below $5~\text{nm}$ despite the rounding effects that could be expected from metallisation,\cite{Smith2014} indicating that the approach offers a facile method to realise nanometre-scale features. Furthermore, since the photolithography step is only required to define relatively large feature sizes ($>{}100$'s nanometres), equipment specifications are relaxed and widely accessible.

\subsection{Nanoimprint lithography}

Nanoimprint lithography (NIL) represents an exceptionally low-cost, high-throughput fabrication method that may be used to manufacture plasmonic devices.\cite{Guo2007a,Boltasseva2009} It essentially involves the creation of a hard, reusable, nanostructured master stamp that is pressed into relatively soft polymer material. Importantly, the NIL process is highly repeatable, which leads to production expenses that are scarcely more than the energy and material costs involved. NIL relies on mechanical deformation to transfer the pattern from stamp to the polymer, a physical process that enables structure dimensions to be defined well beyond the limitations of light diffraction or beam scattering; demonstrations of NIL routinely achieve ${\sim}10~\text{nm}$ features\cite{Schift2008} and $1-2~\text{nm}$\cite{Malyarchuk2005} has been reached. Furthermore, NIL-based techniques offer new paths to improve structured-metal surface roughness through the use of atomically-smooth stamps and thermal relaxation, and can employ diverse mould types such as metal direct nanoimprinting (embossing) for the fabrication of plasmonic components and other metal-containing optical (and electronic) devices.\cite{Buzzi2008,Radha2013}

The ability to mass-produce plasmonic components via NIL is of significant consequence and potentially renders a host of plasmonics-based applications economically viable. The implications for integrated optics alone are tremendous, with NIL representing the only established method to form complex plasmonic components on a wafer scale. In addition, large surface-area technologies such as black metals, structurally-coloured surfaces and enhanced light filters, as well as disposable devices such as sensitive diagnostic tools, threat detectors and environmental monitors are all poised to benefit from the high-throughput production offered by NIL.

\begin{figure}[t!] 
   \centering
    \includegraphics[width=8.6cm,height=6.57cm]{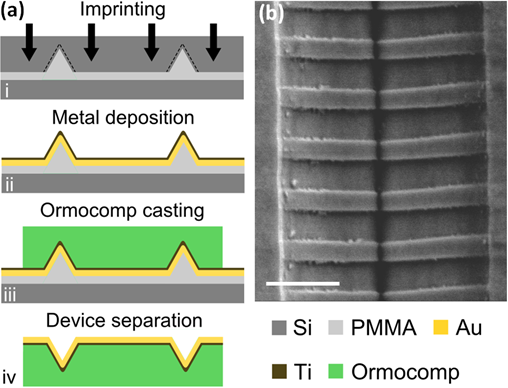}
    \captionsetup{justification=justified}
    \caption{(a)[i-iv] The NIL fabrication scheme for producing gold V-grooves. [i] A silicon stamp containing of the V-grooves (and optional Bragg gratings) is imprinted into PMMA polymer resist, resulting in an inverted polymer replica of the stamp. [ii] Gold $(200~\text{nm})$ is deposited on the patterned polymer followed by a $(5~\text{nm})$ titanium adhesion-promoting layer. [iii] Ormocomp is cast over the metal and defined by UV-lithography to form device-specific polymer substrates. [iv] Ion-beam-etching with the Ormocomp layer as an etch-mask removes the unwanted metal. The PMMA layer is then dissolved in acetone, releasing the stand-alone plasmonic components. (b) SEM of a gold V-groove produced by NIL consisting of a Bragg grating. The scale bar is $1~\mu\text{m}$. (Adapted with permission from ref.\cite{Smith2012} Copyright \copyright{} 2012 Optical Society of America.)}
    \label{fgr:VgrvNIL}
\end{figure} 

Plasmonic V-grooves have been fabricated by NIL\cite{Fernandez-Cuesta2007a,Nielsen2008a} using a double pattern transfer method that replicated the structures of a stamp into metal. The technique relied on the greater adhesion of titanium to Ormocomp than the gold to poly(methylmethacrylate) (PMMA), thereby allowing for the double pattern transfer. This fabrication process was an extension of previous results that produced sharp metal wedges via controlled profiling of a silicon stamp followed by a single pattern transfer.\cite{Bilenberg2005} The double-transfer NIL method for making V-grooves has been further built upon with the introduction of Bragg gratings added to the V-grooves of the stamp by electron-beam lithography; the NIL process is illustrated in Fig.~\ref{fgr:VgrvNIL}(a) and the SEM image of one such V-groove consisting of Bragg gratings shown in Fig.~\ref{fgr:VgrvNIL}(b).\cite{Smith2012} 

\section{Excitation and characterisation}
\label{sec:Excite_&_Characterise}

A variety of configurations have been developed to excite and characterise SPPs based on either photonic or electronic approaches. In both cases, the conversion efficiency represents a major topic to overcome due to otherwise inherent mismatch involved in the photon-plasmon or electron-plasmon couplings. Furthermore, the characterisation of the generated plasmons must account for the non-radiative and potentially nanoscale aspect they possess. Despite these challenges, recent reports using tapered grooves cite significant progress regarding high photon-plasmon coupling efficiencies,\cite{Smith2014,Beermann2014} and super-resolved characterisation of GSP modes has been demonstrated.\cite{Raza2014a} In this section, we review the methods that have been established to excite and characterise plasmons in tapered grooves, focusing on recent achievements and the opportunities they present.

\subsection{Photon-plasmon techniques}

Methods to promote efficient photon-plasmon coupling generally fall under one of two categories: overcoming the wavevector (momentum) mismatch,\cite{Raether1988,Maier2007} or maximising the field amplitude overlap integral.\cite{Briggs2010} Wavevector matching may be achieved by prism- or resonant-based couplers, which under specific configurations can supply the additional momentum necessary for photons to match plasmons (and vice versa). While prism-based approaches may be regarded as bulky or limited to many plasmonic systems due to the requirement of additional high refractive index materials, resonant-based schemes have been shown to offer both compactness and high coupling efficiencies, with prominent examples including Yagi--Uda antennas,\cite{Kosako2010} phase-engineered elements\cite{Vercruysse2013} and grating couplers.\cite{Baron2011} For tapered-groove based devices, resonant coupling is partly responsible for the photon-plasmon conversion in enhanced transmission filters.\cite{Søndergaard2010c} Despite the success of resonant-based couplers to plasmonics, maximising the field overlap is the prevailing technique for exciting plasmons in tapered grooves due to its practical simplicity and effectiveness. This holds true even for most resonator-type configurations of tapered grooves, since the plasmon excitation is still initiated by the field overlap at the interface between the source and plasmonic mode.\cite{Søndergaard2010b,Søndergaard2010c,Zhang2011} As a result, the most common configurations to facilitate photon-plasmon exchange in tapered grooves are end--fire coupling and direct illumination, both of which can promote large, or at least sufficient, field overlap integrals.

\subsubsection{End--fire coupling. }

End--fire coupling consists of a tapered polarisation-maintaining optical fibre aligned to the edge of a plasmonic structure. It is a common plasmon-excitation technique for tapered-grooves due to the wide availability of the required laboratory equipment while also corresponding to the light paths of devices designed for planar integration. The photon-plasmon coupling efficiency, in practice reaching around several percentage points,\cite{Smith2012} is given by the electric field distribution match (including polarisation) between the focused Gaussian beam exiting the fibre and the plasmonic mode of the waveguide. The configuration may be paired with the cutback method, coupling to fluorescent beads,\cite{Fernandez-Cuesta2009} or near-field probing\cite{Zenin2011a} (discussed below) to characterise parameters of interest.\cite{Zenin2012a} 

Despite the prevalence of end--fire coupling, one should be mindful of two important drawbacks that are especially the case for tapered grooves. First, the requirement of a cleaved sample end-facet, for which the cleaving process itself is cumbersome, deteriorates the waveguide entrance quality and leaves it prone to further damage. Second, the coupling efficiency is critically and uniquely sensitive to fibre position and orientation, which causes routine measurements on similar devices to be challenging and imprecise. These drawbacks, while often acceptable, have nevertheless spurred investigation into other field-overlap-based techniques.

\subsubsection{Direct illumination. }

\begin{figure}[t!] 
   \centering
    \includegraphics[width=8.6cm,height=7.41cm]{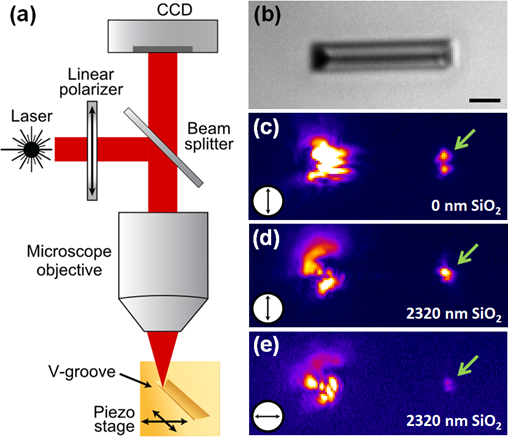}
    \captionsetup{justification=justified}
    \caption{(a) Setup for direct illumination of CPPs via nanomirrors. Light from a laser diode is linearly polarised before impinging onto the sample through a microscope objective. (b) Bright-field image of a V-groove device with waveguide termination nanomirrors. The scale bar is $4~\mu\text{m}$. (c)-(e) Experimentally observed radiation from the out-coupling termination mirrors at the ends of the waveguides. The larger spots on the left are the direct reflection from the incident beam and the right spots (green arrows) are out-coupled light from the termination mirrors. Insets at the bottom left of the images represent the polarisation of the incident electric field. (c) A pair of intensity peaks corresponding to the wedge-based pair of modes occurs for the case without oxidation. (d) A single intensity peak corresponding to a CPP mode is out-coupled from a V-groove device with a $2320~\text{nm}$ \ce{SiO2} layer. The incident polarisation closely matches the electric field of the CPP and leads to efficient in-coupling. (e) A reduced out-coupled intensity occurs for the same device in (d) when the incident light polarisation is rotated $90\degree$ and no longer matches the CPP electric field. (Adapted with permission from ref.\cite{Smith2014} Copyright \copyright{} 2014 American Chemical Society.)}
    \label{fgr:NanomirrExpt}
\end{figure} 

Photon-plasmon coupling by direct illumination involves the conversion of light from an out-of-plane source into the supported mode of a plasmonic device. The light paths given by this configuration coincide with a wide range of tapered-groove-based applications, such as solar harvesters,\cite{Søndergaard2012} sensors,\cite{Søndergaard2010b} boosted EOT filters,\cite{Søndergaard2010c} coloured surfaces,\cite{Zhang2011} and multi-level photonic circuits. Additionally, direct illumination is crucial to the development of practical LOC devices since it allows plasmonic components to be simultaneously excited and characterised under microscope settings by colinearly delivering the light source via the imaging optics.\cite{Smith2014} 

There are two established mechanisms to excite plasmonic modes in tapered grooves using the field overlap from direct illumination. First, the plasmon excitation can be initiated by scattering of the incident light off from the outer edges of the grooves (wedges), where the coupling efficiency is determined by the mode match at the interface. This scattering mechanism is responsible for light-SPP coupling in the cases of plasmonic black metals,\cite{Søndergaard2012} coloured and field-enhanced surfaces,\cite{Søndergaard2010b,Zhang2011} and boosted EOT filters.\cite{Søndergaard2010c} Larger groove openings generally correspond to larger field overlap integrals and therefore promote higher photon-plasmon conversion ratios, with plasmonic black metal designs achieving efficiencies at least as high as $97\%$.\cite{Beermann2014}

The second mechanism involves the use of nanomirrors (Fig.~\ref{fgr:NanomirrExpt}), where normally incident light is redirected along the direction of propagating CPPs and the resulting field overlap with the plasmonic mode facilitates the photon-plasmon coupling.\cite{Radko2011a,Smith2014} The dependence of the incident electric field orientation on the coupling efficiency has been shown to allow for confirmation of the photon-plasmon coupling via nanomirrors [Fig.~\ref{fgr:NanomirrExpt}(d),(e)] due to the strongly polarised nature of the supported plasmonic modes. The most efficient nanomirror inclination angle has been found to be $45\degree$,\cite{Radko2011a} corresponding to the maximum redirection of light into the propagation axis. Initial demonstrations of FIB-milled nanomirrors provided maximum coupling efficiencies up to $10\%$,\cite{Radko2011a} with more recent devices fabricated by photolithography and anisotropic etching indicating $>50\%$ photon-plasmon coupling due to an increased light collection area of the mirror and smoother metal surfaces.\cite{Smith2014} Nanomirror designs that further optimise this collection area, together with the use of inclination angles at or close to $45\degree$, could expect to realise even higher photon-plasmon coupling efficiencies.

\subsubsection{Integrated waveguide coupling. }

Integrated waveguide coupling represents the photon-plasmon exchange at the interface of silicon and metallic waveguides. It is a highly promising approach since it enables the advantages of planar photonic devices to be straightforwardly combined with sub-diffraction-limited plasmonics.\cite{Kinsey2015} Its coupling efficiency, determined by the field amplitude overlap, may potentially reach large values (e.g. ${\sim}80\%$)\cite{Briggs2010} and is the most compact photon-plasmon coupling arrangement available.  

\begin{figure}[t!] 
   \centering
    \includegraphics[width=8.6cm,height=6.27cm]{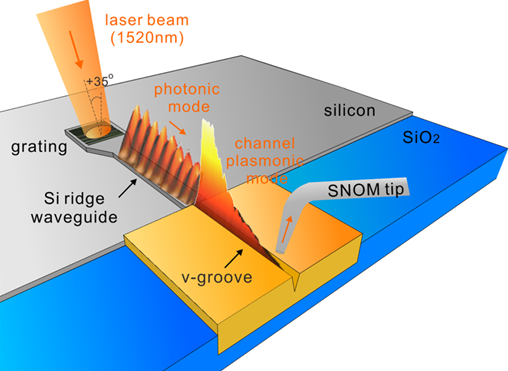}
    \captionsetup{justification=justified}
    \caption{Schematic of the configuration for coupling to V-groove plasmonic modes from a silicon ridge waveguide. (Adapted with permission from ref.\cite{Burgos2014} Copyright \copyright{} 2014 American Chemical Society.)}
    \label{fgr:SiWGCoup}
\end{figure}

Recently, integrated waveguide coupling was implemented for exciting plasmonic modes in tapered grooves via the photonic modes of silicon ridge waveguides (Fig.~\ref{fgr:SiWGCoup}).\cite{Burgos2014} The source light (telecommunications wavelengths) was initially coupled into the silicon waveguide by means of a grating coupler that could select either the transverse electric (TE) (electric field parallel to the substrate) or the transverse magnetic (TM) photonic modes (electric field perpendicular to the substrate) depending on the angle of excitation. Due to the strongly polarised nature of the plasmons supported in tapered grooves (section~\ref{sec:Plasmons_in_V-shaped_grooves}), the choice of TE or TM modes in the silicon waveguide subsequently determined whether groove-based CPP modes or wedge-based SPP modes were launched at the ridge-groove interface. By means of this selective capability, the coupling to highly confined CPP modes could be confirmed and therefore enabled significant progress to be made towards the realisation of advanced integrated silicon/plasmonic nanocircuits. The scheme's initial coupling efficiency into the CPP mode, calculated to be ${\sim}22.2$\%,\cite{Burgos2014} was well sufficient for demonstration purposes and may be improved upon by further optimisation of the field overlap integral.

\subsubsection{Near-field scanning. }

Near-field scanning optical microscopy (NSOM) consists of a nanoscopic optical probe that enters the evanescent near-field and exploits the resulting perturbation to glean information about the local amplitude (and possibly phase). The major advantage of NSOM is that it enables SPP modes to be probed with resolutions down to the 10's of nanometres, which is particularly important to plasmonics due to the challenges associated with theoretical modelling and the intimate dependence the electromagnetic field has on the nanoscale geometry. For these reasons, experimental mapping of nanoscopic light fields represents a significant topic of present-day research, where a rapidly increasing variety of approaches and functionalities have become available.\cite{Rotenberg2014}  

\begin{figure}[t!] 
   \centering
    \includegraphics[width=8.6cm,height=6.54cm]{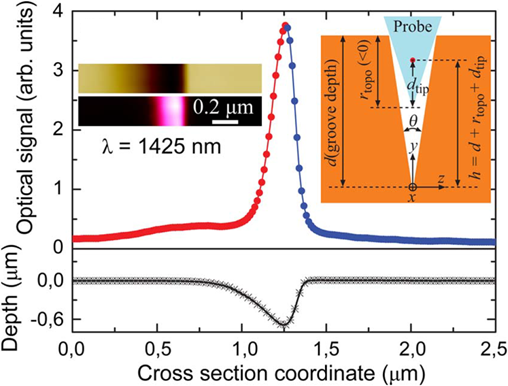}
    \captionsetup{justification=justified}
    \caption{Transverse profile of a propagating CPP mode (filled circles) and the V-groove topography (asterisks) as taken by an NSOM probe. Left insets: images corresponding to the near-field and topographical scans. Right inset: cross-section schematic of the V-groove / NSOM probe configuration under consideration. The probe cannot reach the bottom of the groove. (Adapted with permission from ref.\cite{Zenin2011a} Copyright \copyright{} 2011 Optical Society of America.)}
    \label{fgr:NSOMexpt}
\end{figure}

The NSOM technique has been used extensively to directly map the properties of CPPs in plasmonic V-grooves, including propagation length,\cite{Bozhevolnyi2005a,Zenin2011a} subwavelength mode confinement\cite{Volkov2009a,Zenin2011a} and mode effective refractive indices,\cite{Zenin2011a} as well as the operation of novel components.\cite{Bozhevolnyi2006c,Volkov2007a,Smith2012,Burgos2014} However, accurate NSOM measurements require deep knowledge of the interaction between the near-field probe and the sample under test;\cite{Novotny2007} in the case of tapered grooves, it is crucial to account for the spatial convolution caused by the near-field probe profile and the V-shaped geometry (Fig.~\ref{fgr:NSOMexpt}).\cite{Zenin2011a} Furthermore, inherent problems related to instabilities of the probe-surface distance have proven to be difficult to avoid, especially in narrow and deep V-grooves, which can complicate precise CPP characterisation.\cite{Zenin2011a} Nevertheless, NSOM measurements have provided a wealth of information of plasmons in tapered grooves, and, due to their expediency, can provide immediate explicit insight that would otherwise only be indirectly available through time-consuming full field simulations.

\subsection{Electron-plasmon techniques}

The excitation of plasmons by electrons is a potentially enabling approach to integrate electronics with photonics\cite{Gimzewski1988,Bharadwaj2011} that additionally offers the possibility to retrieve unprecedented spatial information of the optical modes in a plasmonic device.\cite{GarciadeAbajo2010,Nicoletti2011,Raza2014a} A range of electron-plasmon techniques are available that may be categorised based on whether the acceleration voltage of the electron source is low ($<10$~V) or high ($100$'s~kV). Low-voltage-source electrons, e.g. delivered by scanning tunnelling microscopes,\cite{Gimzewski1988,Bharadwaj2011} are appropriate for the exploration of effective electronic-photonic integration schemes since the associated configurations could be feasibly miniaturised and are not demanding in terms of power consumption. A low-voltage scheme for exciting plasmons in tapered grooves by electrons is yet to be demonstrated, but would represent an important advance towards their technological development. High-voltage-source electrons, e.g. fired in a tightly focused beam at a metal surface,\cite{Ritchie1957} have been used to map plasmonic modes with sub-nanometre spatial resolution and allow for investigations into the properties of extremely confined plasmons that are otherwise inaccessible.\cite{Raza2014a} In the following, we discuss the results of such a high-voltage electron-plasmon technique, from which extremely confined GSP modes in tapered grooves have been studied.

\begin{figure}[t!] 
   \centering
    \includegraphics[width=6.5cm,height=8.56cm]{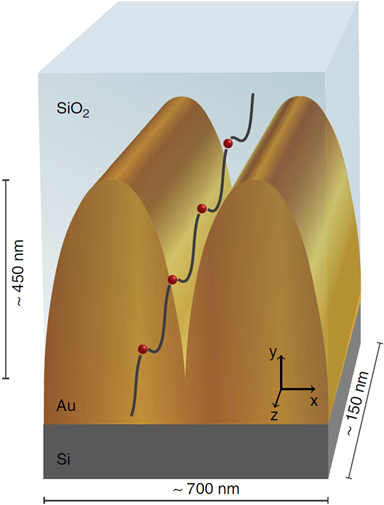}
    \captionsetup{justification=justified}
    \caption{Artistic impression of a gold nanogroove with a swift electron beam moving parallel to the groove axis. The groove is filled with \ce{SiO2} and the substrate is silicon. (Reprinted with permission from ref.\cite{Raza2014a} Copyright \copyright{} 2014 Macmillan Publishers Ltd.)}
    \label{fgr:EELS}
\end{figure} 

\subsubsection{Electron energy-loss spectroscopy. }

EELS involves firing a beam of high energy electrons (acceleration voltage on the order of 100 kV) with a scanning transmission electron microscope, such that electrons impinging on metallic nanostructures can transfer their energy by inelastic scattering into plasmons.\cite{GarciadeAbajo2010} EELS has been widely used to characterize localized surface plasmons in single metallic nanoparticles\cite{Nelayah2007,Bosman2007,Koh2009} since it can explore their properties with a spatial resolution of a few angstroms only\cite{Scholl2012,Raza2013a} due to the short de Broglie wavelength of electrons. Such a resolving power is far beyond the means of diffraction-limited optical characterization methods and even surpasses state-of-the-art NSOM techniques\cite{Eisele2014} by more than a factor of 10. Moreover, the momentum of electrons moving at roughly half the speed of light is ~500 times greater than the corresponding photons of the same energy, thus enabling the excitation of surface plasmons far from the light line where they are tightly confined at the metal-dielectric interface. 

EELS-based investigations have enabled the unique opportunity to probe GSP modes in the crevice of nanogrooves with structural feature sizes below $5~\text{nm}$.\cite{Raza2014a} The first use of electron-based techniques to excite propagating plasmons occurred in the 1970’s to measure the dispersion of SPPs on thin aluminium films,\cite{Pettit1975} yet aside from otherwise preliminary investigations of GSPs in MIM structures ($10~\text{nm}$ dielectric gaps) with cathodoluminescence,\cite{Kuttge2009} the first nanoscale exploration of GSP modes using electrons happened only very recently.\cite{Raza2014a} The cause of this 40-year delay resided mainly in the difficulty to prepare an appropriate (e.g. ${\sim}150~\text{nm}$-thick) cross-sectional slice of the plasmonic device suitable for measurements (Fig.~\ref{fgr:EELS}). Particular cross-section thicknesses are required for EELS in order to render the samples sufficiently transparent for the electron beam to pass through and also reduce the influence of Cherenkov radiation, while not being too thin compared to the propagation length of the GSP modes under investigation. For the recent EELS-based investigation of nanogrooves,\cite{Raza2014a} procuring a ${\sim}150~\text{nm}$-thick cross-section required a challenging process that involved first protecting the device via \ce{SiO2} and platinum deposition, cutting out a section from a larger sample by FIB techniques, welding the section to a lift-out grid, and further refining the lamella thickness carefully by additional FIB profiling.\cite{Raza2014a} While we note that only very lossy modes with propagation lengths on the order of a few tens of nanometres [i.e. the antisymmetric GSP mode (section~\ref{sec:Challenges})] could be investigated here\cite{Raza2014a} as a result of the lamella's thinness (albeit offering significant insights), we foresee in the near future that EELS may be employed to precisely investigate modes of longer propagation length by considering other protective dielectric materials or configurations and thicker lamellas (several 100's of nanometres). Furthermore, we anticipate the use of EELS-based methods to study the presence of non-local effects (section~\ref{sec:Challenges}) where the maximum possible confinement of plasmonic modes is limited by quantum-wave phenomena.

\section{Challenges}
\label{sec:Challenges}

In this section we discuss several challenges critical to the design of tapered-groove-based devices. In general, our relatively newfound ability to control light on the nanoscale has led us to interact with otherwise unnoticed effects, such as the non-local response or unintended device asymmetries, which can profoundly influence system behaviour. Here we focus on the challenges such effects present to the implementation of tapered grooves, and follow with a discussion of the issues that remain for integrating tapered grooves into practical device platforms. We remark that the general and otherwise important topic of circumventing plasmonic losses\cite{Noginov2008,Leon2010,Khurgin2014,Khurgin2015} is beyond the scope of this review, which as of this writing remains unexplored for tapered grooves.

\subsection{Non-local effects}

In our considerations so far, we have discussed the properties of plasmons in the context of classical electrodynamics and the Drude local-response approximation (LRA) of the polarization, where the current density is locally related to the electrical field through Ohm's law.\cite{Maier2007} However, extreme subwavelength mode confinement draws attention to non-local effects beyond the local Drude description due to the quantum-wave nature of the electron gas at the nanoscale.\cite{Raza2014,Manuscript2014} In noble-metal systems, the relevance of non-local corrections to the LRA picture can be estimated from the non-local lengthscale $\delta_{\text{nl}}{\,}{\sim}{\,}\sqrt{(\beta/\omega)^2-i(D/\omega)}$, where $\beta{\,}{\propto}{\,}v_F$ is the speed of longitudinal pressure waves in the plasma and $D$ is a constant for charge-carrier diffusion that accounts phenomenologically for e.g. surface scattering and Landau damping.\cite{Mortensen2014} Typically, the non-local lengthscale is in the sub-nanometre regime.\cite{Raza2014}

\begin{figure}[t!] 
   \centering
    \includegraphics[width=8.6cm,height=8.6cm]{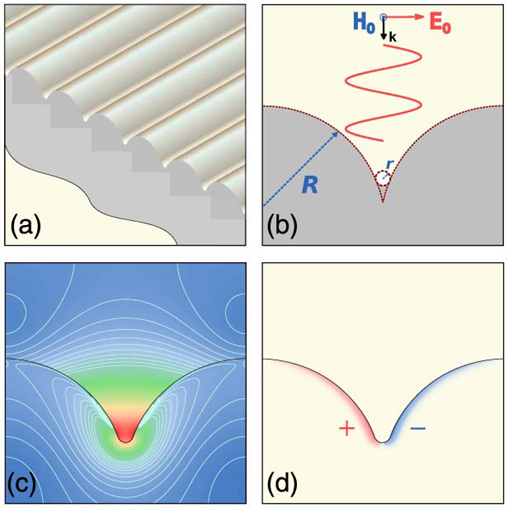}
    \captionsetup{justification=justified}
    \caption{Schematic of the system used for studying non-local effects in tapered grooves. (a) Array of grooves formed by half-cylindrical nanorods. (b) Cross-section of the unit cell. (c) and (d) typical electric-field intensity and charge distributions for a dipole mode. (Reprinted with permission from ref.\cite{Toscano2012} Copyright \copyright{} 2012 Optical Society of America.)}
    \label{fgr:NonLocal}
\end{figure} 

Tapered groove structures have been used as an initial model system\cite{Garcia-Vidal1996} to appreciate the enormous surface-enhanced Raman scattering (SERS) effect on metal surfaces with nanoscale roughness.\cite{Moskovits1985} Studies in this regard have since illustrated the need for going beyond the LRA,\cite{Xiao2008a} where the intrinsic length scale $\delta_{\text{nl}}$ of the electron gas has been shown to smear out otherwise-assumed field singularities.\cite{Toscano2012} As a consequence, the SERS enhancement factor remains finite (Fig.~\ref{fgr:NonLocal}) even for geometries with infinitely sharp features. The qualitative aspects of non-locality seem to have been experimentally confirmed in a configuration of graphene on the rough silver surfaces of tapered grooves,\cite{Zhao2014} where the SERS enhancement was more accurately predicted by accounting for non-local effects. Numerical studies have shown that the properties of waveguides are also affected by non-local responses, with tapered-groove waveguides exhibiting a fundamental limit on the achievable mode confinement.\cite{Toscano2013} This translates into a maximum ceiling on the Purcell enhancement of dipole-emitter decay systems that can be reached, and may have important implications in quantum plasmonics (section~\ref{sec:Applications}).

\subsection{Antisymmetric GSPs}

GSP modes may be classified according to the symmetry of the electric field component oriented across the MIM gap, which can be either symmetric -- i.e. those discussed so far in this Review -- or antisymmetric. Unlike symmetric GSPs, antisymmetric GSPs (aGSPs) feature very strong absorption (propagation lengths of order $10~\text{nm}$)\cite{Raza2014a} due to their symmetric induced-charge patterns. Such a rapid decay of energy implies that any coupling contributions into aGSP modes may be regarded as loss, yet this also makes useful investigations into their occurrence a non-trivial task. 

\begin{figure}[t!] 
   \centering
    \includegraphics[width=8.6cm,height=5.39cm]{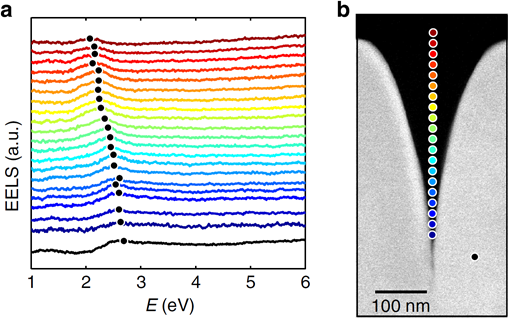}
    \captionsetup{justification=justified}
    \caption{(a) Waterfall plot of experimental EELS measurements at the corresponding positions indicated on the groove image in (b). (Reprinted with permission from ref.\cite{Raza2014a} Copyright \copyright{} 2014 Macmillan Publishers Ltd.)}
    \label{fgr:AGSP}
\end{figure}

Recently, the presence of aGSPs in tapered grooves was confirmed experimentally using high spatial and energy resolution EELS.\cite{Raza2014a} The technique involved scanning the electron probe from the top towards the bottom of a thin convex groove lamella while monitoring the EELS response (Fig.~\ref{fgr:AGSP}). The blue-shift of the resonance peak indicated the onset of an aGSP mode in the narrowing crevice of the groove (down to ${\sim}5~\text{nm}$ width) rather than the exclusive presence of (global) groove GSP modes whose peak positions would otherwise not depend on the electron location in the groove.

While the EELS technique generated aGSP modes by symmetric charge depletion in the groove from the electron probe, an important insight of photon-plasmon coupling into aGSP modes was obtained in the study regarding the supporting structure's \textit{asymmetry}. For example, black gold based on convex groove arrays showed experimentally greater-than-expected light absorption,\cite{Søndergaard2012} since they were initially simulated by perfectly symmetric features and normally incident light such that the absorption was thought to be based solely on the dissipation of symmetric GSP modes. For the symmetric case, incident plane waves generate charged dipole moments that may be conceptually decomposed into two SPPs forming at each wedge in antiphase propagating downwards into the groove. However, the configurations in practice consisted of slight structural asymmetries and inclined light, which in turn allowed for the SPPs excited at opposite wedges to meet in the interior of the groove with relative antiphase-shifts and thereby contribute energy into lossy aGSP modes. It should be noted that ideal-case symmetrical systems -- i.e. the condition considered in most numerical simulations -- do not support the generation of aGSP modes by light, leaving their influence on real devices otherwise unnoticed. Accordingly, the presence of aGSPs represents an easily-overlooked yet potentially crucial factor to consider when practically implementing nanoplasmonic devices that could be affected by additional channels of energy dissipation, e.g. in black metals,\cite{Beermann2014} quantum emitter systems,\cite{Martin-Cano2010} or logic circuits with junctions/bends.\cite{Burgos2014}

\subsection{Integration}

Plasmonic systems require methods to interface with the external world that are not only favourable in terms of fabrication (section~\ref{sec:Design_&_Fabrication}) and coupling efficiency (section~\ref{sec:Excite_&_Characterise}), but are also able to simultaneously maintain the advantages uniquely offered by the nature of plasmons, such as circuit miniaturisation, high information bandwidth and large field enhancement. Integrating devices so as to address this ensemble of requirements represents a major pursuit towards the realisation of useful plasmonics-based technologies. In the following, we discuss our thoughts on several specific issues that are important to the integration of tapered plasmonic grooves.

\subsubsection{Electrically-driven circuits. }

The development of nanoplasmonic components that combine the superior traits of fast electronics with high bandwidth photonics is a potential route to realise the next generation of information processors. The challenge here is to achieve an effective and compact interface between the electronic domain and SPP modes. Nanoplasmonic lasers are a possible answer, although the realisation of electrically-driven, easily-integrable, power efficient and room temperature devices has, despite progress in the area,\cite{Bergman2003,Hill2007,Oulton2009,Noginov2009,Khajavikhan2012} proven difficult to achieve. A recent demonstration that integrated nanoscale plasmonic light-emitting diodes with nanometallic circuits represents a possible alternative.\cite{Huang2014} In any case, the integration of an electrically-driven source to tapered plasmonic grooves is a topic that remains to be explored. Efforts to demonstrate such a system must consider fabrication aspects: e.g. the inclusion of electrodes to drive the device should keep the production wafer-scale and avoid (overly) disturbing the plasmonic modes. Additionally, precise and symmetric positioning of an electrically-driven element to within the groove may be required in order to sufficiently couple to CPPs, which could be frustrated by the large van der Waals forces present at the nanoscale and prevent controlled alignment. Overall, the task to realise electrically-driven tapered plasmonic groove circuits can expect to encounter major hurdles, but practical solutions would be of significant consequence. 

\subsubsection{Corner rounding. }

The finite rounding of corners in tapered grooves, particularly at the bottom groove tip, must be considered in real-world devices -- especially those seeking to exploit plasmon-based field enhancement. Recent advances in the groove formation techniques of FIB milling\cite{Raza2014a} and anisotropic etching\cite{Smith2014} now enable previously unattainable sharp tips with curvature radii below $10~\text{nm}$ to be achieved. Nevertheless, finite corner rounding is unavoidable and is a result of the groove-defining fabrication step or the deposition of a metal layer. Since the ensuing CPP mode distributions are sensitive to the groove tip's sharpness down to the single nanometre level,\cite{Vernon2008a} usefully accurate knowledge of these specifications can be challenging to acquire. Few SEMs are capable of observing features much below $10~\text{nm}$, although a possible alternative is to use scanning transmission electron microscopes capable of sub-nm resolution. Indeed, such a procedure has been explored recently, with initial studies able to characterise the gaps of ultra-sharp convex grooves in a specially prepared lamellar down to values of ${\sim}5~\text{nm}$;\cite{Raza2014a} finer resolutions should be possible under appropriate conditions. Despite the importance of considering tip rounding, we note that non-local effects described earlier in this section impose a practical limit beyond which no additional confinement occurs, and should be kept in mind so as to avoid unnecessary attempts at producing arbitrarily sharp groove tips.

\subsubsection{Devices made on the wafer-scale. }

Tapered groove devices produced by wafer-scale techniques can be exceptionally low-cost, representing sizeable opportunities to the topic areas explored with plasmonics. However, while individual components produced in parallel may have been demonstrated using anisotropic etching\cite{Smith2014} or NIL replication,\cite{Nielsen2008a,Smith2012} the development of cost-effective and wholly integrated devices must address a number of already-known challenges. First, anisotropically-etched grooves are inherently straight, yet must be able to turn to be useful for sophisticated circuits. A possible solution could be based on realising right-angled turns, which should be feasible based on the alignment of the silicon crystal planes and the minimal propagation losses that occur around sharp bends.\cite{Bozhevolnyi2006c} Second, the planar contours that form as a result of the \ce{SiO2} growth step for tailoring CPP modes [Fig.~\ref{fgr:VgrvUVlith}(c)] must be considered when interfacing such grooves in multilayer photonic circuits or bonding a flat sealing lid to a fluidic-based LOC device. Third, integrating NIL-replicated components supported by a polymer substrate into existing technology platforms will require methods such as extending the imprint process to the entire circuit or employing otherwise appropriate imprint materials. In all these cases, the challenges are not insurmountable but nevertheless critical to the potentially affordable integration of tapered grooves.

\section{Outlook}
\label{sec:Conclusion_&_Outlook}

Research into plasmons of tapered grooves has achieved many important milestones to-date that have been central to the development of an assortment of applications and scientific investigations. Progress continues across an increasingly diverse number of fronts that currently includes nanophotonic circuits, quantum plasmonics, light harvesters, high-definition coloured surfaces, advanced optical filters, LOC components, and the fundamental understanding of plasmon excitations and their interactions. The motives and successes behind the advances are largely owed to both the unique and common properties that GSPs and CPPs can possess across a variety of configurations and groove profiles. In particular, tapered grooves offer the principle advantages enabled by the nature of plasmons -- light confinement beyond the diffraction limit -- together with the flexibility of device design and fabrication-method-selection that is unmatched in the field of plasmonics.

The first demonstration of plasmonic modes supported by tapered grooves was reported 10 years ago,\cite{Bozhevolnyi2005a} yet the extensive list of results presented throughout our Review suggests that the topic area still remains in its ``golden age''. We further remark that there are numerous research directions of potentially major consequence that are yet to be explored. Fully-functioning nanoscale light sources integrated with tapered grooves would open many doors regarding information-based and LOC-type technologies. Turning plasmonic losses around so as to efficiently convert electromagnetic energy into heat inside the grooves could open up new possibilities for realising thermally-based optical switches or nano-object control. The plasmonic modes located at the outer wedges rather than the groove tips possess relatively large propagation lengths and unique polarisation properties that may also facilitate novel device schemes. In any case, the many promising applications offered by plasmonic tapered grooves and the unique possibilities they enable for investigating fundamental light-matter interactions point to exciting research years ahead.

\section{Acknowledgements}
\label{sec:Acknowledgements}

C.L.C.S. acknowledges financial support from the Danish Council for Independent Research (Grant No. 12-126601). N.S. acknowledges financial support from the Lundbeck
Foundation (Grant No. R95-A10663). A.K. acknowledges partial financial support from the Innovation Fund Denmark (Grant No. 10-092322, PolyNano). N.A.M. acknowledges funding from the Danish Council for Independent Research, (Grant No. 1323-00087). S.I.B. acknowledges financial support from the European Research Council (Grant No. 341054, PLAQNAP). The Center for Nanostructured Graphene is sponsored by the Danish National Research Foundation (Grant No. DNRF58).

\balance


\footnotesize{
\bibliography{BibliotekNSRev}

\providecommand*{\mcitethebibliography}{\thebibliography}
\csname @ifundefined\endcsname{endmcitethebibliography}
{\let\endmcitethebibliography\endthebibliography}{}
\begin{mcitethebibliography}{219}
\providecommand*{\natexlab}[1]{#1}
\providecommand*{\mciteSetBstSublistMode}[1]{}
\providecommand*{\mciteSetBstMaxWidthForm}[2]{}
\providecommand*{\mciteBstWouldAddEndPuncttrue}
  {\def\EndOfBibitem{\unskip.}}
\providecommand*{\mciteBstWouldAddEndPunctfalse}
  {\let\EndOfBibitem\relax}
\providecommand*{\mciteSetBstMidEndSepPunct}[3]{}
\providecommand*{\mciteSetBstSublistLabelBeginEnd}[3]{}
\providecommand*{\EndOfBibitem}{}
\mciteSetBstSublistMode{f}
\mciteSetBstMaxWidthForm{subitem}
{(\emph{\alph{mcitesubitemcount}})}
\mciteSetBstSublistLabelBeginEnd{\mcitemaxwidthsubitemform\space}
{\relax}{\relax}

\bibitem[Maier \emph{et~al.}(2001)Maier, Brongersma, Kik, Meltzer, Requicha,
  and Atwater]{Maier2001}
S.~A. Maier, M.~L. Brongersma, P.~G. Kik, S.~Meltzer, A.~A.~G. Requicha and
  H.~A. Atwater, \emph{Adv. Mater.}, 2001, \textbf{13}, 1501--1505\relax
\mciteBstWouldAddEndPuncttrue
\mciteSetBstMidEndSepPunct{\mcitedefaultmidpunct}
{\mcitedefaultendpunct}{\mcitedefaultseppunct}\relax
\EndOfBibitem
\bibitem[Zia \emph{et~al.}(2006)Zia, Schuller, Chandran, and
  Brongersma]{Zia2006}
R.~Zia, J.~A. Schuller, A.~Chandran and M.~L. Brongersma, \emph{Mater. Today},
  2006, \textbf{9}, 20--27\relax
\mciteBstWouldAddEndPuncttrue
\mciteSetBstMidEndSepPunct{\mcitedefaultmidpunct}
{\mcitedefaultendpunct}{\mcitedefaultseppunct}\relax
\EndOfBibitem
\bibitem[Lal \emph{et~al.}(2007)Lal, Link, and Halas]{Lal2007}
S.~Lal, S.~Link and N.~J. Halas, \emph{Nature Photon.}, 2007, \textbf{1},
  641--648\relax
\mciteBstWouldAddEndPuncttrue
\mciteSetBstMidEndSepPunct{\mcitedefaultmidpunct}
{\mcitedefaultendpunct}{\mcitedefaultseppunct}\relax
\EndOfBibitem
\bibitem[Schuller \emph{et~al.}(2010)Schuller, Barnard, Cai, Jun, White, and
  Brongersma]{Schuller2010}
J.~A. Schuller, E.~S. Barnard, W.~Cai, Y.~C. Jun, J.~S. White and M.~L.
  Brongersma, \emph{Nature Mater.}, 2010, \textbf{9}, 193--204\relax
\mciteBstWouldAddEndPuncttrue
\mciteSetBstMidEndSepPunct{\mcitedefaultmidpunct}
{\mcitedefaultendpunct}{\mcitedefaultseppunct}\relax
\EndOfBibitem
\bibitem[Han and Bozhevolnyi(2013)]{Han2013}
Z.~Han and S.~I. Bozhevolnyi, \emph{Rep. Prog. Phys.}, 2013, \textbf{76},
  016402\relax
\mciteBstWouldAddEndPuncttrue
\mciteSetBstMidEndSepPunct{\mcitedefaultmidpunct}
{\mcitedefaultendpunct}{\mcitedefaultseppunct}\relax
\EndOfBibitem
\bibitem[Raether(1988)]{Raether1988}
H.~Raether, \emph{{Surface plasmons on smooth and rough surfaces and on
  gratings}}, Springer, Berlin, 1988\relax
\mciteBstWouldAddEndPuncttrue
\mciteSetBstMidEndSepPunct{\mcitedefaultmidpunct}
{\mcitedefaultendpunct}{\mcitedefaultseppunct}\relax
\EndOfBibitem
\bibitem[Takahara \emph{et~al.}(1997)Takahara, Yamagishi, Taki, Morimoto, and
  Kobayashi]{Takahara1997}
J.~Takahara, S.~Yamagishi, H.~Taki, A.~Morimoto and T.~Kobayashi, \emph{Opt.
  Lett.}, 1997, \textbf{22}, 475--7\relax
\mciteBstWouldAddEndPuncttrue
\mciteSetBstMidEndSepPunct{\mcitedefaultmidpunct}
{\mcitedefaultendpunct}{\mcitedefaultseppunct}\relax
\EndOfBibitem
\bibitem[Nerkararyan(1997)]{Nerkararyan}
K.~V. Nerkararyan, \emph{Phys. Lett. A}, 1997, \textbf{237}, 103--5\relax
\mciteBstWouldAddEndPuncttrue
\mciteSetBstMidEndSepPunct{\mcitedefaultmidpunct}
{\mcitedefaultendpunct}{\mcitedefaultseppunct}\relax
\EndOfBibitem
\bibitem[Barnes \emph{et~al.}(2003)Barnes, Dereux, and Ebbesen]{Barnes2003}
W.~L. Barnes, A.~Dereux and T.~W. Ebbesen, \emph{Nature}, 2003, \textbf{424},
  824--830\relax
\mciteBstWouldAddEndPuncttrue
\mciteSetBstMidEndSepPunct{\mcitedefaultmidpunct}
{\mcitedefaultendpunct}{\mcitedefaultseppunct}\relax
\EndOfBibitem
\bibitem[Gramotnev and Bozhevolnyi(2010)]{Gramotnev2010}
D.~K. Gramotnev and S.~I. Bozhevolnyi, \emph{Nature Photon.}, 2010, \textbf{4},
  83--91\relax
\mciteBstWouldAddEndPuncttrue
\mciteSetBstMidEndSepPunct{\mcitedefaultmidpunct}
{\mcitedefaultendpunct}{\mcitedefaultseppunct}\relax
\EndOfBibitem
\bibitem[Gramotnev and Bozhevolnyi(2013)]{Gramotnev2013}
D.~K. Gramotnev and S.~I. Bozhevolnyi, \emph{Nature Photon.}, 2013, \textbf{8},
  13--22\relax
\mciteBstWouldAddEndPuncttrue
\mciteSetBstMidEndSepPunct{\mcitedefaultmidpunct}
{\mcitedefaultendpunct}{\mcitedefaultseppunct}\relax
\EndOfBibitem
\bibitem[Bozhevolnyi \emph{et~al.}(2006)Bozhevolnyi, Volkov, Devaux, Laluet,
  and Ebbesen]{Bozhevolnyi2006c}
S.~I. Bozhevolnyi, V.~S. Volkov, E.~Devaux, J.-Y. Laluet and T.~W. Ebbesen,
  \emph{Nature}, 2006, \textbf{440}, 508--11\relax
\mciteBstWouldAddEndPuncttrue
\mciteSetBstMidEndSepPunct{\mcitedefaultmidpunct}
{\mcitedefaultendpunct}{\mcitedefaultseppunct}\relax
\EndOfBibitem
\bibitem[Chen \emph{et~al.}(2006)Chen, Shakya, and Lipson]{Chen2006}
L.~Chen, J.~Shakya and M.~Lipson, \emph{Opt. Lett.}, 2006, \textbf{31},
  2133--5\relax
\mciteBstWouldAddEndPuncttrue
\mciteSetBstMidEndSepPunct{\mcitedefaultmidpunct}
{\mcitedefaultendpunct}{\mcitedefaultseppunct}\relax
\EndOfBibitem
\bibitem[Pyayt \emph{et~al.}(2008)Pyayt, Wiley, Xia, Chen, and
  Dalton]{Pyayt2008}
A.~L. Pyayt, B.~Wiley, Y.~Xia, A.~Chen and L.~Dalton, \emph{Nature Nanotech.},
  2008, \textbf{3}, 660--5\relax
\mciteBstWouldAddEndPuncttrue
\mciteSetBstMidEndSepPunct{\mcitedefaultmidpunct}
{\mcitedefaultendpunct}{\mcitedefaultseppunct}\relax
\EndOfBibitem
\bibitem[Papaioannou \emph{et~al.}(2011)Papaioannou, Vyrsokinos, Tsilipakos,
  Pitilakis, Hassan, Weeber, Markey, Dereux, Bozhevolnyi, Miliou, Kriezis, and
  Pleros]{Papaioannou2011}
S.~Papaioannou, K.~Vyrsokinos, O.~Tsilipakos, A.~Pitilakis, K.~Hassan,
  J.~Weeber, L.~Markey, A.~Dereux, S.~I. Bozhevolnyi, A.~Miliou, E.~E. Kriezis
  and N.~Pleros, \emph{J. Lightwave Technol.}, 2011, \textbf{29},
  3185--3195\relax
\mciteBstWouldAddEndPuncttrue
\mciteSetBstMidEndSepPunct{\mcitedefaultmidpunct}
{\mcitedefaultendpunct}{\mcitedefaultseppunct}\relax
\EndOfBibitem
\bibitem[Kauranen and Zayats(2012)]{Kauranen2012}
M.~Kauranen and A.~V. Zayats, \emph{Nature Photon.}, 2012, \textbf{6},
  737--748\relax
\mciteBstWouldAddEndPuncttrue
\mciteSetBstMidEndSepPunct{\mcitedefaultmidpunct}
{\mcitedefaultendpunct}{\mcitedefaultseppunct}\relax
\EndOfBibitem
\bibitem[Bergman and Stockman(2003)]{Bergman2003}
D.~Bergman and M.~Stockman, \emph{Phys. Rev. Lett.}, 2003, \textbf{90},
  027402\relax
\mciteBstWouldAddEndPuncttrue
\mciteSetBstMidEndSepPunct{\mcitedefaultmidpunct}
{\mcitedefaultendpunct}{\mcitedefaultseppunct}\relax
\EndOfBibitem
\bibitem[Hill \emph{et~al.}(2007)Hill, Oei, Smalbrugge, Zhu, de~Vries, van
  Veldhoven, van Otten, Eijkemans, Turkiewicz, de~Waardt, Geluk, Kwon, Lee,
  N\"{o}tzel, and Smit]{Hill2007}
M.~T. Hill, Y.-S. Oei, B.~Smalbrugge, Y.~Zhu, T.~de~Vries, P.~J. van Veldhoven,
  F.~W.~M. van Otten, T.~J. Eijkemans, J.~P. Turkiewicz, H.~de~Waardt, E.~J.
  Geluk, S.-H. Kwon, Y.-H. Lee, R.~N\"{o}tzel and M.~K. Smit, \emph{Nature
  Photon.}, 2007, \textbf{1}, 589--594\relax
\mciteBstWouldAddEndPuncttrue
\mciteSetBstMidEndSepPunct{\mcitedefaultmidpunct}
{\mcitedefaultendpunct}{\mcitedefaultseppunct}\relax
\EndOfBibitem
\bibitem[Oulton \emph{et~al.}(2009)Oulton, Sorger, Zentgraf, Ma, Gladden, Dai,
  Bartal, and Zhang]{Oulton2009}
R.~F. Oulton, V.~J. Sorger, T.~Zentgraf, R.-M. Ma, C.~Gladden, L.~Dai,
  G.~Bartal and X.~Zhang, \emph{Nature}, 2009, \textbf{461}, 629--32\relax
\mciteBstWouldAddEndPuncttrue
\mciteSetBstMidEndSepPunct{\mcitedefaultmidpunct}
{\mcitedefaultendpunct}{\mcitedefaultseppunct}\relax
\EndOfBibitem
\bibitem[Noginov \emph{et~al.}(2009)Noginov, Zhu, Belgrave, Bakker, Shalaev,
  Narimanov, Stout, Herz, Suteewong, and Wiesner]{Noginov2009}
M.~A. Noginov, G.~Zhu, A.~M. Belgrave, R.~Bakker, V.~M. Shalaev, E.~E.
  Narimanov, S.~Stout, E.~Herz, T.~Suteewong and U.~Wiesner, \emph{Nature},
  2009, \textbf{460}, 1110--2\relax
\mciteBstWouldAddEndPuncttrue
\mciteSetBstMidEndSepPunct{\mcitedefaultmidpunct}
{\mcitedefaultendpunct}{\mcitedefaultseppunct}\relax
\EndOfBibitem
\bibitem[Khajavikhan \emph{et~al.}(2012)Khajavikhan, Simic, Katz, Lee, Slutsky,
  Mizrahi, Lomakin, and Fainman]{Khajavikhan2012}
M.~Khajavikhan, A.~Simic, M.~Katz, J.~H. Lee, B.~Slutsky, A.~Mizrahi,
  V.~Lomakin and Y.~Fainman, \emph{Nature}, 2012, \textbf{482}, 204--7\relax
\mciteBstWouldAddEndPuncttrue
\mciteSetBstMidEndSepPunct{\mcitedefaultmidpunct}
{\mcitedefaultendpunct}{\mcitedefaultseppunct}\relax
\EndOfBibitem
\bibitem[Altewischer \emph{et~al.}(2002)Altewischer, van Exter, and
  Woerdman]{Altewischer2002a}
E.~Altewischer, M.~P. van Exter and J.~P. Woerdman, \emph{Nature}, 2002,
  \textbf{418}, 304--6\relax
\mciteBstWouldAddEndPuncttrue
\mciteSetBstMidEndSepPunct{\mcitedefaultmidpunct}
{\mcitedefaultendpunct}{\mcitedefaultseppunct}\relax
\EndOfBibitem
\bibitem[Akimov \emph{et~al.}(2007)Akimov, Mukherjee, Yu, Chang, Zibrov,
  Hemmer, Park, and Lukin]{Akimov2007a}
A.~V. Akimov, A.~Mukherjee, C.~L. Yu, D.~E. Chang, A.~S. Zibrov, P.~R. Hemmer,
  H.~Park and M.~D. Lukin, \emph{Nature}, 2007, \textbf{450}, 402--6\relax
\mciteBstWouldAddEndPuncttrue
\mciteSetBstMidEndSepPunct{\mcitedefaultmidpunct}
{\mcitedefaultendpunct}{\mcitedefaultseppunct}\relax
\EndOfBibitem
\bibitem[Kumar \emph{et~al.}(2013)Kumar, Huck, and Andersen]{Kumar2013}
S.~Kumar, A.~Huck and U.~L. Andersen, \emph{Nano Lett.}, 2013, \textbf{13},
  1221--1225\relax
\mciteBstWouldAddEndPuncttrue
\mciteSetBstMidEndSepPunct{\mcitedefaultmidpunct}
{\mcitedefaultendpunct}{\mcitedefaultseppunct}\relax
\EndOfBibitem
\bibitem[Tame \emph{et~al.}(2013)Tame, McEnery, \"{O}zdemir, Lee, Maier, and
  Kim]{Tame2013}
M.~S. Tame, K.~R. McEnery, S.~K. \"{O}zdemir, J.~Lee, S.~A. Maier and M.~S.
  Kim, \emph{Nature Phys.}, 2013, \textbf{9}, 329--340\relax
\mciteBstWouldAddEndPuncttrue
\mciteSetBstMidEndSepPunct{\mcitedefaultmidpunct}
{\mcitedefaultendpunct}{\mcitedefaultseppunct}\relax
\EndOfBibitem
\bibitem[Ebbesen \emph{et~al.}(1998)Ebbesen, Lezec, Ghaemi, Thio, and
  Wolff]{Martin-Moreno2001}
T.~W. Ebbesen, H.~J. Lezec, H.~F. Ghaemi, T.~Thio and P.~A. Wolff,
  \emph{Nature}, 1998, \textbf{391}, 667--9\relax
\mciteBstWouldAddEndPuncttrue
\mciteSetBstMidEndSepPunct{\mcitedefaultmidpunct}
{\mcitedefaultendpunct}{\mcitedefaultseppunct}\relax
\EndOfBibitem
\bibitem[Genet and Ebbesen(2007)]{Genet2007}
C.~Genet and T.~W. Ebbesen, \emph{Nature}, 2007, \textbf{445}, 39--46\relax
\mciteBstWouldAddEndPuncttrue
\mciteSetBstMidEndSepPunct{\mcitedefaultmidpunct}
{\mcitedefaultendpunct}{\mcitedefaultseppunct}\relax
\EndOfBibitem
\bibitem[Laux \emph{et~al.}(2008)Laux, Genet, Skauli, and Ebbesen]{Laux2008}
E.~Laux, C.~Genet, T.~Skauli and T.~W. Ebbesen, \emph{Nature Photon.}, 2008,
  \textbf{2}, 161--164\relax
\mciteBstWouldAddEndPuncttrue
\mciteSetBstMidEndSepPunct{\mcitedefaultmidpunct}
{\mcitedefaultendpunct}{\mcitedefaultseppunct}\relax
\EndOfBibitem
\bibitem[Roberts \emph{et~al.}(2014)Roberts, Pors, Albrektsen, and
  Bozhevolnyi]{Roberts2014}
A.~S. Roberts, A.~Pors, O.~Albrektsen and S.~I. Bozhevolnyi, \emph{Nano Lett.},
  2014, \textbf{14}, 783--7\relax
\mciteBstWouldAddEndPuncttrue
\mciteSetBstMidEndSepPunct{\mcitedefaultmidpunct}
{\mcitedefaultendpunct}{\mcitedefaultseppunct}\relax
\EndOfBibitem
\bibitem[Clausen \emph{et~al.}(2014)Clausen, H{\o}jlund-Nielsen, Christiansen,
  Yazdi, Grajower, Taha, Levy, Kristensen, and Mortensen]{Clausen2014}
J.~S. Clausen, E.~H{\o}jlund-Nielsen, A.~B. Christiansen, S.~Yazdi,
  M.~Grajower, H.~Taha, U.~Levy, A.~Kristensen and N.~A. Mortensen, \emph{Nano
  Lett.}, 2014, \textbf{14}, 4499--4504\relax
\mciteBstWouldAddEndPuncttrue
\mciteSetBstMidEndSepPunct{\mcitedefaultmidpunct}
{\mcitedefaultendpunct}{\mcitedefaultseppunct}\relax
\EndOfBibitem
\bibitem[Kneipp \emph{et~al.}(2002)Kneipp, Kneipp, Itzkan, Dasari, and
  Feld]{Kneipp2002}
K.~Kneipp, H.~Kneipp, I.~Itzkan, R.~R. Dasari and M.~S. Feld, \emph{J. Phys.:
  Condens. Matter}, 2002, \textbf{14}, 597--624\relax
\mciteBstWouldAddEndPuncttrue
\mciteSetBstMidEndSepPunct{\mcitedefaultmidpunct}
{\mcitedefaultendpunct}{\mcitedefaultseppunct}\relax
\EndOfBibitem
\bibitem[Anker \emph{et~al.}(2008)Anker, Hall, Lyandres, Shah, Zhao, and {Van
  Duyne}]{Anker2008}
J.~N. Anker, W.~P. Hall, O.~Lyandres, N.~C. Shah, J.~Zhao and R.~P. {Van
  Duyne}, \emph{Nature Mater.}, 2008, \textbf{7}, 442--53\relax
\mciteBstWouldAddEndPuncttrue
\mciteSetBstMidEndSepPunct{\mcitedefaultmidpunct}
{\mcitedefaultendpunct}{\mcitedefaultseppunct}\relax
\EndOfBibitem
\bibitem[Mayer \emph{et~al.}(2010)Mayer, Hao, Lee, Nordlander, and
  Hafner]{Mayer2010}
K.~M. Mayer, F.~Hao, S.~Lee, P.~Nordlander and J.~H. Hafner,
  \emph{Nanotechnology}, 2010, \textbf{21}, 255503\relax
\mciteBstWouldAddEndPuncttrue
\mciteSetBstMidEndSepPunct{\mcitedefaultmidpunct}
{\mcitedefaultendpunct}{\mcitedefaultseppunct}\relax
\EndOfBibitem
\bibitem[Chung \emph{et~al.}(2011)Chung, Lee, Song, Chun, and Lee]{Chung2011a}
T.~Chung, S.-Y. Lee, E.~Y. Song, H.~Chun and B.~Lee, \emph{Sensors}, 2011,
  \textbf{11}, 10907--29\relax
\mciteBstWouldAddEndPuncttrue
\mciteSetBstMidEndSepPunct{\mcitedefaultmidpunct}
{\mcitedefaultendpunct}{\mcitedefaultseppunct}\relax
\EndOfBibitem
\bibitem[Feng \emph{et~al.}(2012)Feng, Siu, Roelke, Mehta, Rhieu, Palmore, and
  Pacifici]{Feng2012}
J.~Feng, V.~S. Siu, A.~Roelke, V.~Mehta, S.~Y. Rhieu, G.~T.~R. Palmore and
  D.~Pacifici, \emph{Nano Lett.}, 2012, \textbf{12}, 602--9\relax
\mciteBstWouldAddEndPuncttrue
\mciteSetBstMidEndSepPunct{\mcitedefaultmidpunct}
{\mcitedefaultendpunct}{\mcitedefaultseppunct}\relax
\EndOfBibitem
\bibitem[Zijlstra \emph{et~al.}(2012)Zijlstra, Paulo, and Orrit]{Zijlstra2012}
P.~Zijlstra, P.~M.~R. Paulo and M.~Orrit, \emph{Nature Nanotech.}, 2012,
  \textbf{7}, 379--82\relax
\mciteBstWouldAddEndPuncttrue
\mciteSetBstMidEndSepPunct{\mcitedefaultmidpunct}
{\mcitedefaultendpunct}{\mcitedefaultseppunct}\relax
\EndOfBibitem
\bibitem[Fang \emph{et~al.}(2005)Fang, Lee, Sun, and Zhang]{Fang2005}
N.~Fang, H.~Lee, C.~Sun and X.~Zhang, \emph{Science}, 2005, \textbf{308},
  534--7\relax
\mciteBstWouldAddEndPuncttrue
\mciteSetBstMidEndSepPunct{\mcitedefaultmidpunct}
{\mcitedefaultendpunct}{\mcitedefaultseppunct}\relax
\EndOfBibitem
\bibitem[Hell(2007)]{Hell2007}
S.~W. Hell, \emph{Science}, 2007, \textbf{316}, 1153--8\relax
\mciteBstWouldAddEndPuncttrue
\mciteSetBstMidEndSepPunct{\mcitedefaultmidpunct}
{\mcitedefaultendpunct}{\mcitedefaultseppunct}\relax
\EndOfBibitem
\bibitem[Xiong \emph{et~al.}(2007)Xiong, Liu, Sun, and Zhang]{Xiong2007}
Y.~Xiong, Z.~Liu, C.~Sun and X.~Zhang, \emph{Nano Lett.}, 2007, \textbf{7},
  3360--5\relax
\mciteBstWouldAddEndPuncttrue
\mciteSetBstMidEndSepPunct{\mcitedefaultmidpunct}
{\mcitedefaultendpunct}{\mcitedefaultseppunct}\relax
\EndOfBibitem
\bibitem[Liu \emph{et~al.}(2011)Liu, Hentschel, Weiss, Alivisatos, and
  Giessen]{Liu2011}
N.~Liu, M.~Hentschel, T.~Weiss, A.~P. Alivisatos and H.~Giessen,
  \emph{Science}, 2011, \textbf{332}, 1407--10\relax
\mciteBstWouldAddEndPuncttrue
\mciteSetBstMidEndSepPunct{\mcitedefaultmidpunct}
{\mcitedefaultendpunct}{\mcitedefaultseppunct}\relax
\EndOfBibitem
\bibitem[Schuck \emph{et~al.}(2013)Schuck, Weber-Bargioni, Ashby, Ogletree,
  Schwartzberg, and Cabrini]{Schuck2013}
P.~J. Schuck, A.~Weber-Bargioni, P.~D. Ashby, D.~F. Ogletree, A.~Schwartzberg
  and S.~Cabrini, \emph{Adv. Funct. Mater.}, 2013, \textbf{23},
  2539--2553\relax
\mciteBstWouldAddEndPuncttrue
\mciteSetBstMidEndSepPunct{\mcitedefaultmidpunct}
{\mcitedefaultendpunct}{\mcitedefaultseppunct}\relax
\EndOfBibitem
\bibitem[Grigorenko \emph{et~al.}(2008)Grigorenko, Roberts, Dickinson, and
  Zhang]{Grigorenko2008}
A.~N. Grigorenko, N.~W. Roberts, M.~R. Dickinson and Y.~Zhang, \emph{Nature
  Photon.}, 2008, \textbf{2}, 365--370\relax
\mciteBstWouldAddEndPuncttrue
\mciteSetBstMidEndSepPunct{\mcitedefaultmidpunct}
{\mcitedefaultendpunct}{\mcitedefaultseppunct}\relax
\EndOfBibitem
\bibitem[Juan \emph{et~al.}(2011)Juan, Righini, and Quidant]{Juan2011}
M.~L. Juan, M.~Righini and R.~Quidant, \emph{Nature Photon.}, 2011, \textbf{5},
  349--356\relax
\mciteBstWouldAddEndPuncttrue
\mciteSetBstMidEndSepPunct{\mcitedefaultmidpunct}
{\mcitedefaultendpunct}{\mcitedefaultseppunct}\relax
\EndOfBibitem
\bibitem[Erickson \emph{et~al.}(2011)Erickson, Serey, Chen, and
  Mandal]{Erickson2011}
D.~Erickson, X.~Serey, Y.-F. Chen and S.~Mandal, \emph{Lab Chip}, 2011,
  \textbf{11}, 995--1009\relax
\mciteBstWouldAddEndPuncttrue
\mciteSetBstMidEndSepPunct{\mcitedefaultmidpunct}
{\mcitedefaultendpunct}{\mcitedefaultseppunct}\relax
\EndOfBibitem
\bibitem[Pang and Gordon(2012)]{Pang2012a}
Y.~Pang and R.~Gordon, \emph{Nano Lett.}, 2012, \textbf{12}, 402--6\relax
\mciteBstWouldAddEndPuncttrue
\mciteSetBstMidEndSepPunct{\mcitedefaultmidpunct}
{\mcitedefaultendpunct}{\mcitedefaultseppunct}\relax
\EndOfBibitem
\bibitem[Marag\`{o} \emph{et~al.}(2013)Marag\`{o}, Jones, Gucciardi, Volpe, and
  Ferrari]{Marago2013}
O.~M. Marag\`{o}, P.~H. Jones, P.~G. Gucciardi, G.~Volpe and A.~C. Ferrari,
  \emph{Nature Nanotech.}, 2013, \textbf{8}, 807--819\relax
\mciteBstWouldAddEndPuncttrue
\mciteSetBstMidEndSepPunct{\mcitedefaultmidpunct}
{\mcitedefaultendpunct}{\mcitedefaultseppunct}\relax
\EndOfBibitem
\bibitem[Maier and Atwater(2005)]{Maier2005}
S.~A. Maier and H.~A. Atwater, \emph{J. Appl. Phys.}, 2005, \textbf{98},
  011101\relax
\mciteBstWouldAddEndPuncttrue
\mciteSetBstMidEndSepPunct{\mcitedefaultmidpunct}
{\mcitedefaultendpunct}{\mcitedefaultseppunct}\relax
\EndOfBibitem
\bibitem[Barnes(2006)]{Barnes2006}
W.~L. Barnes, \emph{J. Opt. A: Pure Appl. Opt.}, 2006, \textbf{8},
  S87--S93\relax
\mciteBstWouldAddEndPuncttrue
\mciteSetBstMidEndSepPunct{\mcitedefaultmidpunct}
{\mcitedefaultendpunct}{\mcitedefaultseppunct}\relax
\EndOfBibitem
\bibitem[Ebbesen \emph{et~al.}(2008)Ebbesen, Genet, and
  Bozhevolnyi]{Ebbesen2008}
T.~W. Ebbesen, C.~Genet and S.~I. Bozhevolnyi, \emph{Phys. Today}, 2008,
  \textbf{61(5)}, 44--50\relax
\mciteBstWouldAddEndPuncttrue
\mciteSetBstMidEndSepPunct{\mcitedefaultmidpunct}
{\mcitedefaultendpunct}{\mcitedefaultseppunct}\relax
\EndOfBibitem
\bibitem[Boltasseva(2009)]{Boltasseva2009}
A.~Boltasseva, \emph{J. Opt. A: Pure Appl. Opt.}, 2009, \textbf{11},
  114001\relax
\mciteBstWouldAddEndPuncttrue
\mciteSetBstMidEndSepPunct{\mcitedefaultmidpunct}
{\mcitedefaultendpunct}{\mcitedefaultseppunct}\relax
\EndOfBibitem
\bibitem[Moreno \emph{et~al.}(2006)Moreno, Garc\'{\i}a-Vidal, Rodrigo,
  Mart\'{\i}n-Moreno, and Bozhevolnyi]{Moreno2006a}
E.~Moreno, F.~J. Garc\'{\i}a-Vidal, S.~G. Rodrigo, L.~Mart\'{\i}n-Moreno and
  S.~I. Bozhevolnyi, \emph{Opt. Lett.}, 2006, \textbf{31}, 3447--9\relax
\mciteBstWouldAddEndPuncttrue
\mciteSetBstMidEndSepPunct{\mcitedefaultmidpunct}
{\mcitedefaultendpunct}{\mcitedefaultseppunct}\relax
\EndOfBibitem
\bibitem[Economou(1969)]{Economou1969}
E.~N. Economou, \emph{Phys. Rev.}, 1969, \textbf{182}, 539--554\relax
\mciteBstWouldAddEndPuncttrue
\mciteSetBstMidEndSepPunct{\mcitedefaultmidpunct}
{\mcitedefaultendpunct}{\mcitedefaultseppunct}\relax
\EndOfBibitem
\bibitem[Wang and Wang(2004)]{Wang2004}
B.~Wang and G.~P. Wang, \emph{Opt. Lett.}, 2004, \textbf{29}, 1992--4\relax
\mciteBstWouldAddEndPuncttrue
\mciteSetBstMidEndSepPunct{\mcitedefaultmidpunct}
{\mcitedefaultendpunct}{\mcitedefaultseppunct}\relax
\EndOfBibitem
\bibitem[Veronis and Fan(2005)]{Veronis2005}
G.~Veronis and S.~Fan, \emph{Opt. Lett.}, 2005, \textbf{30}, 3359--61\relax
\mciteBstWouldAddEndPuncttrue
\mciteSetBstMidEndSepPunct{\mcitedefaultmidpunct}
{\mcitedefaultendpunct}{\mcitedefaultseppunct}\relax
\EndOfBibitem
\bibitem[Tanaka \emph{et~al.}(2005)Tanaka, Tanaka, and Sugiyama]{Tanaka2005}
K.~Tanaka, M.~Tanaka and T.~Sugiyama, \emph{Opt. Express}, 2005, \textbf{13},
  256--66\relax
\mciteBstWouldAddEndPuncttrue
\mciteSetBstMidEndSepPunct{\mcitedefaultmidpunct}
{\mcitedefaultendpunct}{\mcitedefaultseppunct}\relax
\EndOfBibitem
\bibitem[Dionne \emph{et~al.}(2006)Dionne, Sweatlock, Atwater, and
  Polman]{Dionne2006}
J.~Dionne, L.~Sweatlock, H.~Atwater and A.~Polman, \emph{Phys. Rev. B}, 2006,
  \textbf{73}, 035407\relax
\mciteBstWouldAddEndPuncttrue
\mciteSetBstMidEndSepPunct{\mcitedefaultmidpunct}
{\mcitedefaultendpunct}{\mcitedefaultseppunct}\relax
\EndOfBibitem
\bibitem[Nerkararyan \emph{et~al.}(2011)Nerkararyan, Nerkararyan, and
  Bozhevolnyi]{Nerkararyan2011}
K.~V. Nerkararyan, S.~K. Nerkararyan and S.~I. Bozhevolnyi, \emph{Opt. Lett.},
  2011, \textbf{36}, 4311--3\relax
\mciteBstWouldAddEndPuncttrue
\mciteSetBstMidEndSepPunct{\mcitedefaultmidpunct}
{\mcitedefaultendpunct}{\mcitedefaultseppunct}\relax
\EndOfBibitem
\bibitem[Babadjanyan \emph{et~al.}(2000)Babadjanyan, Margaryan, and
  Nerkararyan]{Babadjanyan2000}
A.~J. Babadjanyan, N.~L. Margaryan and K.~V. Nerkararyan, \emph{J. Appl.
  Phys.}, 2000, \textbf{87}, 3785\relax
\mciteBstWouldAddEndPuncttrue
\mciteSetBstMidEndSepPunct{\mcitedefaultmidpunct}
{\mcitedefaultendpunct}{\mcitedefaultseppunct}\relax
\EndOfBibitem
\bibitem[Volkov \emph{et~al.}(2009)Volkov, Bozhevolnyi, Rodrigo,
  Garc\'{\i}a-Vidal, Ebbesen, and Alle]{Volkov2009}
V.~S. Volkov, S.~I. Bozhevolnyi, S.~G. Rodrigo, F.~J. Garc\'{\i}a-Vidal, T.~W.
  Ebbesen and N.~B. Alle, \emph{Nano Lett.}, 2009,  1278--1282\relax
\mciteBstWouldAddEndPuncttrue
\mciteSetBstMidEndSepPunct{\mcitedefaultmidpunct}
{\mcitedefaultendpunct}{\mcitedefaultseppunct}\relax
\EndOfBibitem
\bibitem[Gramotnev and Vogel(2011)]{Gramotnev2011}
D.~K. Gramotnev and M.~W. Vogel, \emph{Phys. Lett. A}, 2011, \textbf{375},
  3464--3468\relax
\mciteBstWouldAddEndPuncttrue
\mciteSetBstMidEndSepPunct{\mcitedefaultmidpunct}
{\mcitedefaultendpunct}{\mcitedefaultseppunct}\relax
\EndOfBibitem
\bibitem[Desiatov \emph{et~al.}(2011)Desiatov, Goykhman, and
  Levy]{Desiatov2011}
B.~Desiatov, I.~Goykhman and U.~Levy, \emph{Opt. Express}, 2011, \textbf{19},
  13150--7\relax
\mciteBstWouldAddEndPuncttrue
\mciteSetBstMidEndSepPunct{\mcitedefaultmidpunct}
{\mcitedefaultendpunct}{\mcitedefaultseppunct}\relax
\EndOfBibitem
\bibitem[Stockman(2004)]{Stockman2004}
M.~Stockman, \emph{Phys. Rev. Lett.}, 2004, \textbf{93}, 137404\relax
\mciteBstWouldAddEndPuncttrue
\mciteSetBstMidEndSepPunct{\mcitedefaultmidpunct}
{\mcitedefaultendpunct}{\mcitedefaultseppunct}\relax
\EndOfBibitem
\bibitem[Pile and Gramotnev(2006)]{Pile2006a}
D.~F.~P. Pile and D.~K. Gramotnev, \emph{Appl. Phys. Lett.}, 2006, \textbf{89},
  041111\relax
\mciteBstWouldAddEndPuncttrue
\mciteSetBstMidEndSepPunct{\mcitedefaultmidpunct}
{\mcitedefaultendpunct}{\mcitedefaultseppunct}\relax
\EndOfBibitem
\bibitem[Issa and Guckenberger(2006)]{Issa2006}
N.~A. Issa and R.~Guckenberger, \emph{Plasmonics}, 2006, \textbf{2},
  31--37\relax
\mciteBstWouldAddEndPuncttrue
\mciteSetBstMidEndSepPunct{\mcitedefaultmidpunct}
{\mcitedefaultendpunct}{\mcitedefaultseppunct}\relax
\EndOfBibitem
\bibitem[Verhagen \emph{et~al.}(2009)Verhagen, Spasenovi\'{c}, Polman, and
  Kuipers]{Verhagen2009}
E.~Verhagen, M.~Spasenovi\'{c}, A.~Polman and L.~Kuipers, \emph{Phys. Rev.
  Lett.}, 2009, \textbf{102}, 203904\relax
\mciteBstWouldAddEndPuncttrue
\mciteSetBstMidEndSepPunct{\mcitedefaultmidpunct}
{\mcitedefaultendpunct}{\mcitedefaultseppunct}\relax
\EndOfBibitem
\bibitem[Bozhevolnyi and Nerkararyan(2010)]{Bozhevolnyi2010}
S.~I. Bozhevolnyi and K.~V. Nerkararyan, \emph{Opt. Lett.}, 2010, \textbf{35},
  541--543\relax
\mciteBstWouldAddEndPuncttrue
\mciteSetBstMidEndSepPunct{\mcitedefaultmidpunct}
{\mcitedefaultendpunct}{\mcitedefaultseppunct}\relax
\EndOfBibitem
\bibitem[Oulton \emph{et~al.}(2008)Oulton, Sorger, Genov, Pile, and
  Zhang]{Oulton2008}
R.~F. Oulton, V.~J. Sorger, D.~A. Genov, D.~F.~P. Pile and X.~Zhang,
  \emph{Nature Photon.}, 2008, \textbf{2}, 496--500\relax
\mciteBstWouldAddEndPuncttrue
\mciteSetBstMidEndSepPunct{\mcitedefaultmidpunct}
{\mcitedefaultendpunct}{\mcitedefaultseppunct}\relax
\EndOfBibitem
\bibitem[West \emph{et~al.}(2010)West, Ishii, Naik, Emani, Shalaev, and
  Boltasseva]{West2010}
P.~West, S.~Ishii, G.~Naik, N.~Emani, V.~Shalaev and A.~Boltasseva, \emph{Laser
  Photon. Rev.}, 2010, \textbf{4}, 795--808\relax
\mciteBstWouldAddEndPuncttrue
\mciteSetBstMidEndSepPunct{\mcitedefaultmidpunct}
{\mcitedefaultendpunct}{\mcitedefaultseppunct}\relax
\EndOfBibitem
\bibitem[Goykhman \emph{et~al.}(2010)Goykhman, Desiatov, and
  Levy]{Goykhman2010}
I.~Goykhman, B.~Desiatov and U.~Levy, \emph{Appl. Phys. Lett.}, 2010,
  \textbf{97}, 141106\relax
\mciteBstWouldAddEndPuncttrue
\mciteSetBstMidEndSepPunct{\mcitedefaultmidpunct}
{\mcitedefaultendpunct}{\mcitedefaultseppunct}\relax
\EndOfBibitem
\bibitem[Yang \emph{et~al.}(2011)Yang, Liu, Oulton, Yin, and Zhang]{Yang2011}
X.~Yang, Y.~Liu, R.~F. Oulton, X.~Yin and X.~Zhang, \emph{Nano Lett.}, 2011,
  \textbf{11}, 321--8\relax
\mciteBstWouldAddEndPuncttrue
\mciteSetBstMidEndSepPunct{\mcitedefaultmidpunct}
{\mcitedefaultendpunct}{\mcitedefaultseppunct}\relax
\EndOfBibitem
\bibitem[Naik \emph{et~al.}(2013)Naik, Shalaev, and Boltasseva]{Naik2013}
G.~V. Naik, V.~M. Shalaev and A.~Boltasseva, \emph{Adv. Mater.}, 2013,
  \textbf{25}, 3264--94\relax
\mciteBstWouldAddEndPuncttrue
\mciteSetBstMidEndSepPunct{\mcitedefaultmidpunct}
{\mcitedefaultendpunct}{\mcitedefaultseppunct}\relax
\EndOfBibitem
\bibitem[Quinten \emph{et~al.}(1998)Quinten, Leitner, Krenn, and
  Aussenegg]{Quinten1998}
M.~Quinten, A.~Leitner, J.~R. Krenn and F.~R. Aussenegg, \emph{Opt. Lett.},
  1998, \textbf{23}, 1331--3\relax
\mciteBstWouldAddEndPuncttrue
\mciteSetBstMidEndSepPunct{\mcitedefaultmidpunct}
{\mcitedefaultendpunct}{\mcitedefaultseppunct}\relax
\EndOfBibitem
\bibitem[Maier \emph{et~al.}(2003)Maier, Kik, Atwater, Meltzer, Harel, Koel,
  and Requicha]{Maier2003}
S.~A. Maier, P.~G. Kik, H.~A. Atwater, S.~Meltzer, E.~Harel, B.~E. Koel and
  A.~A.~G. Requicha, \emph{Nature Mater.}, 2003, \textbf{2}, 229--32\relax
\mciteBstWouldAddEndPuncttrue
\mciteSetBstMidEndSepPunct{\mcitedefaultmidpunct}
{\mcitedefaultendpunct}{\mcitedefaultseppunct}\relax
\EndOfBibitem
\bibitem[Xu \emph{et~al.}(2010)Xu, Wu, Luo, and Guo]{Xu2010}
T.~Xu, Y.-K. Wu, X.~Luo and L.~J. Guo, \emph{Nat. Commun.}, 2010, \textbf{1},
  59\relax
\mciteBstWouldAddEndPuncttrue
\mciteSetBstMidEndSepPunct{\mcitedefaultmidpunct}
{\mcitedefaultendpunct}{\mcitedefaultseppunct}\relax
\EndOfBibitem
\bibitem[Henzie \emph{et~al.}(2007)Henzie, Lee, and Odom]{Henzie2007}
J.~Henzie, M.~H. Lee and T.~W. Odom, \emph{Nature Nanotech.}, 2007, \textbf{2},
  549--54\relax
\mciteBstWouldAddEndPuncttrue
\mciteSetBstMidEndSepPunct{\mcitedefaultmidpunct}
{\mcitedefaultendpunct}{\mcitedefaultseppunct}\relax
\EndOfBibitem
\bibitem[Puscasu and Schaich(2008)]{Puscasu2008}
I.~Puscasu and W.~L. Schaich, \emph{Appl. Phys. Lett.}, 2008, \textbf{92},
  233102\relax
\mciteBstWouldAddEndPuncttrue
\mciteSetBstMidEndSepPunct{\mcitedefaultmidpunct}
{\mcitedefaultendpunct}{\mcitedefaultseppunct}\relax
\EndOfBibitem
\bibitem[Qu and Nie(2013)]{Qu2013}
S.-W. Qu and Z.-P. Nie, \emph{Sci. Rep.}, 2013, \textbf{3}, 3172\relax
\mciteBstWouldAddEndPuncttrue
\mciteSetBstMidEndSepPunct{\mcitedefaultmidpunct}
{\mcitedefaultendpunct}{\mcitedefaultseppunct}\relax
\EndOfBibitem
\bibitem[Bryant \emph{et~al.}(2008)Bryant, {Garc\'{\i}a de Abajo}, and
  Aizpurua]{Bryant2008}
G.~W. Bryant, F.~J. {Garc\'{\i}a de Abajo} and J.~Aizpurua, \emph{Nano Lett.},
  2008, \textbf{8}, 631--6\relax
\mciteBstWouldAddEndPuncttrue
\mciteSetBstMidEndSepPunct{\mcitedefaultmidpunct}
{\mcitedefaultendpunct}{\mcitedefaultseppunct}\relax
\EndOfBibitem
\bibitem[Kawata \emph{et~al.}(2009)Kawata, Inouye, and Verma]{Kawata2009}
S.~Kawata, Y.~Inouye and P.~Verma, \emph{Nature Photon.}, 2009, \textbf{3},
  388--394\relax
\mciteBstWouldAddEndPuncttrue
\mciteSetBstMidEndSepPunct{\mcitedefaultmidpunct}
{\mcitedefaultendpunct}{\mcitedefaultseppunct}\relax
\EndOfBibitem
\bibitem[Schnell \emph{et~al.}(2009)Schnell, Garc\'{\i}a-Etxarri, Huber,
  Crozier, Aizpurua, and Hillenbrand]{Schnell2009}
M.~Schnell, A.~Garc\'{\i}a-Etxarri, A.~J. Huber, K.~Crozier, J.~Aizpurua and
  R.~Hillenbrand, \emph{Nature Photon.}, 2009, \textbf{3}, 287--291\relax
\mciteBstWouldAddEndPuncttrue
\mciteSetBstMidEndSepPunct{\mcitedefaultmidpunct}
{\mcitedefaultendpunct}{\mcitedefaultseppunct}\relax
\EndOfBibitem
\bibitem[Novotny and van Hulst(2011)]{Novotny2011}
L.~Novotny and N.~van Hulst, \emph{Nature Photon.}, 2011, \textbf{5},
  83--90\relax
\mciteBstWouldAddEndPuncttrue
\mciteSetBstMidEndSepPunct{\mcitedefaultmidpunct}
{\mcitedefaultendpunct}{\mcitedefaultseppunct}\relax
\EndOfBibitem
\bibitem[Huck \emph{et~al.}(2011)Huck, Kumar, Shakoor, and Andersen]{Huck2011}
A.~Huck, S.~Kumar, A.~Shakoor and U.~L. Andersen, \emph{Phys. Rev. Lett.},
  2011, \textbf{106}, 096801\relax
\mciteBstWouldAddEndPuncttrue
\mciteSetBstMidEndSepPunct{\mcitedefaultmidpunct}
{\mcitedefaultendpunct}{\mcitedefaultseppunct}\relax
\EndOfBibitem
\bibitem[Zenin \emph{et~al.}(2011)Zenin, Volkov, Han, Bozhevolnyi, Devaux, and
  Ebbesen]{Zenin2011a}
V.~A. Zenin, V.~S. Volkov, Z.~Han, S.~I. Bozhevolnyi, E.~Devaux and T.~W.
  Ebbesen, \emph{J. Opt. Soc. Am. B}, 2011, \textbf{28}, 1596--1602\relax
\mciteBstWouldAddEndPuncttrue
\mciteSetBstMidEndSepPunct{\mcitedefaultmidpunct}
{\mcitedefaultendpunct}{\mcitedefaultseppunct}\relax
\EndOfBibitem
\bibitem[Raza \emph{et~al.}(2014)Raza, Stenger, Pors, Holmgaard, Kadkhodazadeh,
  Wagner, Pedersen, Wubs, Bozhevolnyi, and Mortensen]{Raza2014a}
S.~Raza, N.~Stenger, A.~Pors, T.~Holmgaard, S.~Kadkhodazadeh, J.~B. Wagner,
  K.~Pedersen, M.~Wubs, S.~I. Bozhevolnyi and N.~A. Mortensen, \emph{Nat.
  Commun.}, 2014, \textbf{5}, 4125\relax
\mciteBstWouldAddEndPuncttrue
\mciteSetBstMidEndSepPunct{\mcitedefaultmidpunct}
{\mcitedefaultendpunct}{\mcitedefaultseppunct}\relax
\EndOfBibitem
\bibitem[Skovsen \emph{et~al.}(2013)Skovsen, S{\o}ndergaard, Lemke, Holmgaard,
  Lei{\ss}ner, Eriksen, Beermann, Bauer, Pedersen, and
  Bozhevolnyi]{Skovsen2013a}
E.~Skovsen, T.~S{\o}ndergaard, C.~Lemke, T.~Holmgaard, T.~Lei{\ss}ner, R.~L.
  Eriksen, J.~Beermann, M.~Bauer, K.~Pedersen and S.~I. Bozhevolnyi,
  \emph{Appl. Phys. Lett.}, 2013, \textbf{103}, 211102\relax
\mciteBstWouldAddEndPuncttrue
\mciteSetBstMidEndSepPunct{\mcitedefaultmidpunct}
{\mcitedefaultendpunct}{\mcitedefaultseppunct}\relax
\EndOfBibitem
\bibitem[Beermann \emph{et~al.}(2011)Beermann, S{\o}ndergaard, Novikov,
  Bozhevolnyi, Devaux, and Ebbesen]{Beermann2011}
J.~Beermann, T.~S{\o}ndergaard, S.~M. Novikov, S.~I. Bozhevolnyi, E.~Devaux and
  T.~W. Ebbesen, \emph{New J. Phys.}, 2011, \textbf{13}, 063029\relax
\mciteBstWouldAddEndPuncttrue
\mciteSetBstMidEndSepPunct{\mcitedefaultmidpunct}
{\mcitedefaultendpunct}{\mcitedefaultseppunct}\relax
\EndOfBibitem
\bibitem[Smith \emph{et~al.}(2014)Smith, Thilsted, Garcia-Ortiz, Radko, Marie,
  Jeppesen, Vannahme, Bozhevolnyi, and Kristensen]{Smith2014}
C.~L.~C. Smith, A.~H. Thilsted, C.~E. Garcia-Ortiz, I.~P. Radko, R.~Marie,
  C.~Jeppesen, C.~Vannahme, S.~I. Bozhevolnyi and A.~Kristensen, \emph{Nano
  Lett.}, 2014, \textbf{14}, 1659--1664\relax
\mciteBstWouldAddEndPuncttrue
\mciteSetBstMidEndSepPunct{\mcitedefaultmidpunct}
{\mcitedefaultendpunct}{\mcitedefaultseppunct}\relax
\EndOfBibitem
\bibitem[Smith \emph{et~al.}(2012)Smith, Desiatov, Goykmann, Fernandez-Cuesta,
  Levy, and Kristensen]{Smith2012}
C.~L.~C. Smith, B.~Desiatov, I.~Goykmann, I.~Fernandez-Cuesta, U.~Levy and
  A.~Kristensen, \emph{Opt. Express}, 2012, \textbf{20}, 5696--706\relax
\mciteBstWouldAddEndPuncttrue
\mciteSetBstMidEndSepPunct{\mcitedefaultmidpunct}
{\mcitedefaultendpunct}{\mcitedefaultseppunct}\relax
\EndOfBibitem
\bibitem[Volkov \emph{et~al.}(2007)Volkov, Bozhevolnyi, Devaux, Laluet, and
  Ebbesen]{Volkov2007a}
V.~S. Volkov, S.~I. Bozhevolnyi, E.~Devaux, J.-Y. Laluet and T.~W. Ebbesen,
  \emph{Nano Lett.}, 2007, \textbf{7}, 880--4\relax
\mciteBstWouldAddEndPuncttrue
\mciteSetBstMidEndSepPunct{\mcitedefaultmidpunct}
{\mcitedefaultendpunct}{\mcitedefaultseppunct}\relax
\EndOfBibitem
\bibitem[Nielsen \emph{et~al.}(2008)Nielsen, Fernandez-Cuesta, Boltasseva,
  Volkov, Bozhevolnyi, Klukowska, and Kristensen]{Nielsen2008a}
R.~B. Nielsen, I.~Fernandez-Cuesta, A.~Boltasseva, V.~S. Volkov, S.~I.
  Bozhevolnyi, A.~Klukowska and A.~Kristensen, \emph{Opt. Lett.}, 2008,
  \textbf{33}, 2800--2\relax
\mciteBstWouldAddEndPuncttrue
\mciteSetBstMidEndSepPunct{\mcitedefaultmidpunct}
{\mcitedefaultendpunct}{\mcitedefaultseppunct}\relax
\EndOfBibitem
\bibitem[Radko \emph{et~al.}(2011)Radko, Holmgaard, Han, Pedersen, and
  Bozhevolnyi]{Radko2011a}
I.~P. Radko, T.~Holmgaard, Z.~Han, K.~Pedersen and S.~I. Bozhevolnyi,
  \emph{Appl. Phys. Lett.}, 2011, \textbf{99}, 213109\relax
\mciteBstWouldAddEndPuncttrue
\mciteSetBstMidEndSepPunct{\mcitedefaultmidpunct}
{\mcitedefaultendpunct}{\mcitedefaultseppunct}\relax
\EndOfBibitem
\bibitem[S{\o}ndergaard \emph{et~al.}(2012)S{\o}ndergaard, Novikov, Holmgaard,
  Eriksen, Beermann, Han, Pedersen, and Bozhevolnyi]{Søndergaard2012}
T.~S{\o}ndergaard, S.~M. Novikov, T.~Holmgaard, R.~L. Eriksen, J.~Beermann,
  Z.~Han, K.~Pedersen and S.~I. Bozhevolnyi, \emph{Nat. Commun.}, 2012,
  \textbf{3}, 969\relax
\mciteBstWouldAddEndPuncttrue
\mciteSetBstMidEndSepPunct{\mcitedefaultmidpunct}
{\mcitedefaultendpunct}{\mcitedefaultseppunct}\relax
\EndOfBibitem
\bibitem[S{\o}ndergaard \emph{et~al.}(2010)S{\o}ndergaard, Bozhevolnyi,
  Beermann, Novikov, Devaux, and Ebbesen]{Søndergaard2010b}
T.~S{\o}ndergaard, S.~I. Bozhevolnyi, J.~Beermann, S.~M. Novikov, E.~Devaux and
  T.~W. Ebbesen, \emph{Nano Lett.}, 2010, \textbf{10}, 291--5\relax
\mciteBstWouldAddEndPuncttrue
\mciteSetBstMidEndSepPunct{\mcitedefaultmidpunct}
{\mcitedefaultendpunct}{\mcitedefaultseppunct}\relax
\EndOfBibitem
\bibitem[S{\o}ndergaard \emph{et~al.}(2010)S{\o}ndergaard, Bozhevolnyi,
  Novikov, Beermann, Devaux, and Ebbesen]{Søndergaard2010c}
T.~S{\o}ndergaard, S.~I. Bozhevolnyi, S.~M. Novikov, J.~Beermann, E.~Devaux and
  T.~W. Ebbesen, \emph{Nano Lett.}, 2010, \textbf{10}, 3123--8\relax
\mciteBstWouldAddEndPuncttrue
\mciteSetBstMidEndSepPunct{\mcitedefaultmidpunct}
{\mcitedefaultendpunct}{\mcitedefaultseppunct}\relax
\EndOfBibitem
\bibitem[Pile and Gramotnev(2004)]{Pile2004a}
D.~F.~P. Pile and D.~K. Gramotnev, \emph{Opt. Lett.}, 2004, \textbf{29},
  1069--71\relax
\mciteBstWouldAddEndPuncttrue
\mciteSetBstMidEndSepPunct{\mcitedefaultmidpunct}
{\mcitedefaultendpunct}{\mcitedefaultseppunct}\relax
\EndOfBibitem
\bibitem[Bozhevolnyi \emph{et~al.}(2005)Bozhevolnyi, Volkov, Devaux, and
  Ebbesen]{Bozhevolnyi2005a}
S.~Bozhevolnyi, V.~Volkov, E.~Devaux and T.~Ebbesen, \emph{Phys. Rev. Lett.},
  2005, \textbf{95}, 046802\relax
\mciteBstWouldAddEndPuncttrue
\mciteSetBstMidEndSepPunct{\mcitedefaultmidpunct}
{\mcitedefaultendpunct}{\mcitedefaultseppunct}\relax
\EndOfBibitem
\bibitem[Bozhevolnyi \emph{et~al.}(2007)Bozhevolnyi, Volkov, Devaux, Laluet,
  and Ebbesen]{Bozhevolnyi2007a}
S.~Bozhevolnyi, V.~Volkov, E.~Devaux, J.-Y. Laluet and T.~Ebbesen, \emph{Appl.
  Phys. A Mater. Sci. Process.}, 2007, \textbf{89}, 225--231\relax
\mciteBstWouldAddEndPuncttrue
\mciteSetBstMidEndSepPunct{\mcitedefaultmidpunct}
{\mcitedefaultendpunct}{\mcitedefaultseppunct}\relax
\EndOfBibitem
\bibitem[Lu and Maradudin(1990)]{Lu1990}
J.~Q. Lu and A.~A. Maradudin, \emph{Phys. Rev. B}, 1990, \textbf{42},
  11159--11165\relax
\mciteBstWouldAddEndPuncttrue
\mciteSetBstMidEndSepPunct{\mcitedefaultmidpunct}
{\mcitedefaultendpunct}{\mcitedefaultseppunct}\relax
\EndOfBibitem
\bibitem[Novikov and Maradudin(2002)]{Novikov2002a}
I.~Novikov and A.~Maradudin, \emph{Phys. Rev. B}, 2002, \textbf{66},
  035403\relax
\mciteBstWouldAddEndPuncttrue
\mciteSetBstMidEndSepPunct{\mcitedefaultmidpunct}
{\mcitedefaultendpunct}{\mcitedefaultseppunct}\relax
\EndOfBibitem
\bibitem[Bozhevolnyi(2006)]{Bozhevolnyi2006b}
S.~I. Bozhevolnyi, \emph{Opt. Express}, 2006, \textbf{14}, 9467--76\relax
\mciteBstWouldAddEndPuncttrue
\mciteSetBstMidEndSepPunct{\mcitedefaultmidpunct}
{\mcitedefaultendpunct}{\mcitedefaultseppunct}\relax
\EndOfBibitem
\bibitem[Vernon \emph{et~al.}(2008)Vernon, Gramotnev, and Pile]{Vernon2008a}
K.~C. Vernon, D.~K. Gramotnev and D.~F.~P. Pile, \emph{J. Appl. Phys.}, 2008,
  \textbf{103}, 034304\relax
\mciteBstWouldAddEndPuncttrue
\mciteSetBstMidEndSepPunct{\mcitedefaultmidpunct}
{\mcitedefaultendpunct}{\mcitedefaultseppunct}\relax
\EndOfBibitem
\bibitem[Bozhevolnyi and Jung(2008)]{Bozhevolnyi2008a}
S.~I. Bozhevolnyi and J.~Jung, \emph{Opt. Express}, 2008, \textbf{16},
  2676--84\relax
\mciteBstWouldAddEndPuncttrue
\mciteSetBstMidEndSepPunct{\mcitedefaultmidpunct}
{\mcitedefaultendpunct}{\mcitedefaultseppunct}\relax
\EndOfBibitem
\bibitem[Dintinger and Martin(2009)]{Dintinger2009a}
J.~Dintinger and O.~J.~F. Martin, \emph{Opt. Express}, 2009, \textbf{17},
  2364--74\relax
\mciteBstWouldAddEndPuncttrue
\mciteSetBstMidEndSepPunct{\mcitedefaultmidpunct}
{\mcitedefaultendpunct}{\mcitedefaultseppunct}\relax
\EndOfBibitem
\bibitem[Lee and Kim(2011)]{Lee2011a}
S.~Lee and S.~Kim, \emph{Opt. Express}, 2011, \textbf{19}, 9836--47\relax
\mciteBstWouldAddEndPuncttrue
\mciteSetBstMidEndSepPunct{\mcitedefaultmidpunct}
{\mcitedefaultendpunct}{\mcitedefaultseppunct}\relax
\EndOfBibitem
\bibitem[Zenin \emph{et~al.}(2012)Zenin, Volkov, Han, Bozhevolnyi, Devaux, and
  Ebbesen]{Zenin2012a}
V.~A. Zenin, V.~S. Volkov, Z.~Han, S.~I. Bozhevolnyi, E.~Devaux and T.~W.
  Ebbesen, \emph{Opt. Express}, 2012, \textbf{20}, 6124--34\relax
\mciteBstWouldAddEndPuncttrue
\mciteSetBstMidEndSepPunct{\mcitedefaultmidpunct}
{\mcitedefaultendpunct}{\mcitedefaultseppunct}\relax
\EndOfBibitem
\bibitem[Bian \emph{et~al.}(2013)Bian, Zheng, Zhao, Liu, Su, Liu, Zhu, and
  Zhou]{Bian2013a}
Y.~Bian, Z.~Zheng, X.~Zhao, L.~Liu, Y.~Su, J.~Liu, J.~Zhu and T.~Zhou, \emph{J.
  Opt.}, 2013, \textbf{15}, 055011\relax
\mciteBstWouldAddEndPuncttrue
\mciteSetBstMidEndSepPunct{\mcitedefaultmidpunct}
{\mcitedefaultendpunct}{\mcitedefaultseppunct}\relax
\EndOfBibitem
\bibitem[Burgos \emph{et~al.}(2014)Burgos, Lee, Feigenbaum, Briggs, and
  Atwater]{Burgos2014}
S.~P. Burgos, H.~W. Lee, E.~Feigenbaum, R.~M. Briggs and H.~A. Atwater,
  \emph{Nano Lett.}, 2014, \textbf{14}, 3284--3292\relax
\mciteBstWouldAddEndPuncttrue
\mciteSetBstMidEndSepPunct{\mcitedefaultmidpunct}
{\mcitedefaultendpunct}{\mcitedefaultseppunct}\relax
\EndOfBibitem
\bibitem[Vesseur \emph{et~al.}(2010)Vesseur, {Garc\'{\i}a de Abajo}, and
  Polman]{Vesseur2010}
E.~J.~R. Vesseur, F.~J. {Garc\'{\i}a de Abajo} and A.~Polman, \emph{Phys. Rev.
  B}, 2010, \textbf{82}, 165419\relax
\mciteBstWouldAddEndPuncttrue
\mciteSetBstMidEndSepPunct{\mcitedefaultmidpunct}
{\mcitedefaultendpunct}{\mcitedefaultseppunct}\relax
\EndOfBibitem
\bibitem[Mart\'{\i}n-Cano \emph{et~al.}(2010)Mart\'{\i}n-Cano,
  Mart\'{\i}n-Moreno, Garc\'{\i}a-Vidal, and Moreno]{Martin-Cano2010}
D.~Mart\'{\i}n-Cano, L.~Mart\'{\i}n-Moreno, F.~J. Garc\'{\i}a-Vidal and
  E.~Moreno, \emph{Nano Lett.}, 2010, \textbf{10}, 3129--34\relax
\mciteBstWouldAddEndPuncttrue
\mciteSetBstMidEndSepPunct{\mcitedefaultmidpunct}
{\mcitedefaultendpunct}{\mcitedefaultseppunct}\relax
\EndOfBibitem
\bibitem[Mart\'{\i}n-Cano \emph{et~al.}(2011)Mart\'{\i}n-Cano,
  Gonz\'{a}lez-Tudela, Mart\'{\i}n-Moreno, Garc\'{\i}a-Vidal, Tejedor, and
  Moreno]{Martin-Cano2011}
D.~Mart\'{\i}n-Cano, A.~Gonz\'{a}lez-Tudela, L.~Mart\'{\i}n-Moreno, F.~J.
  Garc\'{\i}a-Vidal, C.~Tejedor and E.~Moreno, \emph{Phys. Rev. B}, 2011,
  \textbf{84}, 235306\relax
\mciteBstWouldAddEndPuncttrue
\mciteSetBstMidEndSepPunct{\mcitedefaultmidpunct}
{\mcitedefaultendpunct}{\mcitedefaultseppunct}\relax
\EndOfBibitem
\bibitem[Gonzalez-Tudela \emph{et~al.}(2011)Gonzalez-Tudela, Mart\'{\i}n-Cano,
  Moreno, Martin-Moreno, Tejedor, and Garc\'{\i}a-Vidal]{Gonzalez-Tudela2011}
A.~Gonzalez-Tudela, D.~Mart\'{\i}n-Cano, E.~Moreno, L.~Martin-Moreno,
  C.~Tejedor and F.~J. Garc\'{\i}a-Vidal, \emph{Phys. Rev. Lett.}, 2011,
  \textbf{106}, 020501\relax
\mciteBstWouldAddEndPuncttrue
\mciteSetBstMidEndSepPunct{\mcitedefaultmidpunct}
{\mcitedefaultendpunct}{\mcitedefaultseppunct}\relax
\EndOfBibitem
\bibitem[Vernon \emph{et~al.}(2012)Vernon, Tischler, and Kurth]{Vernon2012a}
K.~C. Vernon, N.~Tischler and M.~L. Kurth, \emph{J. Appl. Phys.}, 2012,
  \textbf{111}, 064323\relax
\mciteBstWouldAddEndPuncttrue
\mciteSetBstMidEndSepPunct{\mcitedefaultmidpunct}
{\mcitedefaultendpunct}{\mcitedefaultseppunct}\relax
\EndOfBibitem
\bibitem[Beermann \emph{et~al.}(2013)Beermann, Eriksen, S{\o}ndergaard,
  Holmgaard, Pedersen, and Bozhevolnyi]{Beermann2013a}
J.~Beermann, R.~L. Eriksen, T.~S{\o}ndergaard, T.~Holmgaard, K.~Pedersen and
  S.~I. Bozhevolnyi, \emph{New J. Phys.}, 2013, \textbf{15}, 073007\relax
\mciteBstWouldAddEndPuncttrue
\mciteSetBstMidEndSepPunct{\mcitedefaultmidpunct}
{\mcitedefaultendpunct}{\mcitedefaultseppunct}\relax
\EndOfBibitem
\bibitem[S{\o}ndergaard and Bozhevolnyi(2013)]{Søndergaard2013a}
T.~S{\o}ndergaard and S.~I. Bozhevolnyi, \emph{New J. Phys.}, 2013,
  \textbf{15}, 013034\relax
\mciteBstWouldAddEndPuncttrue
\mciteSetBstMidEndSepPunct{\mcitedefaultmidpunct}
{\mcitedefaultendpunct}{\mcitedefaultseppunct}\relax
\EndOfBibitem
\bibitem[Beermann \emph{et~al.}(2014)Beermann, Eriksen, Holmgaard, Pedersen,
  and Bozhevolnyi]{Beermann2014}
J.~Beermann, R.~L. Eriksen, T.~Holmgaard, K.~Pedersen and S.~I. Bozhevolnyi,
  \emph{Sci. Rep.}, 2014, \textbf{4}, 6904\relax
\mciteBstWouldAddEndPuncttrue
\mciteSetBstMidEndSepPunct{\mcitedefaultmidpunct}
{\mcitedefaultendpunct}{\mcitedefaultseppunct}\relax
\EndOfBibitem
\bibitem[S{\o}ndergaard \emph{et~al.}(2011)S{\o}ndergaard, Bozhevolnyi,
  Beermann, Novikov, Devaux, and Ebbesen]{Søndergaard2011}
T.~S{\o}ndergaard, S.~I. Bozhevolnyi, J.~Beermann, S.~M. Novikov, E.~Devaux and
  T.~W. Ebbesen, \emph{J. Opt. Soc. Am. B}, 2011, \textbf{29}, 130--7\relax
\mciteBstWouldAddEndPuncttrue
\mciteSetBstMidEndSepPunct{\mcitedefaultmidpunct}
{\mcitedefaultendpunct}{\mcitedefaultseppunct}\relax
\EndOfBibitem
\bibitem[Zhang \emph{et~al.}(2011)Zhang, Ou, Papasimakis, Chen, Macdonald, and
  Zheludev]{Zhang2011}
J.~Zhang, J.-Y. Ou, N.~Papasimakis, Y.~Chen, K.~F. Macdonald and N.~I.
  Zheludev, \emph{Opt. Express}, 2011, \textbf{19}, 23279--85\relax
\mciteBstWouldAddEndPuncttrue
\mciteSetBstMidEndSepPunct{\mcitedefaultmidpunct}
{\mcitedefaultendpunct}{\mcitedefaultseppunct}\relax
\EndOfBibitem
\bibitem[Zhang \emph{et~al.}(2012)Zhang, Ou, MacDonald, and
  Zheludev]{Zhang2012b}
J.~Zhang, J.-Y. Ou, K.~F. MacDonald and N.~I. Zheludev, \emph{J. Opt.}, 2012,
  \textbf{14}, 114002\relax
\mciteBstWouldAddEndPuncttrue
\mciteSetBstMidEndSepPunct{\mcitedefaultmidpunct}
{\mcitedefaultendpunct}{\mcitedefaultseppunct}\relax
\EndOfBibitem
\bibitem[Rose \emph{et~al.}(2014)Rose, Wirth, Hatem, Ahmed, Burns, Naughton,
  and Kempa]{Rose2014}
A.~H. Rose, B.~M. Wirth, R.~E. Hatem, a.~P.~R. Ahmed, M.~J. Burns, M.~J.
  Naughton and K.~Kempa, \emph{Opt. Express}, 2014, \textbf{22},
  5228--5233\relax
\mciteBstWouldAddEndPuncttrue
\mciteSetBstMidEndSepPunct{\mcitedefaultmidpunct}
{\mcitedefaultendpunct}{\mcitedefaultseppunct}\relax
\EndOfBibitem
\bibitem[Odgaard \emph{et~al.}(2014)Odgaard, Laursen, and
  S{\o}ndergaard]{Odgaard2014}
M.~Odgaard, M.~G. Laursen and T.~S{\o}ndergaard, \emph{J. Opt. Soc. Am. B},
  2014, \textbf{31}, 1853--1860\relax
\mciteBstWouldAddEndPuncttrue
\mciteSetBstMidEndSepPunct{\mcitedefaultmidpunct}
{\mcitedefaultendpunct}{\mcitedefaultseppunct}\relax
\EndOfBibitem
\bibitem[Shalin \emph{et~al.}(2014)Shalin, Ginzburg, Belov, Kivshar, and
  Zayats]{Shalin2014a}
A.~S. Shalin, P.~Ginzburg, P.~A. Belov, Y.~S. Kivshar and A.~V. Zayats,
  \emph{Laser Photon. Rev.}, 2014, \textbf{8}, 131--136\relax
\mciteBstWouldAddEndPuncttrue
\mciteSetBstMidEndSepPunct{\mcitedefaultmidpunct}
{\mcitedefaultendpunct}{\mcitedefaultseppunct}\relax
\EndOfBibitem
\bibitem[Fernandez-Cuesta \emph{et~al.}(2007)Fernandez-Cuesta, Nielsen,
  Boltasseva, Borrisé, Pérez-Murano, and Kristensen]{Fernandez-Cuesta2007a}
I.~Fernandez-Cuesta, R.~B. Nielsen, A.~Boltasseva, X.~Borrisé,
  F.~Pérez-Murano and A.~Kristensen, \emph{J. Vac. Sci. Technol. B}, 2007,
  \textbf{25}, 2649--2653\relax
\mciteBstWouldAddEndPuncttrue
\mciteSetBstMidEndSepPunct{\mcitedefaultmidpunct}
{\mcitedefaultendpunct}{\mcitedefaultseppunct}\relax
\EndOfBibitem
\bibitem[Toscano \emph{et~al.}(2012)Toscano, Raza, Xiao, Wubs, Jauho,
  Bozhevolnyi, and Mortensen]{Toscano2012}
G.~Toscano, S.~Raza, S.~Xiao, M.~Wubs, A.-P. Jauho, S.~I. Bozhevolnyi and N.~A.
  Mortensen, \emph{Opt. Lett.}, 2012, \textbf{37}, 2538--40\relax
\mciteBstWouldAddEndPuncttrue
\mciteSetBstMidEndSepPunct{\mcitedefaultmidpunct}
{\mcitedefaultendpunct}{\mcitedefaultseppunct}\relax
\EndOfBibitem
\bibitem[Toscano \emph{et~al.}(2013)Toscano, Raza, Yan, Jeppesen, Xiao, Wubs,
  Jauho, Bozhevolnyi, and Mortensen]{Toscano2013}
G.~Toscano, S.~Raza, W.~Yan, C.~Jeppesen, S.~Xiao, M.~Wubs, A.-P. Jauho, S.~I.
  Bozhevolnyi and N.~A. Mortensen, \emph{Nanophotonics}, 2013, \textbf{2},
  161--166\relax
\mciteBstWouldAddEndPuncttrue
\mciteSetBstMidEndSepPunct{\mcitedefaultmidpunct}
{\mcitedefaultendpunct}{\mcitedefaultseppunct}\relax
\EndOfBibitem
\bibitem[Maier(2007)]{Maier2007}
S.~A. Maier, \emph{{Plasmonics: Fundamentals and Applications}}, Springer,
  Berlin, 2007\relax
\mciteBstWouldAddEndPuncttrue
\mciteSetBstMidEndSepPunct{\mcitedefaultmidpunct}
{\mcitedefaultendpunct}{\mcitedefaultseppunct}\relax
\EndOfBibitem
\bibitem[Noginov \emph{et~al.}(2008)Noginov, Podolskiy, Zhu, Mayy, Bahoura,
  Adegoke, Ritzo, and Reynolds]{Noginov2008}
M.~A. Noginov, V.~A. Podolskiy, G.~Zhu, M.~Mayy, M.~Bahoura, J.~A. Adegoke,
  B.~A. Ritzo and K.~Reynolds, \emph{Opt. Express}, 2008, \textbf{16},
  1385--92\relax
\mciteBstWouldAddEndPuncttrue
\mciteSetBstMidEndSepPunct{\mcitedefaultmidpunct}
{\mcitedefaultendpunct}{\mcitedefaultseppunct}\relax
\EndOfBibitem
\bibitem[{De Leon} and Berini(2010)]{Leon2010}
I.~{De Leon} and P.~Berini, \emph{Nature Photon.}, 2010, \textbf{4},
  382--7\relax
\mciteBstWouldAddEndPuncttrue
\mciteSetBstMidEndSepPunct{\mcitedefaultmidpunct}
{\mcitedefaultendpunct}{\mcitedefaultseppunct}\relax
\EndOfBibitem
\bibitem[Khurgin and Sun(2011)]{Khurgin2011}
J.~B. Khurgin and G.~Sun, \emph{Appl. Phys. Lett.}, 2011, \textbf{99},
  211106\relax
\mciteBstWouldAddEndPuncttrue
\mciteSetBstMidEndSepPunct{\mcitedefaultmidpunct}
{\mcitedefaultendpunct}{\mcitedefaultseppunct}\relax
\EndOfBibitem
\bibitem[Khurgin and Sun(2014)]{Khurgin2014}
J.~B. Khurgin and G.~Sun, \emph{Nature Photon.}, 2014, \textbf{8},
  468--473\relax
\mciteBstWouldAddEndPuncttrue
\mciteSetBstMidEndSepPunct{\mcitedefaultmidpunct}
{\mcitedefaultendpunct}{\mcitedefaultseppunct}\relax
\EndOfBibitem
\bibitem[Hirsch \emph{et~al.}(2003)Hirsch, Stafford, Bankson, Sershen, Rivera,
  Price, Hazle, Halas, and West]{Hirsch2003}
L.~R. Hirsch, R.~J. Stafford, J.~A. Bankson, S.~R. Sershen, B.~Rivera, R.~E.
  Price, J.~D. Hazle, N.~J. Halas and J.~L. West, \emph{Proc. Natl. Acad. Sci.
  USA}, 2003, \textbf{100}, 13549--54\relax
\mciteBstWouldAddEndPuncttrue
\mciteSetBstMidEndSepPunct{\mcitedefaultmidpunct}
{\mcitedefaultendpunct}{\mcitedefaultseppunct}\relax
\EndOfBibitem
\bibitem[Ndukaife \emph{et~al.}(2014)Ndukaife, Mishra, Guler, Nnanna, Wereley,
  and Boltasseva]{Ndukaife2014}
J.~C. Ndukaife, A.~Mishra, U.~Guler, A.~G.~A. Nnanna, S.~T. Wereley and
  A.~Boltasseva, \emph{ACS Nano}, 2014, \textbf{8}, 9035--9043\relax
\mciteBstWouldAddEndPuncttrue
\mciteSetBstMidEndSepPunct{\mcitedefaultmidpunct}
{\mcitedefaultendpunct}{\mcitedefaultseppunct}\relax
\EndOfBibitem
\bibitem[Bozhevolnyi(2009)]{Bozhevolnyi2009}
S.~I. Bozhevolnyi, \emph{{Plasmonic Nanoguides and Circuits}}, Pan Stanford,
  Singapore, 2009\relax
\mciteBstWouldAddEndPuncttrue
\mciteSetBstMidEndSepPunct{\mcitedefaultmidpunct}
{\mcitedefaultendpunct}{\mcitedefaultseppunct}\relax
\EndOfBibitem
\bibitem[Pitarke \emph{et~al.}(2007)Pitarke, Silkin, Chulkov, and
  Echenique]{Pitarke2007}
J.~M. Pitarke, V.~M. Silkin, E.~V. Chulkov and P.~M. Echenique, \emph{Rep.
  Prog. Phys.}, 2007, \textbf{70}, 1--87\relax
\mciteBstWouldAddEndPuncttrue
\mciteSetBstMidEndSepPunct{\mcitedefaultmidpunct}
{\mcitedefaultendpunct}{\mcitedefaultseppunct}\relax
\EndOfBibitem
\bibitem[Oulton \emph{et~al.}(2008)Oulton, Bartal, Pile, and
  Zhang]{Oulton2008a}
R.~F. Oulton, G.~Bartal, D.~F.~P. Pile and X.~Zhang, \emph{New J. Phys.}, 2008,
  \textbf{10}, 105018\relax
\mciteBstWouldAddEndPuncttrue
\mciteSetBstMidEndSepPunct{\mcitedefaultmidpunct}
{\mcitedefaultendpunct}{\mcitedefaultseppunct}\relax
\EndOfBibitem
\bibitem[Stockman(2011)]{Stockman2011}
M.~I. Stockman, \emph{Opt. Express}, 2011, \textbf{19}, 22029--22106\relax
\mciteBstWouldAddEndPuncttrue
\mciteSetBstMidEndSepPunct{\mcitedefaultmidpunct}
{\mcitedefaultendpunct}{\mcitedefaultseppunct}\relax
\EndOfBibitem
\bibitem[Dionne \emph{et~al.}(2006)Dionne, Lezec, and Atwater]{Dionne2006a}
J.~A. Dionne, H.~J. Lezec and H.~A. Atwater, \emph{Nano Lett.}, 2006,
  \textbf{6}, 1928--32\relax
\mciteBstWouldAddEndPuncttrue
\mciteSetBstMidEndSepPunct{\mcitedefaultmidpunct}
{\mcitedefaultendpunct}{\mcitedefaultseppunct}\relax
\EndOfBibitem
\bibitem[Pile and Gramotnev(2005)]{Pile2005}
D.~F.~P. Pile and D.~K. Gramotnev, \emph{Opt. Lett.}, 2005, \textbf{30},
  1186--8\relax
\mciteBstWouldAddEndPuncttrue
\mciteSetBstMidEndSepPunct{\mcitedefaultmidpunct}
{\mcitedefaultendpunct}{\mcitedefaultseppunct}\relax
\EndOfBibitem
\bibitem[Khurgin and Sun(2012)]{Khurgin2012}
J.~B. Khurgin and G.~Sun, \emph{Opt. Express}, 2012, \textbf{20},
  28717--23\relax
\mciteBstWouldAddEndPuncttrue
\mciteSetBstMidEndSepPunct{\mcitedefaultmidpunct}
{\mcitedefaultendpunct}{\mcitedefaultseppunct}\relax
\EndOfBibitem
\bibitem[Palik(1985)]{Palik}
E.~D. Palik, \emph{{Handbook of Optical Constants of Solids}}, Academic, San
  Diego, 1985\relax
\mciteBstWouldAddEndPuncttrue
\mciteSetBstMidEndSepPunct{\mcitedefaultmidpunct}
{\mcitedefaultendpunct}{\mcitedefaultseppunct}\relax
\EndOfBibitem
\bibitem[Zia \emph{et~al.}(2004)Zia, Selker, Catrysse, and Brongersma]{Zia2004}
R.~Zia, M.~D. Selker, P.~B. Catrysse and M.~L. Brongersma, \emph{J. Opt. Soc.
  Am. A}, 2004, \textbf{21}, 2442--6\relax
\mciteBstWouldAddEndPuncttrue
\mciteSetBstMidEndSepPunct{\mcitedefaultmidpunct}
{\mcitedefaultendpunct}{\mcitedefaultseppunct}\relax
\EndOfBibitem
\bibitem[Raza \emph{et~al.}(2013)Raza, Christensen, Wubs, Bozhevolnyi, and
  Mortensen]{Raza2013}
S.~Raza, T.~Christensen, M.~Wubs, S.~I. Bozhevolnyi and N.~A. Mortensen,
  \emph{Phys. Rev. B}, 2013, \textbf{88}, 115401\relax
\mciteBstWouldAddEndPuncttrue
\mciteSetBstMidEndSepPunct{\mcitedefaultmidpunct}
{\mcitedefaultendpunct}{\mcitedefaultseppunct}\relax
\EndOfBibitem
\bibitem[Bozhevolnyi and Nerkararyan(2009)]{Bozhevolnyi2009b}
S.~I. Bozhevolnyi and K.~V. Nerkararyan, \emph{Opt. Express}, 2009,
  \textbf{17}, 10327--34\relax
\mciteBstWouldAddEndPuncttrue
\mciteSetBstMidEndSepPunct{\mcitedefaultmidpunct}
{\mcitedefaultendpunct}{\mcitedefaultseppunct}\relax
\EndOfBibitem
\bibitem[Bozhevolnyi \emph{et~al.}(2010)Bozhevolnyi, Nerkararyan, and
  Hovsepyan]{Bozhevolnyi2010a}
S.~I. Bozhevolnyi, K.~V. Nerkararyan and S.~B. Hovsepyan, \emph{J. Contemp.
  Phys.}, 2010, \textbf{45}, 302--306\relax
\mciteBstWouldAddEndPuncttrue
\mciteSetBstMidEndSepPunct{\mcitedefaultmidpunct}
{\mcitedefaultendpunct}{\mcitedefaultseppunct}\relax
\EndOfBibitem
\bibitem[Gramotnev(2005)]{Gramotnev2005a}
D.~K. Gramotnev, \emph{J. Appl. Phys.}, 2005, \textbf{98}, 104302\relax
\mciteBstWouldAddEndPuncttrue
\mciteSetBstMidEndSepPunct{\mcitedefaultmidpunct}
{\mcitedefaultendpunct}{\mcitedefaultseppunct}\relax
\EndOfBibitem
\bibitem[Gramotnev and Vernon(2007)]{Gramotnev2007}
D.~Gramotnev and K.~Vernon, \emph{Appl. Phys. B}, 2007, \textbf{86},
  7--17\relax
\mciteBstWouldAddEndPuncttrue
\mciteSetBstMidEndSepPunct{\mcitedefaultmidpunct}
{\mcitedefaultendpunct}{\mcitedefaultseppunct}\relax
\EndOfBibitem
\bibitem[Vernon \emph{et~al.}(2007)Vernon, Gramotnev, and Pile]{Vernon2007a}
K.~C. Vernon, D.~K. Gramotnev and D.~F.~P. Pile, \emph{J. Appl. Phys.}, 2007,
  \textbf{101}, 104312\relax
\mciteBstWouldAddEndPuncttrue
\mciteSetBstMidEndSepPunct{\mcitedefaultmidpunct}
{\mcitedefaultendpunct}{\mcitedefaultseppunct}\relax
\EndOfBibitem
\bibitem[S{\o}ndergaard and Bozhevolnyi(2009)]{Søndergaard2009a}
T.~S{\o}ndergaard and S.~Bozhevolnyi, \emph{Phys. Rev. B}, 2009, \textbf{80},
  195407\relax
\mciteBstWouldAddEndPuncttrue
\mciteSetBstMidEndSepPunct{\mcitedefaultmidpunct}
{\mcitedefaultendpunct}{\mcitedefaultseppunct}\relax
\EndOfBibitem
\bibitem[Bozhevolnyi and Nerkararyan(2009)]{Bozhevolnyi2009a}
S.~I. Bozhevolnyi and K.~V. Nerkararyan, \emph{Opt. Lett.}, 2009, \textbf{34},
  2039--41\relax
\mciteBstWouldAddEndPuncttrue
\mciteSetBstMidEndSepPunct{\mcitedefaultmidpunct}
{\mcitedefaultendpunct}{\mcitedefaultseppunct}\relax
\EndOfBibitem
\bibitem[Aubry \emph{et~al.}(2010)Aubry, Lei, Fern\'{a}ndez-Dom\'{\i}nguez,
  Sonnefraud, Maier, and Pendry]{Aubry2010}
A.~Aubry, D.~Y. Lei, A.~I. Fern\'{a}ndez-Dom\'{\i}nguez, Y.~Sonnefraud, S.~A.
  Maier and J.~B. Pendry, \emph{Nano Lett.}, 2010, \textbf{10}, 2574--9\relax
\mciteBstWouldAddEndPuncttrue
\mciteSetBstMidEndSepPunct{\mcitedefaultmidpunct}
{\mcitedefaultendpunct}{\mcitedefaultseppunct}\relax
\EndOfBibitem
\bibitem[Fern\'{a}ndez-Dom\'{\i}nguez
  \emph{et~al.}(2010)Fern\'{a}ndez-Dom\'{\i}nguez, Maier, and
  Pendry]{Fernandez-Dominguez2010}
A.~I. Fern\'{a}ndez-Dom\'{\i}nguez, S.~A. Maier and J.~B. Pendry, \emph{Phys.
  Rev. Lett.}, 2010, \textbf{105}, 266807\relax
\mciteBstWouldAddEndPuncttrue
\mciteSetBstMidEndSepPunct{\mcitedefaultmidpunct}
{\mcitedefaultendpunct}{\mcitedefaultseppunct}\relax
\EndOfBibitem
\bibitem[Hocker and Burns(1977)]{Hocker1977}
G.~B. Hocker and W.~K. Burns, \emph{Appl. Opt.}, 1977, \textbf{16},
  113--8\relax
\mciteBstWouldAddEndPuncttrue
\mciteSetBstMidEndSepPunct{\mcitedefaultmidpunct}
{\mcitedefaultendpunct}{\mcitedefaultseppunct}\relax
\EndOfBibitem
\bibitem[Polemi \emph{et~al.}(2011)Polemi, Al\`{u}, and Engheta]{Polemi2011}
A.~Polemi, A.~Al\`{u} and N.~Engheta, \emph{IEEE Antenn. Wireless Propag.
  Lett.}, 2011, \textbf{10}, 199--202\relax
\mciteBstWouldAddEndPuncttrue
\mciteSetBstMidEndSepPunct{\mcitedefaultmidpunct}
{\mcitedefaultendpunct}{\mcitedefaultseppunct}\relax
\EndOfBibitem
\bibitem[Gramotnev and Pile(2004)]{Gramotnev2004a}
D.~K. Gramotnev and D.~F.~P. Pile, \emph{Appl. Phys. Lett.}, 2004, \textbf{85},
  6323\relax
\mciteBstWouldAddEndPuncttrue
\mciteSetBstMidEndSepPunct{\mcitedefaultmidpunct}
{\mcitedefaultendpunct}{\mcitedefaultseppunct}\relax
\EndOfBibitem
\bibitem[Yan and Qiu(2007)]{Yan2007a}
M.~Yan and M.~Qiu, \emph{J. Opt. Soc. Am. B}, 2007, \textbf{24}, 2333\relax
\mciteBstWouldAddEndPuncttrue
\mciteSetBstMidEndSepPunct{\mcitedefaultmidpunct}
{\mcitedefaultendpunct}{\mcitedefaultseppunct}\relax
\EndOfBibitem
\bibitem[Pile \emph{et~al.}(2005)Pile, Ogawa, Gramotnev, Okamoto, Haraguchi,
  Fukui, and Matsuo]{Pile2005a}
D.~F.~P. Pile, T.~Ogawa, D.~K. Gramotnev, T.~Okamoto, M.~Haraguchi, M.~Fukui
  and S.~Matsuo, \emph{Appl. Phys. Lett.}, 2005, \textbf{87}, 061106\relax
\mciteBstWouldAddEndPuncttrue
\mciteSetBstMidEndSepPunct{\mcitedefaultmidpunct}
{\mcitedefaultendpunct}{\mcitedefaultseppunct}\relax
\EndOfBibitem
\bibitem[Moreno \emph{et~al.}(2008)Moreno, Rodrigo, Bozhevolnyi,
  Mart\'{\i}n-Moreno, and Garc\'{\i}a-Vidal]{Moreno2008}
E.~Moreno, S.~Rodrigo, S.~Bozhevolnyi, L.~Mart\'{\i}n-Moreno and
  F.~Garc\'{\i}a-Vidal, \emph{Phys. Rev. Lett.}, 2008, \textbf{100},
  023901\relax
\mciteBstWouldAddEndPuncttrue
\mciteSetBstMidEndSepPunct{\mcitedefaultmidpunct}
{\mcitedefaultendpunct}{\mcitedefaultseppunct}\relax
\EndOfBibitem
\bibitem[Boltasseva \emph{et~al.}(2008)Boltasseva, Volkov, Nielsen, Moreno,
  Rodrigo, and Bozhevolnyi]{Boltasseva2008a}
A.~Boltasseva, V.~S. Volkov, R.~B. Nielsen, E.~Moreno, S.~G. Rodrigo and S.~I.
  Bozhevolnyi, \emph{Opt. Express}, 2008, \textbf{16}, 5252--60\relax
\mciteBstWouldAddEndPuncttrue
\mciteSetBstMidEndSepPunct{\mcitedefaultmidpunct}
{\mcitedefaultendpunct}{\mcitedefaultseppunct}\relax
\EndOfBibitem
\bibitem[Bian \emph{et~al.}(2011)Bian, Zheng, Liu, Liu, Zhu, and
  Zhou]{Bian2011a}
Y.~Bian, Z.~Zheng, Y.~Liu, J.~Liu, J.~Zhu and T.~Zhou, \emph{Opt. Express},
  2011, \textbf{19}, 22417--22\relax
\mciteBstWouldAddEndPuncttrue
\mciteSetBstMidEndSepPunct{\mcitedefaultmidpunct}
{\mcitedefaultendpunct}{\mcitedefaultseppunct}\relax
\EndOfBibitem
\bibitem[Srivastava and Kumar(2009)]{Srivastava2009a}
T.~Srivastava and A.~Kumar, \emph{J. Appl. Phys.}, 2009, \textbf{106},
  043104\relax
\mciteBstWouldAddEndPuncttrue
\mciteSetBstMidEndSepPunct{\mcitedefaultmidpunct}
{\mcitedefaultendpunct}{\mcitedefaultseppunct}\relax
\EndOfBibitem
\bibitem[Volkov \emph{et~al.}(2009)Volkov, Gosciniak, Bozhevolnyi, Rodrigo,
  Mart\'{\i}n-Moreno, Garc\'{\i}a-Vidal, Devaux, and Ebbesen]{Volkov2009a}
V.~S. Volkov, J.~Gosciniak, S.~I. Bozhevolnyi, S.~G. Rodrigo,
  L.~Mart\'{\i}n-Moreno, F.~J. Garc\'{\i}a-Vidal, E.~Devaux and T.~W. Ebbesen,
  \emph{New J. Phys.}, 2009, \textbf{11}, 113043\relax
\mciteBstWouldAddEndPuncttrue
\mciteSetBstMidEndSepPunct{\mcitedefaultmidpunct}
{\mcitedefaultendpunct}{\mcitedefaultseppunct}\relax
\EndOfBibitem
\bibitem[{De Angelis} \emph{et~al.}(2010){De Angelis}, Das, Candeloro, Patrini,
  Galli, Bek, Lazzarino, Maksymov, Liberale, Andreani, and {Di
  Fabrizio}]{DeAngelis2010}
F.~{De Angelis}, G.~Das, P.~Candeloro, M.~Patrini, M.~Galli, A.~Bek,
  M.~Lazzarino, I.~Maksymov, C.~Liberale, L.~C. Andreani and E.~{Di Fabrizio},
  \emph{Nature Nanotech.}, 2010, \textbf{5}, 67--72\relax
\mciteBstWouldAddEndPuncttrue
\mciteSetBstMidEndSepPunct{\mcitedefaultmidpunct}
{\mcitedefaultendpunct}{\mcitedefaultseppunct}\relax
\EndOfBibitem
\bibitem[Ozbay(2006)]{Ozbay2006}
E.~Ozbay, \emph{Science}, 2006, \textbf{311}, 189--93\relax
\mciteBstWouldAddEndPuncttrue
\mciteSetBstMidEndSepPunct{\mcitedefaultmidpunct}
{\mcitedefaultendpunct}{\mcitedefaultseppunct}\relax
\EndOfBibitem
\bibitem[Atwater(2007)]{Atwater2007}
H.~A. Atwater, \emph{Sci. Am.}, 2007, \textbf{296}, 56--63\relax
\mciteBstWouldAddEndPuncttrue
\mciteSetBstMidEndSepPunct{\mcitedefaultmidpunct}
{\mcitedefaultendpunct}{\mcitedefaultseppunct}\relax
\EndOfBibitem
\bibitem[Dicken \emph{et~al.}(2008)Dicken, Sweatlock, Pacifici, Lezec,
  Bhattacharya, and Atwater]{Dicken2008}
M.~J. Dicken, L.~A. Sweatlock, D.~Pacifici, H.~J. Lezec, K.~Bhattacharya and
  H.~A. Atwater, \emph{Nano Lett.}, 2008, \textbf{8}, 4048--4052\relax
\mciteBstWouldAddEndPuncttrue
\mciteSetBstMidEndSepPunct{\mcitedefaultmidpunct}
{\mcitedefaultendpunct}{\mcitedefaultseppunct}\relax
\EndOfBibitem
\bibitem[Cai \emph{et~al.}(2009)Cai, White, and Brongersma]{Cai2009a}
W.~Cai, J.~S. White and M.~L. Brongersma, \emph{Nano Lett.}, 2009, \textbf{9},
  4403--4411\relax
\mciteBstWouldAddEndPuncttrue
\mciteSetBstMidEndSepPunct{\mcitedefaultmidpunct}
{\mcitedefaultendpunct}{\mcitedefaultseppunct}\relax
\EndOfBibitem
\bibitem[Brongersma and Shalaev(2010)]{Brongersma2010a}
M.~L. Brongersma and V.~M. Shalaev, \emph{Science}, 2010, \textbf{328},
  440--1\relax
\mciteBstWouldAddEndPuncttrue
\mciteSetBstMidEndSepPunct{\mcitedefaultmidpunct}
{\mcitedefaultendpunct}{\mcitedefaultseppunct}\relax
\EndOfBibitem
\bibitem[Volkov \emph{et~al.}(2006)Volkov, Bozhevolnyi, Devaux, and
  Ebbesen]{Volkov2006}
V.~S. Volkov, S.~I. Bozhevolnyi, E.~Devaux and T.~W. Ebbesen, \emph{Appl. Phys.
  Lett.}, 2006, \textbf{89}, 143108\relax
\mciteBstWouldAddEndPuncttrue
\mciteSetBstMidEndSepPunct{\mcitedefaultmidpunct}
{\mcitedefaultendpunct}{\mcitedefaultseppunct}\relax
\EndOfBibitem
\bibitem[Volkov \emph{et~al.}(2006)Volkov, Bozhevolnyi, Devaux, and
  Ebbesen]{Volkov2006a}
V.~S. Volkov, S.~I. Bozhevolnyi, E.~Devaux and T.~W. Ebbesen, \emph{Opt.
  Express}, 2006, \textbf{14}, 4494\relax
\mciteBstWouldAddEndPuncttrue
\mciteSetBstMidEndSepPunct{\mcitedefaultmidpunct}
{\mcitedefaultendpunct}{\mcitedefaultseppunct}\relax
\EndOfBibitem
\bibitem[Burke \emph{et~al.}(1986)Burke, Stegeman, and Tamir]{Burke1986}
J.~J. Burke, G.~I. Stegeman and T.~Tamir, \emph{Phys. Rev. B}, 1986,
  \textbf{33}, 5186--5201\relax
\mciteBstWouldAddEndPuncttrue
\mciteSetBstMidEndSepPunct{\mcitedefaultmidpunct}
{\mcitedefaultendpunct}{\mcitedefaultseppunct}\relax
\EndOfBibitem
\bibitem[Sarid(1981)]{Sarid1981}
D.~Sarid, \emph{Phys. Rev. Lett.}, 1981, \textbf{47}, 1927--1930\relax
\mciteBstWouldAddEndPuncttrue
\mciteSetBstMidEndSepPunct{\mcitedefaultmidpunct}
{\mcitedefaultendpunct}{\mcitedefaultseppunct}\relax
\EndOfBibitem
\bibitem[Dionne \emph{et~al.}(2005)Dionne, Sweatlock, Atwater, and
  Polman]{Dionne2005}
J.~Dionne, L.~Sweatlock, H.~Atwater and A.~Polman, \emph{Phys. Rev. B}, 2005,
  \textbf{72}, 075405\relax
\mciteBstWouldAddEndPuncttrue
\mciteSetBstMidEndSepPunct{\mcitedefaultmidpunct}
{\mcitedefaultendpunct}{\mcitedefaultseppunct}\relax
\EndOfBibitem
\bibitem[Boltasseva \emph{et~al.}(2005)Boltasseva, Nikolajsen, Leosson, Kjaer,
  Larsen, and Bozhevolnyi]{Boltasseva2005}
A.~Boltasseva, T.~Nikolajsen, K.~Leosson, K.~Kjaer, M.~Larsen and
  S.~Bozhevolnyi, \emph{J. Lightwave Technol.}, 2005, \textbf{23},
  413--422\relax
\mciteBstWouldAddEndPuncttrue
\mciteSetBstMidEndSepPunct{\mcitedefaultmidpunct}
{\mcitedefaultendpunct}{\mcitedefaultseppunct}\relax
\EndOfBibitem
\bibitem[Wootters(1998)]{Wootters1998}
W.~K. Wootters, \emph{Phys. Rev. Lett.}, 1998, \textbf{80}, 2245--2248\relax
\mciteBstWouldAddEndPuncttrue
\mciteSetBstMidEndSepPunct{\mcitedefaultmidpunct}
{\mcitedefaultendpunct}{\mcitedefaultseppunct}\relax
\EndOfBibitem
\bibitem[Purcell(1946)]{Purcell1946}
E.~M. Purcell, \emph{Phys. Rev.}, 1946, \textbf{69}, 681\relax
\mciteBstWouldAddEndPuncttrue
\mciteSetBstMidEndSepPunct{\mcitedefaultmidpunct}
{\mcitedefaultendpunct}{\mcitedefaultseppunct}\relax
\EndOfBibitem
\bibitem[Huang \emph{et~al.}(2014)Huang, Seo, Sarmiento, Huo, Harris, and
  Brongersma]{Huang2014}
K.~C.~Y. Huang, M.-K. Seo, T.~Sarmiento, Y.~Huo, J.~S. Harris and M.~L.
  Brongersma, \emph{Nature Photon.}, 2014, \textbf{8}, 244--249\relax
\mciteBstWouldAddEndPuncttrue
\mciteSetBstMidEndSepPunct{\mcitedefaultmidpunct}
{\mcitedefaultendpunct}{\mcitedefaultseppunct}\relax
\EndOfBibitem
\bibitem[Pfund(1933)]{Pfund1933}
A.~H. Pfund, \emph{J. Opt. Soc. Am.}, 1933, \textbf{23}, 375--8\relax
\mciteBstWouldAddEndPuncttrue
\mciteSetBstMidEndSepPunct{\mcitedefaultmidpunct}
{\mcitedefaultendpunct}{\mcitedefaultseppunct}\relax
\EndOfBibitem
\bibitem[Vorobyev and Guo(2005)]{Vorobyev2005}
A.~Vorobyev and C.~Guo, \emph{Phys. Rev. B}, 2005, \textbf{72}, 195422\relax
\mciteBstWouldAddEndPuncttrue
\mciteSetBstMidEndSepPunct{\mcitedefaultmidpunct}
{\mcitedefaultendpunct}{\mcitedefaultseppunct}\relax
\EndOfBibitem
\bibitem[Vorobyev and Guo(2008)]{Vorobyev2008}
A.~Y. Vorobyev and C.~Guo, \emph{J. Appl. Phys.}, 2008, \textbf{104},
  053516\relax
\mciteBstWouldAddEndPuncttrue
\mciteSetBstMidEndSepPunct{\mcitedefaultmidpunct}
{\mcitedefaultendpunct}{\mcitedefaultseppunct}\relax
\EndOfBibitem
\bibitem[Verma \emph{et~al.}(2009)Verma, Ichimura, Yano, Saito, and
  Kawata]{Verma2009}
P.~Verma, T.~Ichimura, T.~Yano, Y.~Saito and S.~Kawata, \emph{Laser Photon.
  Rev.}, 2009, \textbf{4}, 548--561\relax
\mciteBstWouldAddEndPuncttrue
\mciteSetBstMidEndSepPunct{\mcitedefaultmidpunct}
{\mcitedefaultendpunct}{\mcitedefaultseppunct}\relax
\EndOfBibitem
\bibitem[Fernandez-Cuesta \emph{et~al.}(2009)Fernandez-Cuesta, Nielsen,
  Boltasseva, Borrise, Pérez-Murano, and Kristensen]{Fernandez-Cuesta2009}
I.~Fernandez-Cuesta, R.~B. Nielsen, A.~Boltasseva, X.~Borrise,
  F.~Pérez-Murano and A.~Kristensen, \emph{Appl. Phys. Lett.}, 2009,
  \textbf{95}, 203102\relax
\mciteBstWouldAddEndPuncttrue
\mciteSetBstMidEndSepPunct{\mcitedefaultmidpunct}
{\mcitedefaultendpunct}{\mcitedefaultseppunct}\relax
\EndOfBibitem
\bibitem[Choi \emph{et~al.}(2009)Choi, Pile, Nam, Bartal, and Zhang]{Choi2009}
H.~Choi, D.~F.~P. Pile, S.~Nam, G.~Bartal and X.~Zhang, \emph{Opt. Express},
  2009, \textbf{17}, 7519--7524\relax
\mciteBstWouldAddEndPuncttrue
\mciteSetBstMidEndSepPunct{\mcitedefaultmidpunct}
{\mcitedefaultendpunct}{\mcitedefaultseppunct}\relax
\EndOfBibitem
\bibitem[Melngailis(1987)]{Melngailis1987}
J.~Melngailis, \emph{J. Vac. Sci. Technol. B}, 1987, \textbf{5}, 469--495\relax
\mciteBstWouldAddEndPuncttrue
\mciteSetBstMidEndSepPunct{\mcitedefaultmidpunct}
{\mcitedefaultendpunct}{\mcitedefaultseppunct}\relax
\EndOfBibitem
\bibitem[Melli \emph{et~al.}(2013)Melli, Polyakov, Gargas, Huynh, Scipioni,
  Bao, Ogletree, Schuck, Cabrini, and Weber-Bargioni]{Melli2013}
M.~Melli, A.~Polyakov, D.~Gargas, C.~Huynh, L.~Scipioni, W.~Bao, D.~F.
  Ogletree, P.~J. Schuck, S.~Cabrini and A.~Weber-Bargioni, \emph{Nano Lett.},
  2013, \textbf{13}, 2687--91\relax
\mciteBstWouldAddEndPuncttrue
\mciteSetBstMidEndSepPunct{\mcitedefaultmidpunct}
{\mcitedefaultendpunct}{\mcitedefaultseppunct}\relax
\EndOfBibitem
\bibitem[Resnik \emph{et~al.}(2005)Resnik, Vrtacnik, Aljancic, Mozek, and
  Amon]{Resnik2005}
D.~Resnik, D.~Vrtacnik, U.~Aljancic, M.~Mozek and S.~Amon, \emph{J. Micromech.
  Microeng.}, 2005, \textbf{15}, 1174--1183\relax
\mciteBstWouldAddEndPuncttrue
\mciteSetBstMidEndSepPunct{\mcitedefaultmidpunct}
{\mcitedefaultendpunct}{\mcitedefaultseppunct}\relax
\EndOfBibitem
\bibitem[Madou(2002)]{Madou2002}
M.~J. Madou, \emph{{Fundamentals of Microfabrication: The Science of
  Miniaturization}}, CRC Press, London, 2002\relax
\mciteBstWouldAddEndPuncttrue
\mciteSetBstMidEndSepPunct{\mcitedefaultmidpunct}
{\mcitedefaultendpunct}{\mcitedefaultseppunct}\relax
\EndOfBibitem
\bibitem[Guo(2007)]{Guo2007a}
L.~J. Guo, \emph{Adv. Mater.}, 2007, \textbf{19}, 495--513\relax
\mciteBstWouldAddEndPuncttrue
\mciteSetBstMidEndSepPunct{\mcitedefaultmidpunct}
{\mcitedefaultendpunct}{\mcitedefaultseppunct}\relax
\EndOfBibitem
\bibitem[Schift(2008)]{Schift2008}
H.~Schift, \emph{J. Vac. Sci. Technol. B}, 2008, \textbf{26}, 458--480\relax
\mciteBstWouldAddEndPuncttrue
\mciteSetBstMidEndSepPunct{\mcitedefaultmidpunct}
{\mcitedefaultendpunct}{\mcitedefaultseppunct}\relax
\EndOfBibitem
\bibitem[Malyarchuk \emph{et~al.}(2005)Malyarchuk, Hua, Mack, Velasquez, White,
  Nuzzo, and Rogers]{Malyarchuk2005}
V.~Malyarchuk, F.~Hua, N.~H. Mack, V.~T. Velasquez, J.~O. White, R.~G. Nuzzo
  and J.~A. Rogers, \emph{Opt. Express}, 2005, \textbf{13}, 5669--5675\relax
\mciteBstWouldAddEndPuncttrue
\mciteSetBstMidEndSepPunct{\mcitedefaultmidpunct}
{\mcitedefaultendpunct}{\mcitedefaultseppunct}\relax
\EndOfBibitem
\bibitem[Buzzi \emph{et~al.}(2008)Buzzi, Robin, Callegari, and
  L\"{o}ffler]{Buzzi2008}
S.~Buzzi, F.~Robin, V.~Callegari and J.~F. L\"{o}ffler, \emph{Microelectron.
  Eng.}, 2008, \textbf{85}, 419--424\relax
\mciteBstWouldAddEndPuncttrue
\mciteSetBstMidEndSepPunct{\mcitedefaultmidpunct}
{\mcitedefaultendpunct}{\mcitedefaultseppunct}\relax
\EndOfBibitem
\bibitem[Radha \emph{et~al.}(2013)Radha, Lim, Saifullah, and
  Kulkarni]{Radha2013}
B.~Radha, S.~H. Lim, M.~S.~M. Saifullah and G.~U. Kulkarni, \emph{Sci. Rep.},
  2013, \textbf{3}, 1078\relax
\mciteBstWouldAddEndPuncttrue
\mciteSetBstMidEndSepPunct{\mcitedefaultmidpunct}
{\mcitedefaultendpunct}{\mcitedefaultseppunct}\relax
\EndOfBibitem
\bibitem[Bilenberg \emph{et~al.}(2005)Bilenberg, Jacobsen, Pastore, Nielsen,
  Enghoff, Jeppesen, Larsen, and Kristensen]{Bilenberg2005}
B.~Bilenberg, S.~Jacobsen, C.~Pastore, T.~Nielsen, S.~R. Enghoff, C.~Jeppesen,
  A.~V. Larsen and A.~Kristensen, \emph{Jpn. J. Appl. Phys.}, 2005,
  \textbf{44}, 5606--5608\relax
\mciteBstWouldAddEndPuncttrue
\mciteSetBstMidEndSepPunct{\mcitedefaultmidpunct}
{\mcitedefaultendpunct}{\mcitedefaultseppunct}\relax
\EndOfBibitem
\bibitem[Briggs \emph{et~al.}(2010)Briggs, Grandidier, Burgos, Feigenbaum, and
  Atwater]{Briggs2010}
R.~M. Briggs, J.~Grandidier, S.~P. Burgos, E.~Feigenbaum and H.~A. Atwater,
  \emph{Nano Lett.}, 2010, \textbf{10}, 4851--7\relax
\mciteBstWouldAddEndPuncttrue
\mciteSetBstMidEndSepPunct{\mcitedefaultmidpunct}
{\mcitedefaultendpunct}{\mcitedefaultseppunct}\relax
\EndOfBibitem
\bibitem[Kosako \emph{et~al.}(2010)Kosako, Kadoya, and Hofmann]{Kosako2010}
T.~Kosako, Y.~Kadoya and H.~F. Hofmann, \emph{Nature Photon.}, 2010,
  \textbf{4}, 312--315\relax
\mciteBstWouldAddEndPuncttrue
\mciteSetBstMidEndSepPunct{\mcitedefaultmidpunct}
{\mcitedefaultendpunct}{\mcitedefaultseppunct}\relax
\EndOfBibitem
\bibitem[Vercruysse \emph{et~al.}(2013)Vercruysse, Sonnefraud, Verellen, Fuchs,
  {Di Martino}, Lagae, Moshchalkov, Maier, and {Van Dorpe}]{Vercruysse2013}
D.~Vercruysse, Y.~Sonnefraud, N.~Verellen, F.~B. Fuchs, G.~{Di Martino},
  L.~Lagae, V.~V. Moshchalkov, S.~A. Maier and P.~{Van Dorpe}, \emph{Nano
  letters}, 2013,  3843--3849\relax
\mciteBstWouldAddEndPuncttrue
\mciteSetBstMidEndSepPunct{\mcitedefaultmidpunct}
{\mcitedefaultendpunct}{\mcitedefaultseppunct}\relax
\EndOfBibitem
\bibitem[Baron \emph{et~al.}(2011)Baron, Devaux, Rodier, Hugonin, Rousseau,
  Genet, Ebbesen, and Lalanne]{Baron2011}
A.~Baron, E.~Devaux, J.-C. Rodier, J.-P. Hugonin, E.~Rousseau, C.~Genet, T.~W.
  Ebbesen and P.~Lalanne, \emph{Nano Lett.}, 2011, \textbf{11}, 4207--12\relax
\mciteBstWouldAddEndPuncttrue
\mciteSetBstMidEndSepPunct{\mcitedefaultmidpunct}
{\mcitedefaultendpunct}{\mcitedefaultseppunct}\relax
\EndOfBibitem
\bibitem[Kinsey \emph{et~al.}(2015)Kinsey, Ferrera, Shalaev, and
  Boltasseva]{Kinsey2015}
N.~Kinsey, M.~Ferrera, V.~M. Shalaev and A.~Boltasseva, \emph{J. Opt. Soc. Am.
  A}, 2015, \textbf{32}, 121--142\relax
\mciteBstWouldAddEndPuncttrue
\mciteSetBstMidEndSepPunct{\mcitedefaultmidpunct}
{\mcitedefaultendpunct}{\mcitedefaultseppunct}\relax
\EndOfBibitem
\bibitem[Rotenberg and Kuipers(2014)]{Rotenberg2014}
N.~Rotenberg and L.~Kuipers, \emph{Nature Photon.}, 2014, \textbf{8},
  919--926\relax
\mciteBstWouldAddEndPuncttrue
\mciteSetBstMidEndSepPunct{\mcitedefaultmidpunct}
{\mcitedefaultendpunct}{\mcitedefaultseppunct}\relax
\EndOfBibitem
\bibitem[Novotny(2007)]{Novotny2007}
L.~Novotny, \emph{{The History of Near-field Optics}}, Elsevier, Amsterdam,
  2007, pp. 137--185\relax
\mciteBstWouldAddEndPuncttrue
\mciteSetBstMidEndSepPunct{\mcitedefaultmidpunct}
{\mcitedefaultendpunct}{\mcitedefaultseppunct}\relax
\EndOfBibitem
\bibitem[Gimzewski \emph{et~al.}(1988)Gimzewski, Reihl, Coombs, and
  Schlittler]{Gimzewski1988}
J.~K. Gimzewski, B.~Reihl, J.~H. Coombs and R.~R. Schlittler, \emph{Phys. B},
  1988, \textbf{72}, 497--501\relax
\mciteBstWouldAddEndPuncttrue
\mciteSetBstMidEndSepPunct{\mcitedefaultmidpunct}
{\mcitedefaultendpunct}{\mcitedefaultseppunct}\relax
\EndOfBibitem
\bibitem[Bharadwaj \emph{et~al.}(2011)Bharadwaj, Bouhelier, and
  Novotny]{Bharadwaj2011}
P.~Bharadwaj, A.~Bouhelier and L.~Novotny, \emph{Phys. Rev. Lett.}, 2011,
  \textbf{106}, 226802\relax
\mciteBstWouldAddEndPuncttrue
\mciteSetBstMidEndSepPunct{\mcitedefaultmidpunct}
{\mcitedefaultendpunct}{\mcitedefaultseppunct}\relax
\EndOfBibitem
\bibitem[{Garc\'{\i}a de Abajo}(2010)]{GarciadeAbajo2010}
F.~J. {Garc\'{\i}a de Abajo}, \emph{Rev. Mod. Phys.}, 2010, \textbf{82},
  209--275\relax
\mciteBstWouldAddEndPuncttrue
\mciteSetBstMidEndSepPunct{\mcitedefaultmidpunct}
{\mcitedefaultendpunct}{\mcitedefaultseppunct}\relax
\EndOfBibitem
\bibitem[Nicoletti \emph{et~al.}(2011)Nicoletti, Wubs, Mortensen, Sigle, van
  Aken, and Midgley]{Nicoletti2011}
O.~Nicoletti, M.~Wubs, N.~A. Mortensen, W.~Sigle, P.~A. van Aken and P.~A.
  Midgley, \emph{Opt. Express}, 2011, \textbf{19}, 9\relax
\mciteBstWouldAddEndPuncttrue
\mciteSetBstMidEndSepPunct{\mcitedefaultmidpunct}
{\mcitedefaultendpunct}{\mcitedefaultseppunct}\relax
\EndOfBibitem
\bibitem[Ritchie(1957)]{Ritchie1957}
R.~H. Ritchie, \emph{Phys. Rev.}, 1957, \textbf{106}, 874--881\relax
\mciteBstWouldAddEndPuncttrue
\mciteSetBstMidEndSepPunct{\mcitedefaultmidpunct}
{\mcitedefaultendpunct}{\mcitedefaultseppunct}\relax
\EndOfBibitem
\bibitem[Nelayah \emph{et~al.}(2007)Nelayah, Kociak, St\'{e}phan, {Garc\'{\i}a
  de Abajo}, Tenc\'{e}, Henrard, Taverna, Pastoriza-Santos, Liz-Marz\'{a}n, and
  Colliex]{Nelayah2007}
J.~Nelayah, M.~Kociak, O.~St\'{e}phan, F.~J. {Garc\'{\i}a de Abajo},
  M.~Tenc\'{e}, L.~Henrard, D.~Taverna, I.~Pastoriza-Santos, L.~M.
  Liz-Marz\'{a}n and C.~Colliex, \emph{Nature Phys.}, 2007, \textbf{3},
  348--353\relax
\mciteBstWouldAddEndPuncttrue
\mciteSetBstMidEndSepPunct{\mcitedefaultmidpunct}
{\mcitedefaultendpunct}{\mcitedefaultseppunct}\relax
\EndOfBibitem
\bibitem[Bosman \emph{et~al.}(2007)Bosman, Keast, Watanabe, Maaroof, and
  Cortie]{Bosman2007}
M.~Bosman, V.~J. Keast, M.~Watanabe, A.~I. Maaroof and M.~B. Cortie,
  \emph{Nanotechnology}, 2007, \textbf{18}, 165505\relax
\mciteBstWouldAddEndPuncttrue
\mciteSetBstMidEndSepPunct{\mcitedefaultmidpunct}
{\mcitedefaultendpunct}{\mcitedefaultseppunct}\relax
\EndOfBibitem
\bibitem[Koh \emph{et~al.}(2009)Koh, Bao, Khan, Smith, Kothleitner, Nordlander,
  Maier, and Mccomb]{Koh2009}
A.~L. Koh, K.~Bao, I.~Khan, W.~E. Smith, G.~Kothleitner, P.~Nordlander, S.~A.
  Maier and D.~W. Mccomb, \emph{ACS Nano}, 2009, \textbf{3}, 3015--3022\relax
\mciteBstWouldAddEndPuncttrue
\mciteSetBstMidEndSepPunct{\mcitedefaultmidpunct}
{\mcitedefaultendpunct}{\mcitedefaultseppunct}\relax
\EndOfBibitem
\bibitem[Scholl \emph{et~al.}(2012)Scholl, Koh, and Dionne]{Scholl2012}
J.~A. Scholl, A.~L. Koh and J.~A. Dionne, \emph{Nature}, 2012, \textbf{483},
  421--7\relax
\mciteBstWouldAddEndPuncttrue
\mciteSetBstMidEndSepPunct{\mcitedefaultmidpunct}
{\mcitedefaultendpunct}{\mcitedefaultseppunct}\relax
\EndOfBibitem
\bibitem[Raza \emph{et~al.}(2013)Raza, Stenger, Kadkhodazadeh, Fischer,
  Kostesha, Jauho, Burrows, Wubs, and Mortensen]{Raza2013a}
S.~Raza, N.~Stenger, S.~Kadkhodazadeh, S.~V. Fischer, N.~Kostesha, A.-P. Jauho,
  A.~Burrows, M.~Wubs and N.~A. Mortensen, \emph{Nanophotonics}, 2013,
  \textbf{2}, 131--138\relax
\mciteBstWouldAddEndPuncttrue
\mciteSetBstMidEndSepPunct{\mcitedefaultmidpunct}
{\mcitedefaultendpunct}{\mcitedefaultseppunct}\relax
\EndOfBibitem
\bibitem[Eisele \emph{et~al.}(2014)Eisele, Cocker, Huber, Plankl, Viti,
  Ercolani, Sorba, Vitiello, and Huber]{Eisele2014}
M.~Eisele, T.~L. Cocker, M.~A. Huber, M.~Plankl, L.~Viti, D.~Ercolani,
  L.~Sorba, M.~S. Vitiello and R.~Huber, \emph{Nature Photon.}, 2014,
  \textbf{8}, 841--845\relax
\mciteBstWouldAddEndPuncttrue
\mciteSetBstMidEndSepPunct{\mcitedefaultmidpunct}
{\mcitedefaultendpunct}{\mcitedefaultseppunct}\relax
\EndOfBibitem
\bibitem[Pettit \emph{et~al.}(1975)Pettit, Silcox, and Vincent]{Pettit1975}
R.~B. Pettit, J.~Silcox and R.~Vincent, \emph{Phys. Rev. B}, 1975, \textbf{11},
  3116--3123\relax
\mciteBstWouldAddEndPuncttrue
\mciteSetBstMidEndSepPunct{\mcitedefaultmidpunct}
{\mcitedefaultendpunct}{\mcitedefaultseppunct}\relax
\EndOfBibitem
\bibitem[Kuttge \emph{et~al.}(2009)Kuttge, Cai, {Garc\'{\i}a De Abajo}, and
  Polman]{Kuttge2009}
M.~Kuttge, W.~Cai, F.~J. {Garc\'{\i}a De Abajo} and A.~Polman, \emph{Phys. Rev.
  B}, 2009, \textbf{80}, 3--6\relax
\mciteBstWouldAddEndPuncttrue
\mciteSetBstMidEndSepPunct{\mcitedefaultmidpunct}
{\mcitedefaultendpunct}{\mcitedefaultseppunct}\relax
\EndOfBibitem
\bibitem[Khurgin(2015)]{Khurgin2015}
J.~B. Khurgin, \emph{Nature Nanotech.}, 2015, \textbf{10}, 2--6\relax
\mciteBstWouldAddEndPuncttrue
\mciteSetBstMidEndSepPunct{\mcitedefaultmidpunct}
{\mcitedefaultendpunct}{\mcitedefaultseppunct}\relax
\EndOfBibitem
\bibitem[Raza \emph{et~al.}(2014)Raza, Bozhevolnyi, Wubs, and
  Mortensen]{Raza2014}
S.~Raza, S.~I. Bozhevolnyi, M.~Wubs and N.~A. Mortensen,
  \emph{arXiv:1412.0942}, 2014,  1--19\relax
\mciteBstWouldAddEndPuncttrue
\mciteSetBstMidEndSepPunct{\mcitedefaultmidpunct}
{\mcitedefaultendpunct}{\mcitedefaultseppunct}\relax
\EndOfBibitem
\bibitem[Khurgin(2014)]{Manuscript2014}
J.~B. Khurgin, \emph{Faraday Discuss.}, 2014,  1--15\relax
\mciteBstWouldAddEndPuncttrue
\mciteSetBstMidEndSepPunct{\mcitedefaultmidpunct}
{\mcitedefaultendpunct}{\mcitedefaultseppunct}\relax
\EndOfBibitem
\bibitem[Mortensen \emph{et~al.}(2014)Mortensen, Raza, Wubs, S{\o}ndergaard,
  and Bozhevolnyi]{Mortensen2014}
N.~A. Mortensen, S.~Raza, M.~Wubs, T.~S{\o}ndergaard and S.~I. Bozhevolnyi,
  \emph{Nat. Commun.}, 2014, \textbf{5}, 3809\relax
\mciteBstWouldAddEndPuncttrue
\mciteSetBstMidEndSepPunct{\mcitedefaultmidpunct}
{\mcitedefaultendpunct}{\mcitedefaultseppunct}\relax
\EndOfBibitem
\bibitem[Garc\'{\i}a-Vidal and Pendry(1996)]{Garcia-Vidal1996}
F.~J. Garc\'{\i}a-Vidal and J.~B. Pendry, \emph{Phys. Rev. Lett.}, 1996,
  \textbf{77}, 1163--1166\relax
\mciteBstWouldAddEndPuncttrue
\mciteSetBstMidEndSepPunct{\mcitedefaultmidpunct}
{\mcitedefaultendpunct}{\mcitedefaultseppunct}\relax
\EndOfBibitem
\bibitem[Moskovits(1985)]{Moskovits1985}
M.~Moskovits, \emph{Rev. Mod. Phys.}, 1985, \textbf{57}, 783--828\relax
\mciteBstWouldAddEndPuncttrue
\mciteSetBstMidEndSepPunct{\mcitedefaultmidpunct}
{\mcitedefaultendpunct}{\mcitedefaultseppunct}\relax
\EndOfBibitem
\bibitem[Xiao \emph{et~al.}(2008)Xiao, Mortensen, and Jauho]{Xiao2008a}
S.~Xiao, N.~A. Mortensen and A.-P. Jauho, \emph{J. Eur. Opt. Soc, Rapid Publ.},
  2008, \textbf{3}, 08022\relax
\mciteBstWouldAddEndPuncttrue
\mciteSetBstMidEndSepPunct{\mcitedefaultmidpunct}
{\mcitedefaultendpunct}{\mcitedefaultseppunct}\relax
\EndOfBibitem
\bibitem[Zhao \emph{et~al.}(2014)Zhao, Liu, Lei, and Chai]{Zhao2014}
Y.~Zhao, X.~Liu, D.~Y. Lei and Y.~Chai, \emph{Nanoscale}, 2014, \textbf{6},
  1311--7\relax
\mciteBstWouldAddEndPuncttrue
\mciteSetBstMidEndSepPunct{\mcitedefaultmidpunct}
{\mcitedefaultendpunct}{\mcitedefaultseppunct}\relax
\EndOfBibitem
\end{mcitethebibliography}
\bibliographystyle{rsc} 
}

\end{document}